 \documentclass[aps, twocolumn, showpacs, amsmath]{revtex4-2}
\usepackage{dcolumn}
\usepackage{bm}
\usepackage{graphicx}
\usepackage{color}
\usepackage{hyperref}
\usepackage{amsthm}
\usepackage{amssymb}

\begin{document}
\def\be{\begin{equation}}
\def\ee{\end{equation}}

\def\bc{\begin{center}}
\def\ec{\end{center}}
\def\bea{\begin{eqnarray}}
\def\eea{\end{eqnarray}}
\newcommand{\avg}[1]{\langle{#1}\rangle}
\newcommand{\Avg}[1]{\left\langle{#1}\right\rangle}

\def\ie{\textit{i.e.}}
\def\etal{\textit{et al.}}
\def\m{\vec{m}}
\def\G{\mathcal{G}}

\newcommand{\davide}[1]{{\bf\color{blue}#1}}
\newcommand{\gin}[1]{{\bf\color{green}#1}}

\title{Pairwise correlations of global times in one-dimensional Brownian motion under stochastic resetting}
\author{Yihao Wang}
\author{Hanshuang Chen}\email{chenhshf@ahu.edu.cn}
\affiliation{School of Physics and Optoelectronic Engineering, Anhui University, Hefei 230601, China}
\begin{abstract}
Brownian motion with stochastic resetting—a process combining standard diffusion with random returns to a fixed position—has emerged as a powerful framework with applications spanning statistical physics, chemical kinetics, biology, and finance. In this study, we investigate the mutual correlations among three global characteristic times for  one-dimensional resetting Brownian motion $x(\tau)$ over the interval $\tau \in \left[ 0, t\right] $: the occupation time $t_o$ spent on the positive semi-axis, the time $t_m$ at which $x(\tau)$ attains its global maximum, and the last-passage time $t_{\ell}$ when the process crosses the origin. For the process starting from the origin and undergoing Poissonian resetting back to the origin, we analytically compute the pairwise joint distributions of these three times (in the Laplace domain) and derive their pairwise correlation coefficients. Our results reveal that these global times display rich correlations, with a non-trivial dependence on the resetting rate $r$. Specifically, we find that (i) While $t_{o}$ and $t_{\ell}^m$ are uncorrelated for any positive integer $m$, $t_{o}^2$ and $t_{\ell}^m$ display  anti-correlation; (ii) A positive correlation exists between $t_{o}$ and $t_{m}$, which decays toward zero following a logarithmically corrected power-law way with an exponent of $-2$ as $r \to \infty$; (iii) The correlation between $t_{m}$ and $t_{\ell}$ shifts from positive to negative as $r$ increases. All analytical predictions are validated by extensive numerical simulations.
\end{abstract}
\maketitle

\section{Introduction}
Brownian motion—along with its discrete-time counterpart, the random walk—stands as the simplest yet most ubiquitous stochastic process \cite{feller1971introduction,morters2010brownian,klafter2011first}, with widespread applications across physics and chemistry \cite{mazo2008brownian}, electrical engineering \cite{doyle1984random}, economics \cite{baz2004financial}, and social sciences \cite{schweitzer2003brownian}. Despite their conceptual simplicity, analytically computations of trajectory-based observables remains highly non-trivial, owing to the strong temporal correlations in the particle’s position \cite{majumdar2007brownian}. As early as 1940, L\'evy discovered the well-known arcsine law \cite{Levy1940ArcsineLaw}. The law states that three times depending on a global trajectory of a one-dimensional Brownian motion $x(\tau)$—starting from the origin over the interval $\tau \in \left[ 0, t\right] $—obey the same distribution. These global times are: the occupation time $t_o$ spent on the positive semi-axis, the time $t_m$ at which $x(\tau)$ attains its global maximum, and the last time $t_{\ell}$ that the process crosses the origin. The common distribution is given by \cite{feller1971introduction,majumdar2007brownian}
\begin{eqnarray}\label{eq0.0}
P_0(t_c|t)=\frac{1}{\pi \sqrt{t_c(t-t_c)}},
\end{eqnarray}
where the subscript $c$ denotes $o$, $m$ or $\ell$. The name of the law stems from the fact that the cumulative distribution, which reads $F(z)=\int_{0}^{z} dt_c P_0(t_c|t)=\frac{2}{\pi} {\rm{arcsin}}(\sqrt{\frac{z}{t}})$. This finding is non-trivial, as the U-shaped distribution in Eq.(\ref{eq0.0}) is counterintuitive: 
the average value $\langle t_c\rangle=t/2$ corresponds to the minimum of the distribution, i.e., the less probable outcome, whereas values close to the extrema $t_c=0$ and $t_c=t$ are far more likely.

Extensions of the arcsine law have been explored across diverse systems. Occupation time statistics appear in practical systems, including phase ordering dynamics in magnetic systems \cite{dornic1998large,drouffe1998stationary,baldassarri1999statistics} and financial time series \cite{bouchaud2003theory}. Experimentally, it has been investigated in colloidal
quantum dots \cite{brokmann2003statistical,stefani2009beyond}, stochastic thermodynamics \cite{barato2018arcsine,dey2022experimental}, and coherently driven optical resonators \cite{ramesh2024arcsine}. Theoretically, it has been extensively studied in numerous systems, including Brownian motion with drift \cite{nyawo2017minimal,nyawo2018dynamical}, Gaussian stationary process \cite{majumdar2002large,ehrhardt2004persistence}, fractional Brownian motion \cite{sadhu2018generalized,sadhu2021functionals}, run-and-tumble particle \cite{SinghArcsinelaws_RTP,bressloff2020occupation,mukherjee2024large}, random acceleration process \cite{boutcheng2016occupation}, renewal processes \cite{godreche2001statistics,PhysRevLett.107.170601}, trap model \cite{burov2007occupation}, continuous-time random walk \cite{bel2005weak,barkai2006residence,del2025generalized}, heterogeneous diffusion \cite{singh2022extreme}, diffusion in a disordered potential \cite{PhysRevLett.89.060601,PhysRevE.73.051102} or under confinement \cite{grebenkov2007residence,kaldasch2022stiffness,kay2023extreme,huang2024extremal}, diffusion under stochastic resetting \cite{den2019properties,smith2023striking,yan2023breakdown,tazbierski2025arcsine}, noninteracting Brownian gas \cite{burenev2024occupation}, and noncrossing Brownian particles \cite{mukherjee2023dynamical}.  
The maximum of a stochastic process and its corresponding time $t_m$ fall within the domain of extreme-value statistics \cite{schehr2010extreme,majumdar2020extreme,majumdar2024statistics}, with their statistics closely related to first-passage properties of stochastic processes \cite{redner2001guide}. This topic has also garnered significant attention, with studies covering constrained Brownian motion \cite{majumdar2008time,randon2007distribution}, random acceleration process \cite{majumdar2010time}, run-and-tumble processes \cite{SinghArcsinelaws_RTP}, stochastic resetting \cite{singh2021extremal,mori2021distribution,mori2022time,huang2024extreme,guo2023extremal,guo2024extremal,huang2024extreme}, among others \cite{majumdar2010hitting,sadhu2021functionals,singh2022extreme,del2025generalized,huang2024extremal}.  Notably, the statistics of $t_m$ have found applications in convex hull problems \cite{PhysRevLett.103.140602,dumonteil2013spatial,chupeau2015convex,PhysRevE.103.022135,majumdar2021mean} and in determining  whether a stationary process is in equilibrium  \cite{mori2021distribution,mori2022time}. The last-passage time $t_{\ell}$ has also been studied for diffusion in external potentials \cite{bao2006last,comtet2020last} or in the presence of boundaries \cite{kay2023extreme,huang2024extremal}, fractional Brownian motion \cite{sadhu2018generalized,sadhu2021functionals}, inhomogeneous environments \cite{singh2022extreme,del2025generalized},  permutation-generated random walks \cite{fang2021arcsine}, run-and-tumble process \cite{SinghArcsinelaws_RTP}, and resetting systems \cite{tazbierski2025arcsine}. Applications include electronic ring oscillators \cite{leung2004novel,robson2014truly}, fission investigations \cite{bao2004determination}, and developing Monte Carlo methods for capacitance calculations in flat/spherical surface systems \cite{hwang2006last,yu2021last}.

While the marginal distributions of these three global times $t_o$, $t_m$ and $t_{\ell}$ have been studied across various stochastic systems, their mutual correlations—encoded in pairwise joint distributions—have received far less attention. Until recently, Hartmann and Majumdar exactly computed the pairwise joint distributions of these three times for standard Brownian motion (SBM), showing that the distributions differ markedly from one another and exhibit rather rich and non-trivial correlations between these times \cite{hartmann2025exact}. As pointed out in \cite{hartmann2025exact}, the joint distributions may hold potential implications for reducing costs in tracking animal movement—such as that of tigers inhabiting mixed habitats of deep woods and open grasslands—using high-resolution satellite footage. However, despite the crucial importance of precisely computing joint distributions between global observables, this challenging task remains largely unexplored in current literature, with very limited existing results available. Another instance involves the exact computation of the joint distribution of 
$t_m$ (the time of the global maximum) and $t_{\min}$
(the time of the global minimum) \cite{mori2019time,mori2020distribution}. The time difference between $t_m$ and  $t_{\min}$ is valuable for stock investors to maximize their profits; it has also been applied to studying the positional difference between the minimum and maximum heights of a fluctuating interface. In this work, we present a compelling new example—resetting Brownian motion (RBM) \cite{evans2011diffusion}—that enables the successful analytical computation of pairwise correlations between the three global times: $t_o$, $t_m$, and $t_{\ell}$.

Stochastic resetting describes a renewal process wherein dynamics are stochastically interrupted and then restarted from the beginning. This topic has recently garnered significant attention, owing to its broad applications in search problems \cite{kusmierz2014first,kusmierz2015optimal,PhysRevLett.112.240601,chechkin2018random,JPA2022.55.274005}, the optimization of computer algorithms \cite{montanari2002optimizing,keidar2025adaptive}, and biophysics \cite{reuveni2014role,rotbart2015michaelis} (see \cite{evans2020stochastic,Gupta2022Review} for two recent reviews). Evans and Majumdar investigated a paradigmatic model of one-dimensional Brownian motion: the particle is instantaneously reset to its initial position at a constant rate, with free diffusion between consecutive resets. Such resetting induces diverse intriguing phenomena, including a nonequilibrium stationary state with non-Gaussian positional fluctuations, and a finite mean first-passage time to a target—one that can be minimized by tuning the resetting rate \cite{evans2011diffusion}. Extensions in this field include spatially \cite{evans2011diffusion2} or temporally \cite{pal2016diffusion,NJP2016.18.033006,PhysRevE.93.060102,PhysRevE.96.012126,PhysRevE.100.032110,chen2022random} dependent resetting rates, higher dimensions \cite{Evans2014_Reset_Highd}, complex geometries/networks \cite{Christou2015,PhysRevE.99.032123,PhysRevResearch.2.033027,BressloffJSTAT2021,PhysRevE.105.034109,PhysRevE.101.062147,huang2021random,JSM2022.053201}, non-instantaneous resetting \cite{EvansJPA2018,PalNJP2019,PhysRevE.101.052130,GuptaJPA2020,mercado2020intermittent,radice2021one,santra2021brownian}, external potentials \cite{GuptaJPA2020,pal2015diffusion,ahmad2019first,ray2020diffusion}, and other Brownian variants such as run-and-tumble particles \cite{evans2018run,santra2020run,bressloff2020occupation}, active particles \cite{scacchi2018mean,kumar2020active}, and constrained Brownian particles \cite{de2022optimal}, among others \cite{basu2019symmetric,wang2021time,vinod2022time,vinod2022nonergodicity,wang2022restoring,barkai2023ergodic}. The resetting has also triggered other intriguing phenomena, such as so-called ``resetting transition'' \cite{PhysRevLett.116.170601,pal2017first,JPA2022.55.021001}, and has found its applications in other fields including stochastic thermodynamics \cite{fuchs2016stochastic,pal2017integral,gupta2020work,mori2023entropy}, record statistics \cite{JStatMech2022.063202,JPA2022.55.034002,kumar2023universal,smith2023striking}, optimal control theory \cite{de2023resetting}, fluctuating interface \cite{gupta2014fluctuating}, and single-particle experiments \cite{tal2020experimental,besga2020optimal,faisant2021optimal}.

In the present work, we focus on the pairwise correlations among the three global times $t_o$, $t_m$ and $t_{\ell}$ (as mentioned earlier) for one-dimensional Brownian motion under Poissonian resetting. Specifically, the particle starts from the origin and is reset instantaneously to the origin at a constant rate $r$. The particle undergoes a free diffusion between consecutive resets. While the marginal distributions of these three global times have already been derived \cite{den2019properties,yan2023breakdown,singh2021extremal,majumdar2021mean,tazbierski2025arcsine} (with the marginal distribution of $t_m$ obtained solely in Laplace space \cite{singh2021extremal}), to the best of our knowledge, their pairwise joint distributions have not yet been reported. Using the Feynman-Kac method \cite{majumdar2007brownian} involved with a renewal formalism \cite{chechkin2018random,evans2020stochastic} and an $\epsilon$-path decomposition method \cite{majumdar2004exact,majumdar2005airy}, we analytically obtain the bivariate joint distributions of $t_o$, $t_m$ and $t_{\ell}$ in Laplace space. These quantities facilitate the computation of correlations between the three times. Our results reveal that these correlations exhibit a complex, non-trivial dependence on the parameter $r$, manifested through three key observations: (i) While $t_{o}$ and $t_{\ell}^m$ (for any positive integer $m$) remain uncorrelated, $t_{o}^2$ and $t_{\ell}^m$ exhibit a significant negative correlation; (ii) A positive correlation emerges between $t_{o}$ and $t_{m}$, where the correlation coefficient decreases toward zero as $r \to \infty$; (iii) Notably, the correlation between $t_{m}$ and $t_{\ell}$ undergoes a sign change from positive to negative as $r$ increases. Finally, we conduct extensive numerical simulations for validating our analytical predictions, demonstrating excellent agreement between them.

\section{Model and OBSERVABLES OF INTEREST}
We consider a one-dimensional resetting Brownian motion (RBM) starting from $x(0)=x_0$. The particle's position $x(t)$ is reset to $x_r$ at random times following a Poisson process with a constant rate $r$. Within a time interval $dt$, the position $x(t)$ evolves  according to the stochastic Langevin dynamics \cite{evans2020stochastic}:
\begin{eqnarray}\label{eq1}
{x}( {t + dt} ) = \left\{ \begin{array}{llc}
{x}( t ) + \sqrt {2 D } {\xi }(t) dt , &  {\rm{with}} \, {\rm{prob.}}  & 1-r dt,  \\
{x}_r, &   {\rm{with}} \, {\rm{prob.}}   & r dt,  \\ 
\end{array}  \right. 
\end{eqnarray}
where $D$ is the diffusion coefficient, and ${\xi }(t)$ is a Gaussian white noise with zero mean $\langle {\xi (t)} \rangle  = 0$ and delta correlator $\langle {\xi (t)}{\xi (t')} \rangle  = \delta ( {t - t'} )$.

\begin{figure}
	\centerline{\includegraphics*[width=1.0\columnwidth]{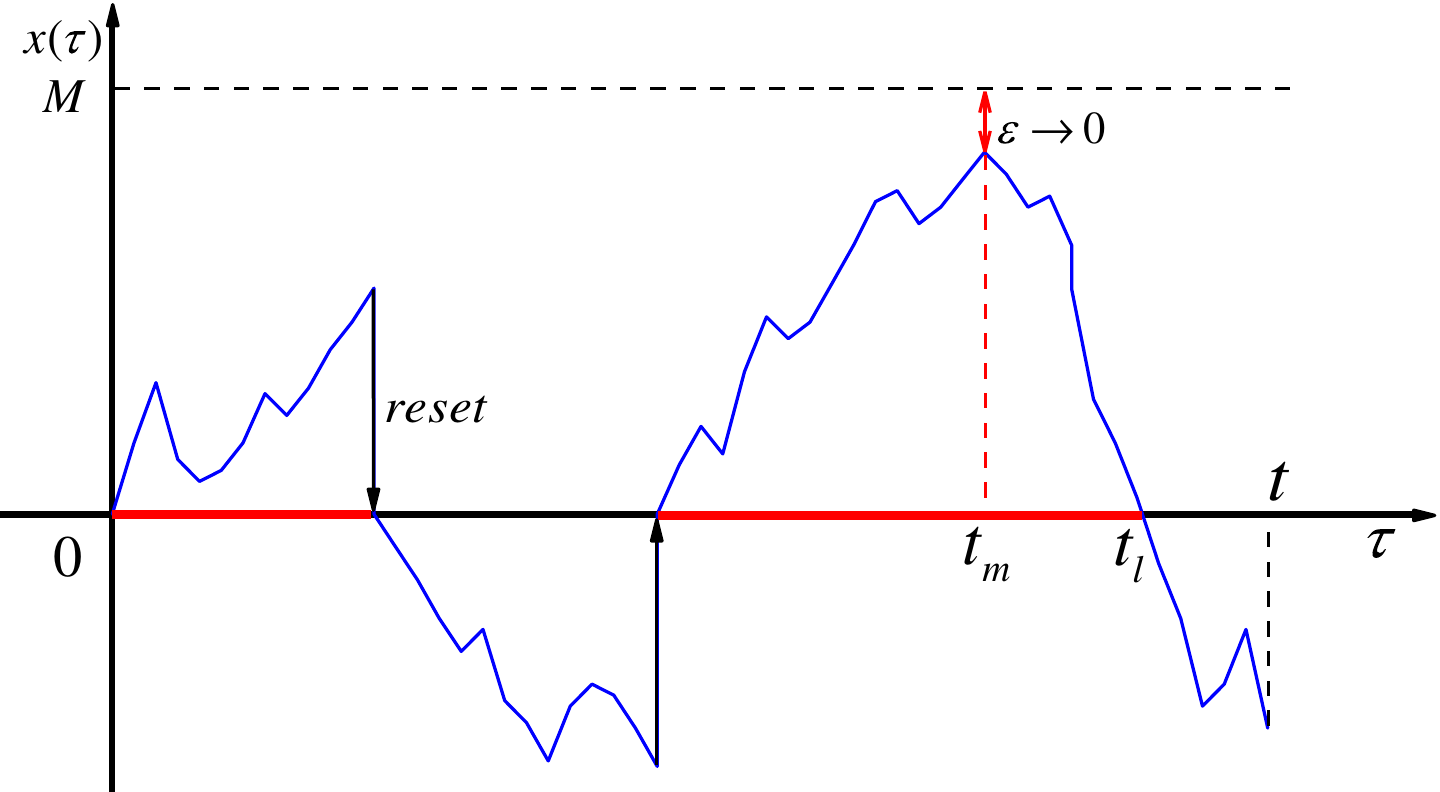}}
	\caption{A schematic plot of one-dimensional resetting Brownian motion $x(\tau)$ over a duration $t$. The particle starts from $x(0)=0$ and is reset to the origin at a constant rate $r$. The particle's position reaches its maximum $M$ at time $t_m$.  Also shown are the last-passage time $t_{\ell}$, and the occupation time $t_o$ corresponding the sum of the red intervals on the time axis where $x(\tau)>0$.   \label{fig_schematic_timeseries}}
\end{figure}

In this work, we focus on three classical observables of RBM over a global time interval: the last time $t_{\ell}$ the process crosses the origin, the occupation time $t_o$ spent on the positive (or negative) semi-axis, and the time $t_m$ at which the process reaches its maximum position (see Fig.\ref{fig_schematic_timeseries} for a demonstration). Their precise definitions are as follows. 

(i) The last crossing time $t_{\ell}$ of the origin is defined as
\begin{eqnarray}\label{eq1.1}
{t_{\ell}} = \mathop {\sup }\limits_\tau  \left\{ {\tau  \in \left[ {0,t} \right]: x( \tau  ) = 0} \right\}.
\end{eqnarray}

(ii) The occupation time $t_o$ on the positive semi-axis is defined as
\begin{eqnarray}\label{eq1.2}
{t_o} = \int_0^t {\theta \left[ {x( \tau  )} \right]} d\tau,
\end{eqnarray}
where $\theta(x)$ is the standard Heaviside function, with $\theta(x)=1$ for $x>0$ and $\theta(x)=0$ for $x<0$.

(iii) The time $t_m$ corresponds to the moment when the process attains its maximum $M$. Thus, $M$ is defined as
\begin{eqnarray}\label{eq1.3}
M = \mathop {\sup }\limits_{\tau  \in \left[ {0,t} \right]} x\left( \tau  \right),
\end{eqnarray}
and $t_m$ is the argument at which this supremum is achieved, i.e., $t_m={\rm{argsup}}_{\tau  \in \left[ {0,t} \right]} x(\tau)$.

More specifically, we consider here the case where the particle's initial position coincides with the resetting position, i.e., $x_0=x_r=0$. Previous studies have demonstrated that the marginal distributions of the three global times are entirely distinct from one another—markedly different from standard Brownian motion, where these distributions coincide and are governed by the arcsine law in Eq.(\ref{eq0.0}). The marginal distribution of $t_{\ell}$ was first obtained in Ref.\cite{tazbierski2025arcsine}. For completeness, we rederive this distribution using a more concise method, as detailed in Appendix \ref{app_tl_rbm}. In the presence of resetting, the arcsine law no longer hold, and the distribution of $t_{\ell}$ lose its symmetry about $t/2$. Notably, $t_{\ell}$ exhibits a higher probability of taking larger values, which aligns with our intuition. To compute the marginal distribution of $t_o$, the Feynman-Kac method \cite{majumdar2007brownian} combined with a renewal formalism for Poissonian resetting \cite{chechkin2018random,evans2020stochastic} was employed  \cite{den2019properties,yan2023breakdown}. Interestingly, the distribution of $t_o$ remains symmetric about $t/2$ for any resetting rate $r$, ensuring that its mean is always $t/2$. However, as $r$ increases, the distribution undergoes a shape transition from a $U$-shaped to a $W$-shaped profile  \cite{yan2023breakdown}. To derive the distribution of $t_m$, one first calculates an extended distribution $P(M,t_m)$ via an $\epsilon$-path decomposition method. The distribution of $t_m$ is then obtained by integrating $P(M,t_m)$  over $M$. This method was originally  proposed in Refs.\cite{majumdar2004exact,majumdar2005airy} and then used in numerous other related problems \cite{majumdar2008time,mori2019time,mori2020distribution,majumdar2021mean,mori2021distribution,mori2022time,huang2024extremal,guo2023extremal,guo2024extremal}. Since the propagator for RBM with an absorbing wall is difficult to obtain in real space, 
only the distribution of $t_m$ was obtained in the Laplace domain, from which the first and second moments of $t_m$ can be derived explicitly \cite{singh2021extremal}. Both numerical simulations and asymptotic analysis indicated that the symmetry breaks down in the presence of resetting \cite{singh2021extremal}.

In the present work, our goal is to derive the mutual joint distributions of the three global times: $t_{\ell}$, $t_o$ and $t_m$. We will derive these joint distributions using the Feynman-Kac method combined with a renewal formalism and an $\epsilon$-path decomposition method. These joint distributions are obtained exclusively in Laplace space, as some quantities about RBM are only available in this space. Nevertheless, it is valuable to derive the joint moments of the three global times and thereby compute the correlations between them.

\section{Joint distribution of $t_{\ell}$ and $t_o$}

\begin{figure}
	\centerline{\includegraphics*[width=1.0\columnwidth]{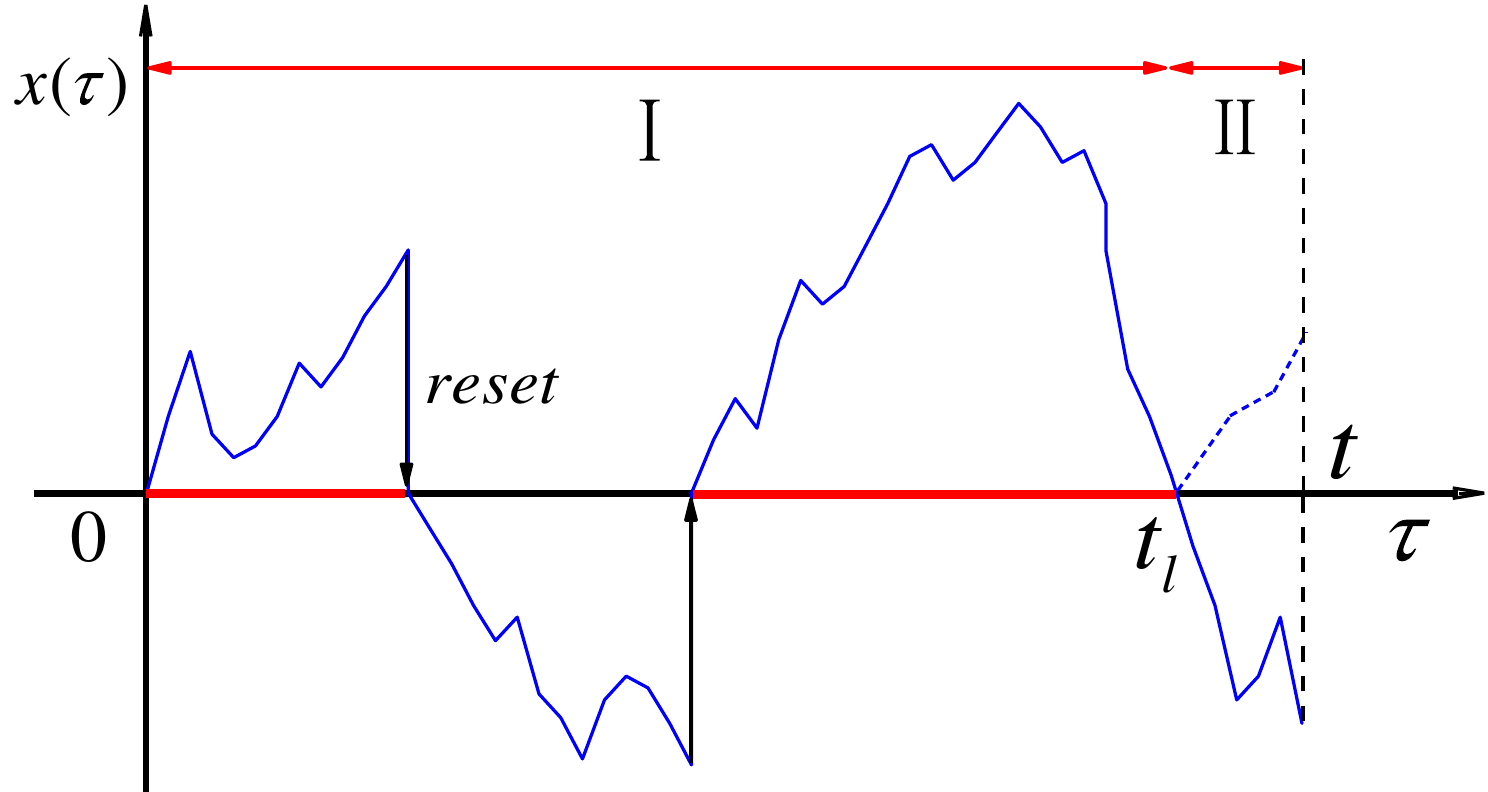}}
	\caption{A schematic plot of one-dimensional resetting Brownian motion $x(\tau)$ over a duration $t$. The particle starts from $x(0)=0$ (resetting to the origin at a constant rate $r$), and passes through the origin for the last time before $t$ at time $t_{\ell}$. Solid red lines show the intervals during which $x(\tau)>0$. The intervals $\left[ 0, t_{\ell}\right]$ and $\left[ t_{\ell}, t\right] $ are marked respectively I and II. \label{fig_schematic_totl}}
\end{figure}

To compute the joint distribution $P_r(t_{\ell},t_o|t)$ of the last-passage time $t_{\ell}$ and the occupation time $t_o$, we first express it as $P_r(t_{\ell},t_o|t)=P_r(t_o|t_{\ell},t)P_r(t_{\ell}|t)$. Here, $P_r(t_o|t_{\ell},t)$ denotes the conditional distribution of $t_o$ given $t_{\ell}$ and $t$, while $P_r(t_{\ell}|t)$ is the marginal distribution of $t_{\ell}$, given in Eq.(\ref{eq8.1}) (see appendix \ref{app_tl_rbm} for the detailed derivations). For fixed of $t_{\ell}$ and $t$, the occupation time $t_o$ can be decomposed into two parts, $t_o=t_o^I+t_o^{II}$. The first part, $t_o^I$, comes from the time interval $\left[ 0,t_{\ell} \right] $ (see Fig.\ref{fig_schematic_totl} for an illustration). In this interval, the particle starts from the origin at time 0 and end at the origin at time $t_{\ell}$, forming a resetting Brownian bridge \cite{de2022optimal}. The statistics of an occupation time for such a bridge are derived in the appendix \ref{app_to_rbb}. The second part, $t_o^{II}$, applies to the interval $\left[ t_{\ell}, t \right] $. During this period, the process either remains entirely above the origin or entirely below it, each with equal probability 
1/2. Thus, the distribution of $t_{o}^{II}$ can be simply written as, $P(t_{o}^{II}|t_{\ell},t)=\frac{1}{2}\delta(t_{o}^{II})+\frac{1}{2}\delta(t_{o}^{II}-t+t_{\ell})$. Since $t_{o}^{I}$ and $t_{o}^{II}$ are independent, the conditional distribution of $t_o$ can be given by their convolution, 
\begin{eqnarray}\label{eq11.1}
{P_r}\left( {{t_o}|{t_{\ell}},t} \right) &=& \int_0^t {dt_o^I} P_r^{bridge}\left( {t_o^I|{t_{\ell}}} \right)P\left( {{t_o} - t_o^I|{t_{\ell}},t} \right) \nonumber \\&=& \frac{1}{2}P_r^{bridge}\left( {{t_o}|{t_{\ell}}} \right) + \frac{1}{2}P_r^{bridge}\left( {{t_o} + {t_{\ell}} - t|{t_{\ell}}} \right). \nonumber \\
\end{eqnarray} 
Hence, we can write down the joint distribution of $t_{\ell}$ and $t_o$,
\begin{eqnarray}\label{eq11.2}
{P_r}( {{t_{\ell}},{t_o}|t} ) &=& {P_r}( {{t_o}|{t_{\ell}},t} ){P_r}( {{t_{\ell}}|t} ) \nonumber \\&=& \frac{1}{2}P_r^{bridge}( {{t_o}|{t_{\ell}}} ){P_r}( {{t_{\ell}}|t} ) \nonumber \\&+& \frac{1}{2}P_r^{bridge}( {{t_o} + {t_{\ell}} -t  |{t_{\ell}}} ){P_r}( {{t_{\ell}}|t} ).
\end{eqnarray}

\textbf{Correlation between $t_{\ell}$ and $t_o$}: Using the joint distribution in Eq.(\ref{eq11.2}), one can compute in principle the joint moments of $t_{\ell}$ and $t_o$
\begin{widetext}
\begin{eqnarray}\label{eq11.3}
{\left\langle {t_{\ell}^mt_o^n} |t \right\rangle _r} &=& \int_0^t {d{t_{\ell}}} \int_0^t {d{t_o}} {P_r}\left( {{t_{\ell}},{t_o}|t} \right)t_{\ell}^mt_o^n  \nonumber \\&=& \frac{1}{2}\int_0^t {d{t_{\ell}}} t_{\ell}^m{P_r}\left( {{t_{\ell}}|t} \right)\int_0^{{t_{\ell}}} {d{t_o}} t_o^nP_r^{bridge}\left( {{t_o}|{t_{\ell}}} \right) + \frac{1}{2}\int_0^t {d{t_{\ell}}} t_{\ell}^m{P_r}\left( {{t_{\ell}}|t} \right)\int_{t - {t_{\ell}}}^t {d{t_o}} t_o^nP_r^{bridge}\left( {{t_o+t_{\ell}-t}|{t_{\ell}}} \right) \nonumber \\&=& \frac{1}{2}\int_0^t {d{t_{\ell}}} t_{\ell}^m{P_r}\left( {{t_{\ell}}|t} \right)\left\langle { {t_o^n} |{t_{\ell}}} \right\rangle _r^{bridge} + \frac{1}{2}\int_0^t {d{t_{\ell}}} t_{\ell}^m{P_r}\left( {{t_{\ell}}|t} \right)\int_0^{{t_{\ell}}} {d{{t'}_o}} {\left( {{{t'}_o} + t - {t_{\ell}}} \right)^n}P_r^{bridge}\left( {{{t'}_o}|{t_{\ell}}} \right).
\end{eqnarray} 
\end{widetext}

For the special case $n=1$ (with $m$ being any non-negative integer), Eq.(\ref{eq11.3}) simplifies to
\begin{eqnarray}\label{eq11.4}
{\left\langle {t_{\ell}^m{t_o}}|t \right\rangle _r} &=& \frac{1}{4}\int_0^t {d{t_{\ell}}} t_{\ell}^{m + 1}{P_r}\left( {{t_{\ell}}|t} \right) \nonumber \\ &+& \frac{1}{2}\int_0^t {d{t_{\ell}}} t_{\ell}^m{P_r}\left( {{t_{\ell}}|t} \right)\left( {t - \frac{{{t_{\ell}}}}{2}} \right) \nonumber \\ &=& \frac{t}{2}{\left\langle { {t_{\ell}^m} |t} \right\rangle _r},
\end{eqnarray} 
where we have used the result of $\left\langle { {{t_o}} |{t_{\ell}}} \right\rangle _r^{bridge} = {t_{\ell}}/2$ from Eq.(\ref{eq10.13a}). Applying the known result, ${\left\langle { {{t_o}} |t} \right\rangle _r} = t/2$ (see Eq.(27a) in Ref.\cite{yan2023breakdown}), to Eq.(\ref{eq11.4}), we obtain the correlation between $t_{\ell}^m$ and $t_o$, 
\begin{eqnarray}\label{eq11.5}
{c_r}\left( {t_{\ell}^m;{t_o}} \right) \triangleq {\left\langle {t_{\ell}^m{t_o}|t} \right\rangle _r} - {\left\langle {t_{\ell}^m|t} \right\rangle _r}{\left\langle {{t_o}|t} \right\rangle _r} = 0.
\end{eqnarray} 
This implies no correlation between $t_{\ell}^m$ and $t_o$, regardless of resetting.

\begin{figure}
	\centerline{\includegraphics*[width=1.0\columnwidth]{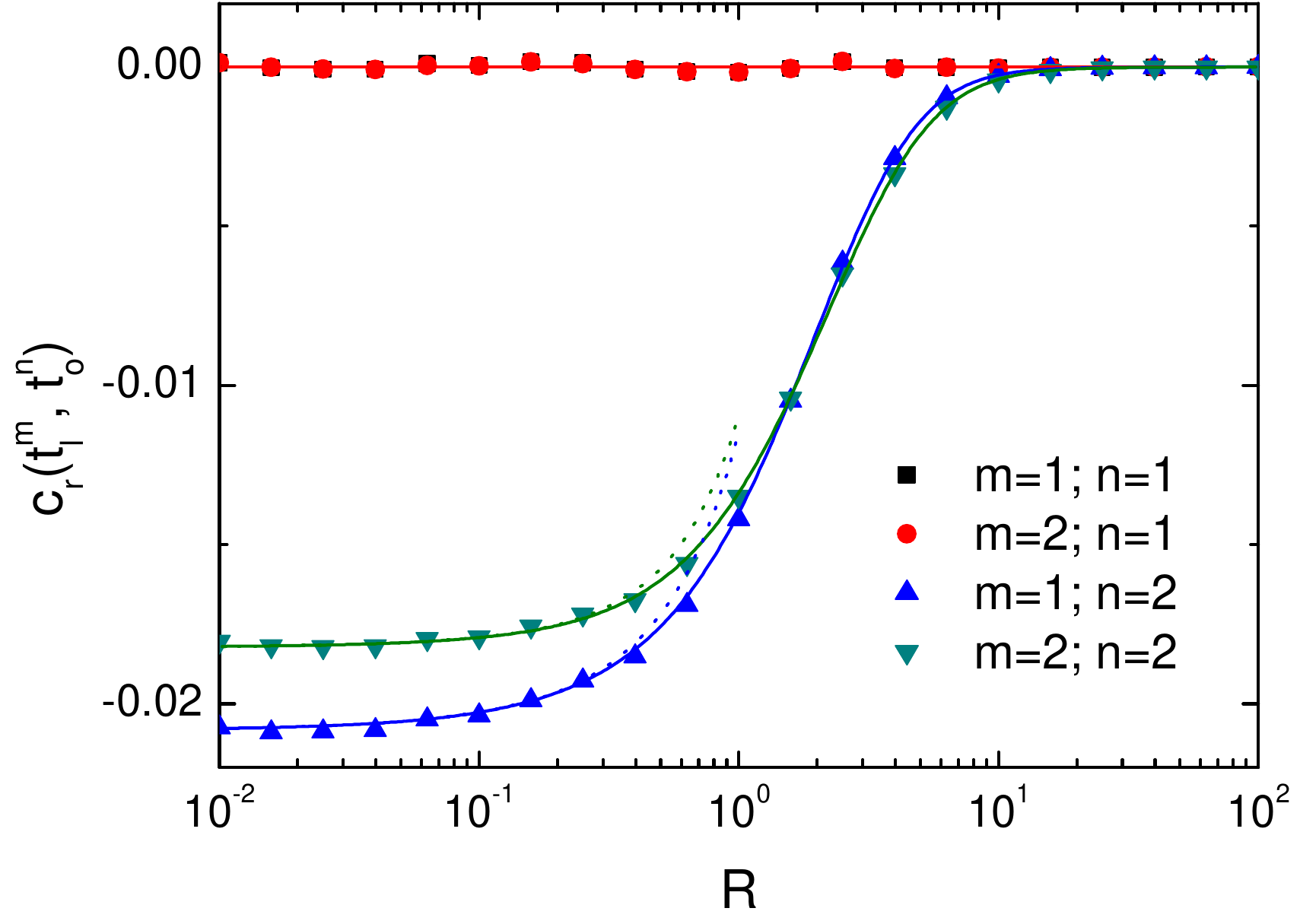}}
	\caption{Correlation coefficients between $t_{\ell}^m$ and $t_{o}^n$ as a function of the resetting rate $r$, for four parameter sets: $(m,n)=(1,1) $, $(2,1)$, $(1,2)$ and $(2,2)$. Solid lines and symbols represent the theoretical and numerical results, respectively. Two dotted lines indicate asymptotic behaviors of $\tilde{c}_r(\tilde{t}_{\ell}^m; \tilde{t}_o^n )$ for $R \ll 1$, see Eqs.(\ref{eq11.7.1}) and (\ref{eq11.7.2}).  \label{fig_corrtotl}}
\end{figure}

For the higher moments of $t_o$, it can turn out to be correlated to $t_{\ell}$. However, it is hard to obtain analytical expressions about the correlations. Here, we focus on the simplest case, i.e., $n=2$. Using the rescaled variables, $\tilde{t}_{\ell,o}=t_{\ell,o}/t$, we have 
\begin{eqnarray}\label{eq11.6}
\tilde{c}_r(\tilde{t}_{\ell}^m; \tilde{t}_o^2 )&=& \frac{1}{t^{m+2}}\left( {\langle {t_{\ell}^mt_o^2|t} \rangle _r} - {\langle {t_{\ell}^m|t} \rangle _r}{\langle {t_o^2|t} \rangle _r}\right)  \nonumber \\&=& \int_0^1 {d{\tilde{t}_{\ell}}} {\tilde{t}_{\ell}^m}\langle {\tilde{t}_o^2} \rangle_r^{bridge}{\tilde{P}_r}( {{\tilde{t}_{\ell}}} ) \nonumber \\ &+ & \frac{{1 - {e^{ - R}} - R + {R^2}}}{{4{R^2}}}{\langle { {\tilde{t}_{\ell}^m} } \rangle _r} - \frac{1}{2}{\langle { {\tilde{t}_{\ell}^{m + 1}} } \rangle _r},
\end{eqnarray} 
where we have used the known result on the second moment of occupation time for RBM (see Eq.(27b) in Ref.\cite{yan2023breakdown}), ${\langle {\tilde{t}_o^2} \rangle _r} = ( {{e^{ - R}} + R + {R^2} - 1} )/( {4{R^2}} )$, and $R=rt$ is the average number of resets over time $t$.

For $R=0$ without resetting, Eq.(\ref{eq11.6}) can be evaluated analytically, 
\begin{eqnarray}\label{eq11.7}
\tilde{c}_0(\tilde{t}_{\ell}^m; \tilde{t}_o^2 )& =& \frac{1}{3}{\langle { {\tilde{t}_{\ell}^{m + 2}} } \rangle _0} + \frac{{{1}}}{8}{\langle { {\tilde{t}_{\ell}^m} } \rangle _0} - \frac{1}{2}{\langle {{\tilde{t}_{\ell}^{m + 1}} } \rangle _0}   \nonumber \\& =&  - \frac{{m\left( {m + 5} \right)\left( {2m} \right)!}}{{6\left( {2m} \right)!!\left( {{\rm{2}}m + 4} \right)!!}} ,
\end{eqnarray} 
where we have utilized ${\langle { {\tilde{t}_{\ell}^m} } \rangle _0} = ( {2m} )!/[ {( {2m} )!!} ]^2$ from the arcsine law for $t_{\ell}$. Notably, $\tilde{c}_0(\tilde{t}_{\ell}; \tilde{t}_o^2 )<0$ for all $m>0$, indicating anti-correlation between $t_{\ell}$ and $t_o^2$.
For the first three values of $m$, we find $\tilde{c}_0(\tilde{t}_{\ell}; \tilde{t}_o^2 )=-\frac{1}{48}$, $\tilde{c}_0(\tilde{t}_{\ell}; \tilde{t}_o^2 )=-\frac{7}{384}$, and $\tilde{c}_0(\tilde{t}_{\ell}; \tilde{t}_o^2 )=-\frac{1}{64}$, consistent with the results in Ref.\cite{hartmann2025exact} (where general expressions for ${c_0}( {t_{\ell}^m;t_o^2} )$ were not reported).

For $R \gtrsim 0$, we can express $\tilde{c}_0(\tilde{t}_{\ell}; \tilde{t}_o^2 )$ in terms of its first two leading-order expansion in $r$,
\begin{eqnarray}\label{eq11.7.1}
\tilde{c}_r(\tilde{t}_{\ell}^m; \tilde{t}_o^2 )&=&  \tilde{c}_0(\tilde{t}_{\ell}^m; \tilde{t}_o^2 )+R \Delta_1 \tilde{c}_r(\tilde{t}_{\ell}^m; \tilde{t}_o^2 ) \nonumber \\ &+&	R^2 \Delta_2 \tilde{c}_r(\tilde{t}_{\ell}^m; \tilde{t}_o^2 ) +o(R^3),
\end{eqnarray}
where
\begin{subequations}\label{eq11.7.2}
	\begin{align}
\Delta_1 \tilde{c}_r (\tilde{t}_{\ell}^m; \tilde{t}_o^2 )&=\frac{(7-m)(2m)!}{24 \cdot 2^{2m} (m-1)! (m+3)!} , \\	
\Delta_2 \tilde{c}_r (\tilde{t}_{\ell}^m; \tilde{t}_o^2 )&=\frac{{m( {526 + 2237m + 554{m^2} + 43{m^3}} )( {2m - 1} )!!}}{{3360 \cdot {2^m}( {m + 4} )!}} .	
	\end{align}		
\end{subequations}	
From Eq.(\ref{eq11.7.2}), we observe that $\Delta_1 \tilde{c}_r(\tilde{t}_{\ell}^m; \tilde{t}_o^2 )>0$ for $m<7$ and $\Delta_1 \tilde{c}_r(\tilde{t}_{\ell}^m; \tilde{t}_o^2 )<0$ for $m>7$. Interestingly, this  first-order correction vanishes when $m=7$. However, the second-order correction $\Delta_2 \tilde{c}_r(\tilde{t}_{\ell}^m; \tilde{t}_o^2 )$ is always positive for any positive integer $m$.

For general values of $r>0$, analytical evaluation of the integral in Eq.(\ref{eq11.6}) is infeasible due to the complexity of the second moment of $t_o$ for the resetting Brownian bridge (Eq.(\ref{eq10.13b})). Instead, we use numerical integration for this term, while the remaining parts of Eq.(\ref{eq11.6}) can be derived analytically. For $m=1$ and $m=2$, using Eq.(\ref{eq8.2}) yields
\begin{widetext}
\begin{subequations}\label{eq11.8}
	\begin{align}
		\tilde{c}_r(\tilde{t}_{\ell}; \tilde{t}_o^2 ) &=\int_0^1 {d{\tilde{t}_{\ell}}} {\tilde{t}_{\ell}}\langle {\tilde{t}_o^2} \rangle_r^{bridge}{\tilde{P}_r}( {{\tilde{t}_{\ell}}} ) + \frac{{2{e^{ - R}} - {e^{ - 2R}} + {R^2} - 2{R^3} - 1}}{{8{R^3}}}, \\
		\tilde{c}_r(\tilde{t}_{\ell}^2; \tilde{t}_o^2 ) &=\int_0^1 {d{\tilde{t}_{\ell}}} {\tilde{t}_{\ell}^2}\langle {\tilde{t}_o^2} \rangle_r^{bridge}{\tilde{P}_r}( {{\tilde{t}_{\ell}}} )  + \frac{{{e^{ - 2R}}\left( {6 - 2R} \right) - {e^{ - R}}\left( {12 + 14R + 10{R^2} + {R^3}} \right) + 6 + 16R - 14{R^2} + 8{R^3} - 8{R^4}}}{{32{R^4}}}.
	\end{align}
\end{subequations}
\end{widetext}

To validate our theory, we performed extensive simulations to collect $10^6$ different realizations of RBM for each $r$. The total time duration was fixed at $t=1$, with $r$ varying from $10^{-2}$ to $10^2$; each realization consisted of $10^4$ time steps with each step $dt=10^{-4}$. From these data, we computed correlation coefficients of $t_{\ell}^m$ and $t_{o}^n$ for $(m,n)=(1,1)$, $(2,1)$, $(1,2)$ and $(2,2)$, shown as symbols in Fig.\ref{fig_corrtotl}. Theoretical results (lines) exhibit excellent agreement with simulations. As predicted, correlations vanish for $n=1$, while for $n=2$, all correlation coefficients are negative for finite $r$, increasing monotonically toward zero as $r \to \infty$.

\section{Joint distribution of $t_m$ and $t_o$}

The joint distribution $P_r(t_o,t_m|t)$ can be computed using a path-decomposition method, originally developed to compute the joint distribution of the maximum displacement $M$ and the time $t_m$ at which the maximum is achieved for RBM \cite{majumdar2021mean,guo2023extremal,guo2024extremal,huang2024extreme}. We first introduce an extended joint distribution $P_r(t_o,t_m,M|t)$ for three random variables $t_o$, $t_m$ and $M$. Integrating over $M$ then yields $P_r(t_o,t_m|t)$, 
\begin{eqnarray}\label{eq6.1}
P_r(t_o,t_m|t)=\int_{0}^{\infty} dM P_r(t_o,t_m|t).
\end{eqnarray}

To derive $P_r(t_o,t_m,M|t)$, we decompose the trajectory into two segments: a left-hand segment for $0 < \tau < t_m$, and a right-hand segment for $t_m < \tau < t$. Due to the Markovian property, the weights of these two segments are completely independent, and the total weight is proportional to their product. For the left segment, the process propagates from $0$ at $\tau=0$ to $M$ at $\tau = t_m$ without touching the absorbing boundary at $x=M$. Furthermore, we denote the occupation time above the origin in this segment as $t_o^L$. The statistical weight of the left segment is given by the propagator $G_r^M(M, t_m, t_o^L|0;0)$. However, it turns out that $G_r^M(M, t_m, t_o^L|0;0)=0$, making its contribution zero. To resolve this problem, we instead compute $G_r^M(M-\epsilon, t_m, t_o^L|0;0)$ and later take the limit $\epsilon \to 0^+$ \cite{majumdar2004exact,randon2007distribution}. For the right segment, the process starts from $M-\epsilon$ and stays below $M$ without crossing the level $M$ between $t_m$ and $t$. 
Its statistical weight is $Q_r^M(M-\epsilon, t-t_m, t_o^R;0)$, where $t_o^R=t_o-t_o^L$ is the occupation time above the origin in the right segment. Combining these, the joint probability density $P_r(t_0,t_m,M|t)$ is the product of the statistical weights of the two segments \cite{hartmann2025exact}
\begin{widetext}
\begin{eqnarray}\label{eq6.2}
P_r(t_o,t_m,M|t)&=&\mathop {\lim }\limits_{\epsilon \to 0^+} \mathcal{N}(\epsilon)\int_{0}^{t} dt_o^L  G_r^M(M-\epsilon, t_m, t_o^L|0;0)  Q_r^M(M-\epsilon, t-t_m, t_o^R;0),
\end{eqnarray} 
where $\mathcal{N}(\epsilon)$ is the normalization constant to be determined later. We take the Laplace transform for Eq.(\ref{eq6.2}) with respect to $t_o$ and $t_m$, and then integrate over $M$,  
\begin{eqnarray}\label{eq6.3}
\tilde{P}_r(p,\lambda|t)&=&\int_{0}^{t} dt_m e^{-\lambda t_m} \int_{0}^{t} dt_o e^{-p t_o} P_r(t_o,t_m|t) =\langle e^{-\lambda t_m-pt_o}\rangle_r \nonumber \\&=&\mathop {\lim }\limits_{\epsilon \to 0^+} \mathcal{N}(\epsilon) \int_{0}^{t} dt_m e^{-\lambda t_m} \int_{0}^{t} dt_o e^{-p t_o} \int_{0}^{\infty} d M \int_{0}^{t} d t_o^L G_r^M(M-\epsilon, t_m, t_o^L|0;0)  Q_r^M(M-\epsilon, t-t_m, t_o^R;0) \nonumber \\&=&\mathop {\lim }\limits_{\epsilon \to 0^+} \mathcal{N}(\epsilon) \int_{0}^{t} dt_m e^{-\lambda t_m}  \int_{0}^{\infty} d M  \tilde{g}_r^M(M-\epsilon, t_m, p|0;0)  \tilde{q}_r^M(M-\epsilon, t-t_m, p;0).
\end{eqnarray} 
where $\tilde{g}_r^M(x, t, p|0;0)$ and $\tilde{q}_r^M(x, t, p;0)$ are the Laplace transformations of $G_r^M(x, t, t_o|0;0)$ and $Q_r^M(x, t, t_o;0)$ with respect to $t_o$, respectively.

We further apply the Laplace transform for Eq.(\ref{eq6.3}) with respect to $t$, 
\begin{eqnarray}\label{eq6.3.1}
\hat{\tilde{P}}_r(p,\lambda,s)&=&\int_{0}^{\infty} dt e^{-s t} \tilde{P}_r(p,\lambda|t) \nonumber \\&=&\mathop {\lim }\limits_{\epsilon \to 0^+} \mathcal{N}(\epsilon)  \int_{0}^{\infty} d M  \tilde{G}_r^M(M-\epsilon, s+\lambda, p|0;0)   \tilde{Q}_r^M(M-\epsilon, s, p;0).
\end{eqnarray} 
Here, $ \tilde{G}_r(M-\epsilon, s, p|0;0)$ denotes the double Laplace transform of the occupation-time-weighted propagator ${G}_r(M-\epsilon, t, t_o|0;0)$ for RMB. It is derived from Eqs.(\ref{eq2.6}) and (\ref{eq5.4.1}) (see appendix \ref{sec_app_pg} for details). Similarly, $\tilde{Q}_r^M(M-\epsilon, s, p;0)$ is the double Laplace transform of the occupation-time-weighted survival probability $Q_r^M(M-\epsilon, t, t_o;0)$ for RMB,  obtained from Eqs.(\ref{eq4.4}) and (\ref{eq5.2}) (see appendix \ref{sec_app_sp} for details). Expanding $ \tilde{G}_r(M-\epsilon, s, p|0;0)$ and $\tilde{Q}_r^M(M-\epsilon, s, p;0)$ to the leading order in $\epsilon$ gives
\begin{eqnarray}\label{eq6.3.2}
\tilde{G}_r(M-\epsilon, s, p|0;0)=\frac{{\sqrt {r + s} ( {p + r + s} )} \epsilon}{{D[ {s\sqrt {p + r + s} \sinh (M\sqrt {(p + r + s)/D} ) + \sqrt {r + s} ( {r + ( {p + s} )\cosh (M\sqrt {(p + r + s)/D} )} )} ]}},
\end{eqnarray} 
and 
\begin{eqnarray}\label{eq6.3.3}
\tilde{Q}_r(M-\epsilon, s, p;0)=\frac{{[ {p + ( {r + s} )\cosh (M\sqrt {(p + r + s)/D} ) + \sqrt {r + s} \sqrt {p + r + s} \sinh (M\sqrt {(p + r + s)/D} )} ]\epsilon}}{{\sqrt D [ {s\sqrt {p + r + s} \sinh (M\sqrt {(p + r + s)/D} ) + \sqrt {r + s} ( {r + ( {p + s} )\cosh (M\sqrt {(p + r + s)/D} )} )} ]}}.
\end{eqnarray}

Setting $\lambda=p=0$, the left-hand side of Eq.(\ref{eq6.3.1}) equals to $1/s$, which allows us to determine the normalization constant, 
\begin{eqnarray}\label{eq7.1}
\mathcal{N}(\epsilon) =\frac{D}{\epsilon^2}.
\end{eqnarray} 
Finally, we obtain the joint distribution of $t_m$ and $t_o$ in the Laplace domain by substituting Eqs.(\ref{eq6.3.2}), (\ref{eq6.3.3}) and (\ref{eq7.1}) into Eq.(\ref{eq6.3.1}), 
\begin{eqnarray}\label{eq6.4}
\hat{\tilde{P}}_r(p,\lambda,s)=\int_{0}^{\infty} d M  {\frac{{\sqrt {r + s} ( {p + r + s} )} }{{[ {s\sqrt {p + r + s} \sinh (M\sqrt {(p + r + s)/D} ) + \sqrt {r + s} ( {r + ( {p + s} )\cosh (M\sqrt {(p + r + s)/D} )} )} ]}}} \nonumber \\ \times \frac{{[ {p + ( {r + s} )\cosh (M\sqrt {(p + r + s)/D} ) + \sqrt {r + s} \sqrt {p + r + s} \sinh (M\sqrt {(p + r + s)/D} )} ]}}{{\sqrt D [ {s\sqrt {p + r + s} \sinh (M\sqrt {(p + r + s)/D} ) + \sqrt {r + s} ( {r + ( {p + s} )\cosh (M\sqrt {(p + r + s)/D} )} )} ]}}.
\end{eqnarray} 
\end{widetext}

The moments in the Laplace domain can be obtained by calculating the following derivative at $\lambda = p = 0$,
\begin{eqnarray}\label{eq7.2}
&&\int_{0}^{\infty} dt e^{-s t}\langle {t_m^{{k_1}}t_o^{{k_2}}} |t\rangle_r  \nonumber \\ &=& { {{{( { - 1} )}^{{k_1} + {k_2}}}{{{\partial ^{{k_1} + {k_2}}}\hat{\tilde{P}}_r(p,\lambda,s)} \over {\partial {\lambda^{{k_1}}}\partial {p^{{k_2}}}}}} |_{\lambda = p = 0}}.
\end{eqnarray}
Subsequently, we perform the inverse Laplace transform with respect to $s$ to obtain the moments $\langle {t_m^{{k_1}}t_o^{{k_2}}} \rangle$.

Setting $k_1=1$ and $k_2=0$ in Eq.(\ref{eq7.2}) gives
\begin{eqnarray}\label{eq7.4}
\int_{0}^{\infty} dt e^{-s t}\langle {t_m}|t \rangle_r  =\frac{{( {2r + s } )}}{{4{s ^2}( {r + s } )}} + \frac{1}{{4rs }}\ln \left(  {\frac{{r + s }}{s }} \right) .
\end{eqnarray}
Inverse Laplace tranformation of Eq.(\ref{eq7.4}) yields the mean value of $t_m$,
\begin{eqnarray}\label{eq7.5}
\langle \tilde{t}_m \rangle_r= \frac{1}{{4R}}\left[ {2R + {e^{ -R}} - 1 + {\gamma _E} + \Gamma ( {0,R} ) + \ln  {R} } \right], \nonumber \\
\end{eqnarray}
where $R=rt$ is the mean number of resets over the time $t$, $\gamma _E= 0.577216...$ is Euler's constant, and $\Gamma ( {0,rt} )$ is the incomplete gamma function. This result has been previously reported in Ref.\cite{singh2021extremal}. Interestingly, $\langle \tilde{t}_m \rangle$ exhibits a nonmonotonic dependence on $R$: $\langle \tilde{t}_m \rangle$ tends to $\frac{1}{2}$ in both limits $R\to 0$ and $R \to \infty$, with a unique maximum $\langle \tilde{t}_m \rangle_{\max} \approx 0.56194$ at $R\approx3.47845$. A similar phenomenon was observed in Ref.\cite{mori2022time}, where the initial position of the particle is not fixed but sampled from a nonequilibrium distribution.

Setting $k_1=0$ and $k_2=1$ in Eq.(\ref{eq7.2}) gives
\begin{eqnarray}\label{eq7.6}
\int_{0}^{\infty} dt e^{-s t}\langle {t_o}|t \rangle_r =\frac{1}{2 s^2}.
\end{eqnarray}
The inversion of Eq.(\ref{eq7.6}) with respect to $s$ yields the mean value of $t_o$, 
\begin{eqnarray}\label{eq7.7}
\langle \tilde{t}_o \rangle_r=\frac{1 }{t}\langle {t_o|t}\rangle_r= \frac{1}{2},
\end{eqnarray}
which is independent of the resetting rate $r$. This is consistent with our previous result \cite{yan2023breakdown}. 

\textbf{Correlation between $t_{m}$ and $t_o$}:  Setting $k_1=k_2=1$ in Eq.(\ref{eq7.2}), we get
\begin{eqnarray}\label{eq7.8}
\int_{0}^{\infty} dt e^{-s t}\langle {t_m t_o}|t \rangle_r  &=&\frac{{8{r^3} + 8{r^2}s  - 5r{s ^2} - 4{s ^3}}}{{16{r^2}{s ^3}\left( {r + s } \right)}} \nonumber \\ &+& \frac{{\left( {r + s } \right)}}{{8{r^2}{s ^2}}}\ln \left( {\frac{{r + s }}{s }} \right) \nonumber \\ &-& \frac{1}{{4{r^3}}}{{\rm{Li}}_2}\left( { - \frac{r}{s }} \right), 
\end{eqnarray}
where ${{\rm{Li}}_2}(x)=\sum_{k=1}^{\infty}x^k/k^2$ is the polylogarithm function. 
Expanding Eq.(\ref{eq7.8}) in powers of $r$ yields
\begin{widetext}
\begin{eqnarray}\label{eq7.9}
\int_{0}^{\infty} dt e^{-s t}\langle {t_m t_o}|t \rangle_r =\frac{{47}}{{72{s ^3}}}  + \sum\limits_{n = 1}^\infty  {{{( { - 1} )}^n}{\left[\frac{1}{{16}} + \frac{1}{{8\left( {n + 1} \right)\left( {n + 2} \right)}} + \frac{1}{{4{{\left( {n + 3} \right)}^3}}} \right] }\frac{{{r^n}}}{{{s ^{n + 3}}}}} .\nonumber \\
\end{eqnarray}
Inverse Laplace transformation of Eq.(\ref{eq7.9}) with respect to $s$ gives 
\begin{eqnarray}\label{eq7.11}
\langle {\tilde{t}_m \tilde{t}_o} \rangle_r=\frac{{47}}{{144}} + \sum\limits_{n = 1}^\infty  {\frac{{{{\left( { - 1} \right)}^n}}}{{\left( {n + 2} \right)!}}\left[ {\frac{1}{{16}} + \frac{1}{{8\left( {n + 1} \right)\left( {n + 2} \right)}} + \frac{1}{{4{{\left( {n + 3} \right)}^3}}}} \right]} {R^n}.
\end{eqnarray}
For $R=0$ (no resetting), the second term on the right-hand side of Eq.(\ref{eq7.11}) vanishes, recovering $\langle {\tilde{t}_m \tilde{t}_o} \rangle_{r=0} =47/144$ as reported in Ref.\cite{hartmann2025exact}. Thus, the second term captures the effect of resetting. 

Eq.(\ref{eq7.11}) can be rewritten using special functions,
\begin{eqnarray}\label{eq7.12}
\langle {\tilde{t}_m \tilde{t}_o} \rangle_r=\frac{{\gamma_E \left( {2 + R + {R^2}} \right)}}{{8{R^3}}} + \frac{{4{R^2} - {e^{ - R}} - 3 - 2R}}{{16{R^2}}} - \frac{{2 + R + {R^2}}}{{8{R^3}}}\left[ {{\rm{Ei}}\left( { - R} \right) - \ln R} \right],
\end{eqnarray}
where ${\rm{Ei}}(x)=\int_{-\infty}^{x}\frac{e^t}{t}dt$ is the exponential integral function. 

Using Eqs.(\ref{eq7.5}), (\ref{eq7.7}), and (\ref{eq7.12}), the correlation between $\tilde{t}_m$ and $\tilde{t}_o$ reads
\begin{eqnarray}\label{eq7.13}
\tilde{c}_r(\tilde{t}_m;\tilde{t}_o)&=&\langle  {\tilde{t}_m \tilde{t}_o}  \rangle_r- \langle  {\tilde{t}_m }  \rangle  \langle  {\tilde{t}_o}  \rangle_r \nonumber \\&=&\frac{{4\gamma_E  + ( {2\gamma_E  - 3} )R - {{\text{e}}^{ - R}}R( {2R + 1} ) + 2( {R + 2} ) [\ln  R-{\rm{Ei}} ({ - R} )]  }}{{16{R^3}}}.
\end{eqnarray}
\end{widetext}

For $R=0$, when resetting is absent,  $\tilde{c}_0(\tilde{t}_m;\tilde{t}_o)=\frac{11}{144}$ \cite{hartmann2025exact}. As $R$ increases, $\tilde{c}_r(\tilde{t}_m;\tilde{t}_o)$ decreases monotonically toward zero. Its asymptotic behaviors are
\begin{eqnarray}\label{eq7.15}
\tilde{c}_r(\tilde{t}_m;\tilde{t}_o) \sim \left\{ \begin{array}{llc}
\frac{11}{144} - \frac{55R}{1152}, &     R \ll 1,  \\
\frac{\ln R}{8 R^2}, &    R \gg 1.  \\ 
\end{array}  \right.
\end{eqnarray}

\begin{figure}
	\centerline{\includegraphics*[width=1.0\columnwidth]{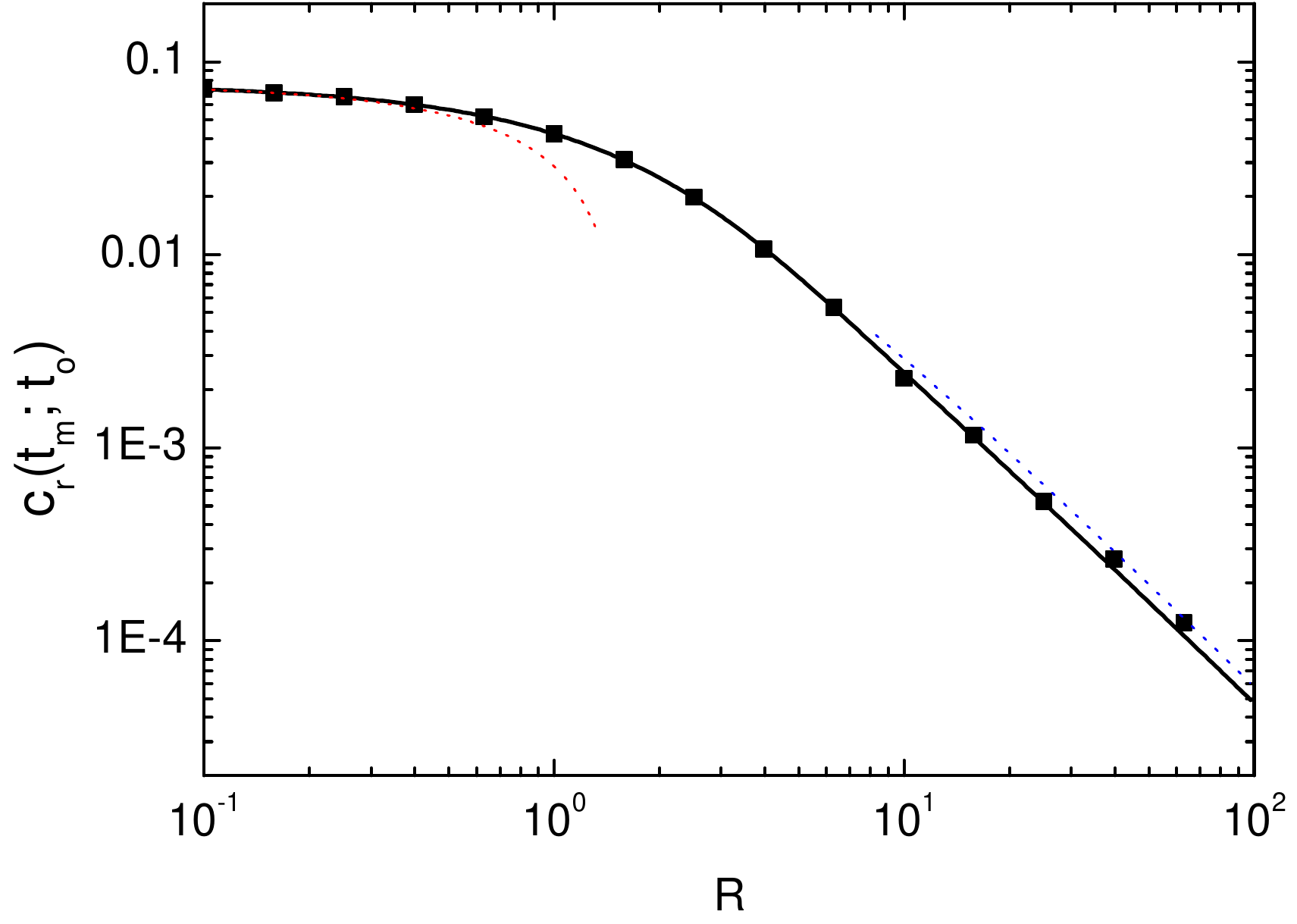}}
	\caption{Correlation coefficient between $t_m$ and $t_{o}$ as a function of the resetting rate $r$. The solid lines and symbols represent theoretical results (from Eq.(\ref{eq7.13})) and simulation results, respectively. The (red and blue) dotted lines indicate   asymptotic behaviors of $\tilde{c}_r(\tilde{t}_m;\tilde{t}_o) $ (see Eq.(\ref{eq7.15})).    \label{fig_corr_tmto}}
\end{figure}

To verify our theoretical predictions, we plot simulation results of $\tilde{c}_r(\tilde{t}_m;\tilde{t}_o)$ as a function of $R$, as indicated by symbols in Fig.\ref{fig_corr_tmto}, and find the excellent agreement between simulations and theory (lines).

\section{Joint distribution of $t_m$ and $t_{\ell}$}
Let us denote by $P_r(t_m,t_{\ell}|t)$ the joint distribution of two random variables $t_m$ and $t_{\ell}$ for the RBM, given a fixed observation time $t$. To get it, it is convenient to introduce an enlarged joint distribution $P_r(M,t_m,t_{\ell}|t)$ of $t_m$, $t_{\ell}$ and $M$ of the global maximum in $\left[0, t \right] $. Once  $P_r(M,t_m,t_{\ell}|t)$ is obtained, one can integrate over $M$ to obtain $P_r(t_m,t_{\ell}|t)$, 
\begin{eqnarray}\label{eq15.1}
P_r(t_m,t_{\ell}|t)=\int_{0}^{\infty} dM P_r(M,t_m,t_{\ell}|t).
\end{eqnarray}

We will use the path decomposition technique to compute $P_r(M,t_m,t_{\ell}|t)$, which depends on whether $t_m<t_{\ell}$ or $t_m>t_{\ell}$ \cite{hartmann2025exact}. We will consider the two cases separately.

\begin{figure}
	\centerline{\includegraphics*[width=1.0\columnwidth]{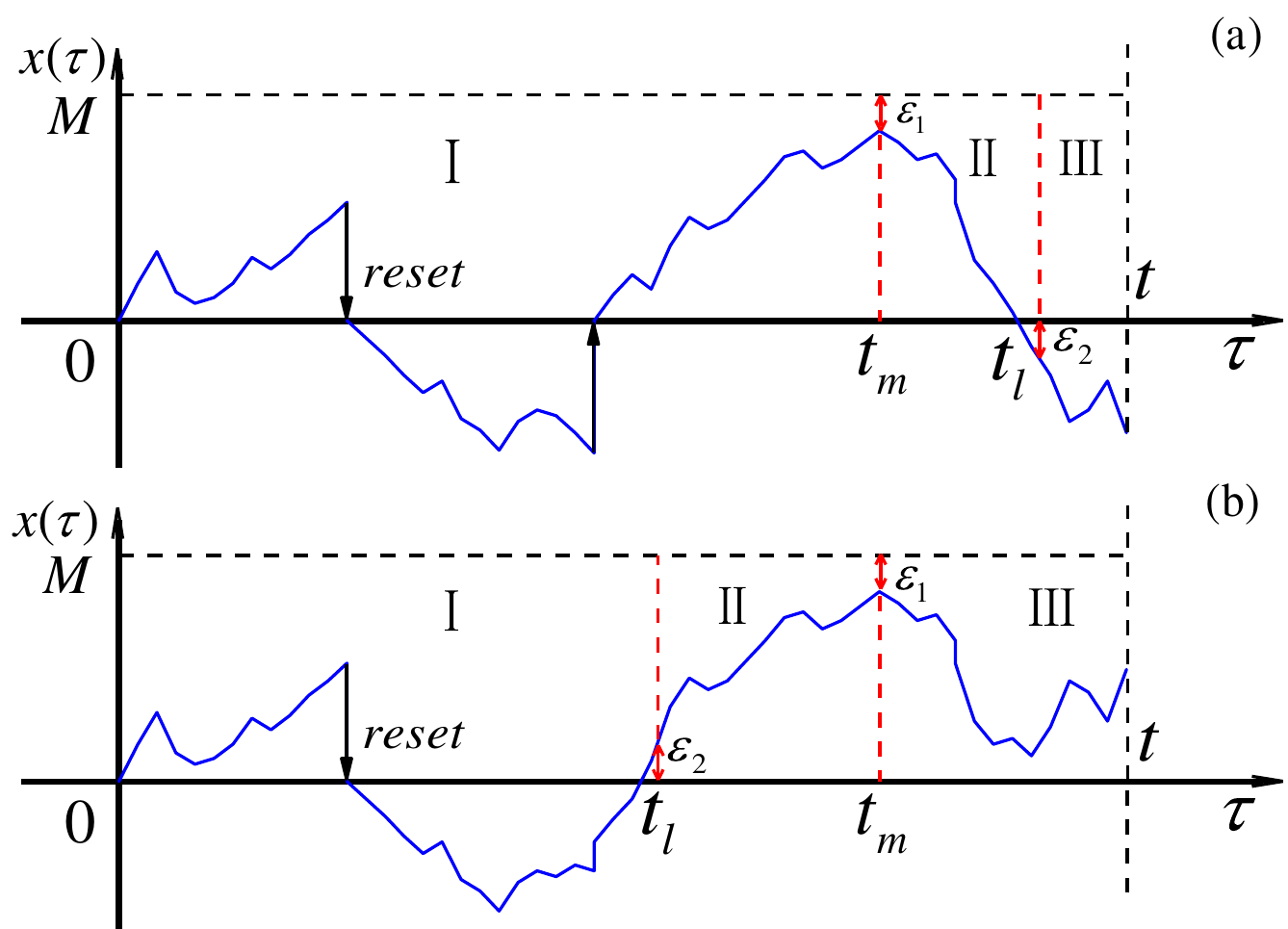}}
	\caption{A schematic plot of one-dimensional resetting Brownian motion $x(\tau)$ over a duration $t$. The particle starts from $x(0)=0$ and is stochastically reset to the origin at a constant rate $r$. Two  cases can be distinguished: (a) The trajectory first reaches its maximum position $M$ at time $t_m$, and then crosses the origin at $t_{\ell}$ for the last time. (b) The trajectory first passes through the origin  for the last time at $t_{\ell}$, and then attains its global maximum $M$. In both cases, we divide the total duration $t$ into three segments based on $t_m$ and $t_{\ell}$. \label{fig_schematic_tmtl}}
\end{figure}

First, we consider the scenario $t_m<t_{\ell}$. Let us examine a typical trajectory in this case (see Fig.\ref{fig_schematic_tmtl}(a) for an illustration): the RBM starts from the origin $x(0)=0$, reaches its maximum value $M$ at time $t=t_m$, and subsequently crosses the origin for the last time at $t=t_{\ell}$. For $M$ to qualify as the maximum, the process must remain below $M$ at all times—a condition enforced by imposing an absorbing boundary at $x=M$.  However, this boundary condition precludes the process from attaining $M$ exactly at $t=t_m$.  Thus, following the approach in the previous section, we introduce a cutoff $\epsilon_1$ at $t=t_m$, requiring the process to reach $M-\epsilon_1$ (where $\epsilon_1 \geq 0$) at this time. Similarly, to enforce the constraint that the process stays either above or below the origin during the interval $\left[ t_{\ell}, t\right] $, we impose an absorbing boundary at $x=0$; this, in turn, prevents the process from arriving at 0 precisely at $t=t_{\ell}$. To address this, we introduce a second cutoff $\epsilon_2$ at $t = t_{\ell}$, stipulating that the process reaches  $\epsilon_2$ at this time and thereafter does not cross zero throughout  $\left[ t_{\ell}, t\right] $. Ultimately, we will take the limits $\epsilon_1 \to 0$ and $\epsilon_2 \to 0$ using the appropriate methods outlined in the previous section.

Following the above procedure, we divide the full time into three segments: (I) the interval $\left[0, t_m \right] $, over which the process originates from the origin, terminates at 
$M-\epsilon_1$, and remains below the level $M$; (II) the interval $\left[t_m, t_{\ell} \right] $, beginning from $M-\epsilon_1$, ending at $\epsilon_2$, while remaining below the level $M$; (III) starting at $\epsilon_2$ the process stays below $M$ and does not cross the origin during $\left[t_{\ell}, t \right] $. For the segment (III), we can distinguish two scenarios: (a) if the process crosses the origin from below, (i.e., $\epsilon_2>0$), and the process stays within $(0,M)$ during $\left[t_{\ell}, t \right]$, and (b) if the process crosses the origin from above, (i.e., $\epsilon_2<0$), the path stays negative throughout  $\left[t_{\ell}, t \right] $. The probabilities of the first and second segments are simply written as the propagator of RBM in the presence of an absorbing boundary $x=M$, i.e., $P_I=G_r^M(M-\epsilon_1,t_m|0;0)$ and $P_{II}=G_r^M(\epsilon_2,t_{\ell}-t_m|M-\epsilon_1;0)$. For the third segment, since no resetting can occur (otherwise, the origin would be crossed again), its probability can be expressed as the probability in the absence of resetting (i.e., the sum of survival probabilities of the two scenarios) multiplied by a factor $e^{-r(t-t_{\ell})}$ quantifying the probability of no resets occurring within $\left[t_{\ell}, t \right] $. Thus, the probability of the third segment is $P_{III}=e^{-r(t-t_{\ell})}\left[ Q_0^{\rm{box}}(\epsilon_2^+,t-t_{\ell})+Q_0^{M=0}(\epsilon_2^-,t-t_{\ell})\right] $, where $Q_0^{M=0}(x_0,t)$ is the survival probability of SBM up to time $t$ with an absorbing wall at $x=0$, starting from $x_0$, and $Q_0^{\rm{box}}(x_0,t)$ is the survival probability of SBM with two absorbing walls at $x=0$ and $x=M$. Putting it all together, we can write $P_r(M,t_m,t_{\ell}|t;t_m<t_{\ell})$ as the product of the probabilities of these three segments,
\begin{widetext}
\begin{eqnarray}\label{eq15.2}
P_r(M,t_m,t_{\ell}|t;t_m<t_{\ell})=\mathop {\lim } \limits_{\epsilon_{1,2} \to 0} \mathcal{N}(\epsilon_1,\epsilon_2) \mathcal{N}G_r^M(M-\epsilon_1,t_m|0;0)G_r^M(\epsilon_2,t_{\ell}-t_m|M-\epsilon_1;0)H_r(\epsilon_2,t-t_{\ell}),
\end{eqnarray}
where we use the notation for brevity
\begin{eqnarray}\label{eq15.3}
H_r(\epsilon_2,t-t_{\ell})=e^{-r(t-t_{\ell})}[ Q_0^{\rm{box}}(\epsilon_2^+,t-t_{\ell})+Q_0^{M=0}(\epsilon_2^-,t-t_{\ell})].
\end{eqnarray}
and the normalization factor $\mathcal{N}(\epsilon_1,\epsilon_2)$ will be determined later.

Performing a triple Laplace transform for Eq.(\ref{eq15.2}) with respect to $t$, $t_{\ell}$, and $t_{m}$, we have
\begin{eqnarray}\label{eq15.4}
\tilde{P}_r(M,\lambda,p,s;t_m<t_{\ell})&=&\int_{0}^{\infty}dt e^{-st} \int_{0}^{t} dt_{\ell} e^{-p t_{\ell}} \int_{0}^{t_{\ell}} dt_m e^{-\lambda t_m}P_r(M,t_m,t_{\ell}|t;t_m<t_{\ell}) \nonumber \\&=&\mathop {\lim } \limits_{\epsilon_{1,2} \to 0} \mathcal{N}(\epsilon_1,\epsilon_2) \tilde{G}_r^M(M-\epsilon_1,s+p+\lambda|0;0)\tilde{G}_r^M(\epsilon_2,s+p|M-\epsilon_1;0)\tilde{H}_0^M(\epsilon_2,r+s),
\end{eqnarray}
\end{widetext}

where 
\begin{eqnarray}\label{eq15.4.1}
\tilde{H}_0^M(\epsilon_2,s)=\tilde{Q}_0^{\rm{box}}(\epsilon_2^+,s)+\tilde{Q}_0^{M=0}(\epsilon_2^-,s).
\end{eqnarray}
In Eq.(\ref{eq15.4}), $\tilde{G}_r^M(M-\epsilon_1,s|0;0)$ and $G_r^M(\epsilon_2,s|M-\epsilon_1;0)$ can be obtained from Eq.(\ref{eqb2.2}) (see appendix \ref{sec_app_rmbpr} for details), given by, up to the leading term in $\epsilon_1$ and $\epsilon_2$, 
\begin{eqnarray}\label{eq15.5}
\tilde{G}_r^M(M-\epsilon_1,s|0;0)=\frac{r+s}{D[r+se^{M\sqrt{(r+s)/D}}]}\epsilon_1,
\end{eqnarray}
and
\begin{eqnarray}\label{eq15.6}
G_r^M(\epsilon_2,s|M-\epsilon_1;0)=\frac{s+r\cosh(M\sqrt{(r+s)/D})}{D(r+se^{M\sqrt{(r+s)/D}})}\epsilon_1. \nonumber \\
\end{eqnarray}
Furthermore, $\tilde{Q}_0^{M=0}(\epsilon_2^-,s)$ and $\tilde{Q}_0^{\rm{box}}(\epsilon_2^+,s)$ can be obtained from Eq.(\ref{eqa1.3}) and Eq.(\ref{eqa2.3}), respectively (see appendix \ref{sec_app_rmbsp} for details), such that $\tilde{H}_0^M(\epsilon_2,s)$ in Eq.(\ref{eq15.4.1}) can be written as
\begin{eqnarray}\label{eq15.7}
\tilde{H}_0^M(\epsilon_2,s)&=&\tilde{Q}_0^{\rm{box}}(\epsilon_2^+,s)+\tilde{Q}_0^{M=0}(\epsilon_2^-,s) \nonumber \\ &=&\frac{\tanh(\frac{M}{2}\sqrt{s/D})}{\sqrt{Ds}}\epsilon_2+\frac{1}{\sqrt{Ds}}\epsilon_2 \nonumber \\&=& \frac{2e^{M\sqrt{s/D}}}{\sqrt{Ds}(1+e^{M\sqrt{s/D}})}\epsilon_2.
\end{eqnarray}
Substituting Eqs.(\ref{eq15.5}), (\ref{eq15.6}), and (\ref{eq15.7}) into Eq.(\ref{eq15.4}), we obtain
\begin{widetext}
\begin{eqnarray}\label{eq15.8}
\tilde{P}_r(M,\lambda,p,s;t_m<t_{\ell})&=&\mathop {\lim } \limits_{\epsilon_{1,2} \to 0} \mathcal{N}(\epsilon_1,\epsilon_2) \frac{r+\lambda+p+s}{D[r+(\lambda+p+s)e^{M\sqrt{(r+\lambda+p+s)/D}}]} \frac{s+p+r\cosh(M\sqrt{(r+s+p)/D})}{D(r+(s+p)e^{M\sqrt{(r+s+p)/D}})} \nonumber \\ &\times& \frac{2e^{M\sqrt{(r+s)/D}}}{\sqrt{D(r+s)}(1+e^{M\sqrt{(r+s)/D}})} \epsilon_1^2 \epsilon_2.
\end{eqnarray}
\end{widetext}

Subsequently, we consider the case where $t_m>t_{\ell}$ (see Fig.\ref{fig_schematic_tmtl}(b) for an illustration).  In this case, the process originates from 0 and reaches $\epsilon_2>0$ at time $t_{\ell}$ while remaining below $M$--this corresponds to the segment I. The probability of the first segment thus equals to the propagator of RBM with an absorbing wall at $x=M$, i.e., $P_{I}=G_r^M(\epsilon_2,t_{\ell}|0;0)$. Next, in the segment II, the process starts from $\epsilon_2$ at $t_{\ell}$, reaches $M-\epsilon_1$ at $t_m$, and stays within the interval $x \in \left[0, M \right] $ throughout the time period $\left[t_{\ell}, t_m \right]$. Hence we can use the propagator of RBM in a box to express the probability of the second segment, i.e., $P_{II}=G_r^{\rm{box}}(M-\epsilon_1,t_m-t_{\ell}|\epsilon_2;0)$. Finally, in region III, the process begins from $M-\epsilon_1$ at $t_m$ and remains inside  $x \in \left[0, M \right] $ for the remaining time interval $\left[t_m, t \right]$ without resetting. Thus, the probability of the third segment is the survival probability of SBM in a box $\left[0, M \right] $ multiplied by a probability factor $e^{-r (t-t_m)}$ without resetting in $\left[t_m, t \right]$, i.e., $P_{III}=e^{-r (t-t_m)} Q_0^{\rm{box}}(M-\epsilon_1,t-t_m;0)$.  As in the previous case, we will then take their product of these three segments to derive the joint distribution $P_r(M,t_m,t_{\ell}|t)$ under the condition $t_m>t_{\ell}$, 
\begin{widetext}
\begin{eqnarray}\label{eq15.9}
P_r(M,t_m,t_{\ell}|t;t_m>t_{\ell})=\mathop {\lim } \limits_{\epsilon_{1,2} \to 0} \mathcal{N}(\epsilon_1,\epsilon_2) G_r^M(\epsilon_2,t_{\ell}|0;0)  G_r^{\rm{box}}(M-\epsilon_1,t_m-t_{\ell}|\epsilon_2;0) e^{-r (t-t_m)} Q_0^{\rm{box}}(M-\epsilon_1,t-t_m;0). \nonumber \\
\end{eqnarray}
Similarly in Eq.(\ref{eq15.4}), we perform a triple Laplace transform for Eq.(\ref{eq15.9}) to have
\begin{eqnarray}\label{eq15.10}
\tilde{P}_r(M,\lambda,p,s;t_m>t_{\ell})=\mathop {\lim } \limits_{\epsilon_{1,2} \to 0} \mathcal{N}(\epsilon_1,\epsilon_2) \tilde{G}_r^M(\epsilon_2,s+p+\lambda|0;0)\tilde{G}_r^{\rm{box}}(M-\epsilon_1,s+\lambda|\epsilon_2;0) \tilde{Q}_0^{\rm{box}}(M-\epsilon_1,r+s;0),
\end{eqnarray}
\end{widetext}
where $\tilde{G}_r^M(\epsilon_2,s|0;0)$, $\tilde{G}_r^{\rm{box}}(M-\epsilon_1,s|\epsilon_2;0)$ and $\tilde{Q}_r^{\rm{box}}(M-\epsilon_1,s;0)$ can be obtained from Eqs.(\ref{eqb2.2}), (\ref{eqb5.1}), and (\ref{eqa2.3}), respectively (see appendix \ref{sec_app_rmbpr} and appendix \ref{sec_app_rmbsp} for details). Up to the leading term in $\epsilon_1$ and $\epsilon_2$, they are
\begin{eqnarray}\label{eq15.11}
\tilde{G}_r^M(\epsilon_2, s|0;0)=\frac{{\sqrt {r + s} \sinh ( {M\sqrt {(r + s)/D} } )}}{{\sqrt D (r + s{e^{M\sqrt {(r + s)/D} }})}},
\end{eqnarray}
\begin{eqnarray}\label{eq15.12}
\tilde{G}_r^{\rm{box}}(M-\epsilon_1,s|\epsilon_2;0)=\frac{{\sqrt {r + s}  {\rm{csch}}( {M\sqrt {(r + s)/D} } )}}{{\sqrt {{D^3}} }}\epsilon_1\epsilon_2, \nonumber \\
\end{eqnarray}
and
\begin{eqnarray}\label{eq15.13}
\tilde{Q}_0^{\rm{box}}(M-\epsilon_1,s;0)=\frac{{\tanh ( {\frac{M}{2}\sqrt {s/D} } )}}{{\sqrt {Ds} }} \epsilon_1.
\end{eqnarray}
Substituting Eqs.(\ref{eq15.11}), (\ref{eq15.12}), and (\ref{eq15.13}) into Eq.(\ref{eq15.10}), we obtain
\begin{widetext}
\begin{eqnarray}\label{eq15.14}
\tilde{P}_r(M,\lambda,p,s;t_m>t_{\ell})&=&\mathop {\lim } \limits_{\epsilon_{1,2} \to 0} \mathcal{N}(\epsilon_1,\epsilon_2) {\frac{{\sqrt {r + s+p+\lambda} \sinh ( {M\sqrt {(r + s+p+\lambda)/D} } )}}{{\sqrt D [r + (s+p+\lambda){e^{M\sqrt {(r + s+p+\lambda)/D} }}]}}} \nonumber \\ &\times& {\frac{{\sqrt {r + s+\lambda}  {\rm{csch}}( {M\sqrt {(r + s+\lambda)/D} } )}}{{\sqrt {{D^3}} }}}  \frac{{\tanh ( {\frac{M}{2}\sqrt {(r+s)/D} } )}}{{\sqrt {D(r+s)} }} \epsilon_1^2 \epsilon_2.
\end{eqnarray}

Adding Eq.(\ref{eq15.8}) and Eq.(\ref{eq15.14}) together, we have
\begin{eqnarray}\label{eq15.15}
\tilde{P}_r(M,\lambda,p,s)&=&\mathop {\lim } \limits_{\epsilon_{1,2} \to 0} \mathcal{N}(\epsilon_1,\epsilon_2) \frac{r+\lambda+p+s}{D[r+(\lambda+p+s)e^{M\sqrt{(r+\lambda+p+s)/D}}]} \frac{s+p+r\cosh(M\sqrt{(r+s+p)/D})}{D(r+(s+p)e^{M\sqrt{(r+s+p)/D}})} \nonumber \\ &\times&  \frac{2e^{M\sqrt{(r+s)/D}}}{\sqrt{D(r+s)}(1+e^{M\sqrt{(r+s)/D}})} \epsilon_1^2 \epsilon_2 + \mathcal{N} {\frac{{\sqrt {r + s+p+\lambda} \sinh ( {M\sqrt {(r + s+p+\lambda)/D} } )}}{{\sqrt D [r + (s+p+\lambda){e^{M\sqrt {(r + s+p+\lambda)/D} }}]}}} \nonumber \\ &\times&  {\frac{{\sqrt {r + s+\lambda}  {\rm{csch}}( {M\sqrt {(r + s+\lambda)/D} } )}}{{\sqrt {{D^3}} }}}  \frac{{\tanh ( {\frac{M}{2}\sqrt {(r+s)/D} } )}}{{\sqrt {D(r+s)} }} \epsilon_1^2 \epsilon_2.
\end{eqnarray}
\end{widetext}

Letting $\lambda \to 0$ and $p \to 0$ in Eq.(\ref{eq15.15}), and integrating over $M$, we get
\begin{eqnarray}\label{eq15.16}
\int_{0}^{\infty}dM \tilde{P}_r(M,\lambda \to 0,p \to 0,s) =\frac{\mathcal{N}(\epsilon_1,\epsilon_2) \epsilon_1^2 \epsilon_2}{D^2 s}.
\end{eqnarray}
In fact, the left-hand side of Eq.(\ref{eq15.16}) equals to the Laplace transform of 1, which just be $1/s$. This leads to the normalization constant,
\begin{eqnarray}\label{eq15.17}
\mathcal{N}(\epsilon_1,\epsilon_2)=\frac{D^2}{\epsilon_1^2 \epsilon_2}.
\end{eqnarray}
Substituting Eq.(\ref{eq15.17}) into Eq.(\ref{eq15.15}), we get the final expression for the enlarged joint distribution in the Laplace domain,
\begin{widetext}
\begin{eqnarray}\label{eq15.18}
\tilde{P}_r(M,\lambda,p,s)&=& \frac{r+\lambda+p+s}{r+(\lambda+p+s)e^{M\sqrt{(r+\lambda+p+s)/D}}} \frac{s+p+r\cosh(M\sqrt{(r+s+p)/D})}{r+(s+p)e^{M\sqrt{(r+s+p)/D}}} \nonumber \\ &\times&  \frac{2e^{M\sqrt{(r+s)/D}}}{\sqrt{D(r+s)}(1+e^{M\sqrt{(r+s)/D}})}  +  {\frac{{\sqrt {r + s+p+\lambda} \sinh ( {M\sqrt {(r + s+p+\lambda)/D} } )}}{{ r + (s+p+\lambda){e^{M\sqrt {(r + s+p+\lambda)/D} }}}}} \nonumber \\ &\times&  {\frac{{\sqrt {r + s+\lambda}  {\rm{csch}}( {M\sqrt {(r + s+\lambda)/D} } )}}{{D }}}  \frac{{\tanh ( {\frac{M}{2}\sqrt {(r+s)/D} } )}}{{\sqrt {r+s} }} .
\end{eqnarray} 
\end{widetext}

\textbf{Correlation between $t_{m}$ and $t_{\ell}$}: While inverting Eq.(\ref{eq15.18}) is challenging, we can instead evaluate the derivative of Eq.(\ref{eq15.18}) at $\lambda=p=0$, integrate over $M$, and finally perform the inverse transform with respect to $s$. This procedure allows us to obtain the (joint) moments of $t_m$ and $t_{\ell}$, 
\begin{eqnarray}\label{eq15.19}
&&\int_{0}^{\infty} dt e^{-s t}\langle {t_m^{{k_1}}t_{\ell}^{{k_2}}} |t\rangle_r  \nonumber \\ &=& { {{{\left( { - 1} \right)}^{{k_1} + {k_2}}}{{{\partial ^{{k_1} + {k_2}}}\tilde{P}_r(\lambda,p,s)} \over {\partial {\lambda^{{k_1}}}\partial {p^{{k_2}}}}}} |_{\lambda = p = 0}},
\end{eqnarray}
where
\begin{eqnarray}\label{eq15.20}
\tilde{P}_r(\lambda,p,s)=\int_{0}^{\infty}dM \tilde{P}_r(M,\lambda,p,s).
\end{eqnarray}
Setting $k_1=1$ and $k_2=0$ yields the mean value of $t_m$ for RBM, which recovers the result in Eq. (\ref{eq7.5}). Setting 
$k_1=0$ and $k_2=1$ gives the mean value of $t_{\ell}$ for RBM, consistent with the result in Eq. (\ref{eq8.2a}). To compute the combined moment $\langle {t_m} { t_{\ell}}\rangle$, we set $k_1=1$ and $k_2=1$,
\begin{eqnarray}\label{eq15.21}
\int_{0}^{\infty} dt e^{-s t}\langle {t_m t_{\ell}} |t\rangle_r  = \tilde{F}_1(s)+\tilde{F}_2(s)+\tilde{F}_3(s)+\tilde{F}_4(s), \nonumber \\
\end{eqnarray}
with
\begin{widetext}
\begin{subequations}\label{eq15.22}
	\begin{align}
		&\tilde{F}_1(s)= - \frac{{( {7{r^2} - 2rs + 7{s^2}} )\zeta ( 3 )}}{{16{{( {r - s} )}^2}{{( {r + s} )}^3}}} - \frac{{24{r^5} - 30{r^4}s - 33{r^3}{s^2} + ( {51 - 2{\pi ^2}} ){r^2}{s^3} - 3r{s^4} - ( {9 + 2{\pi ^2}} ){s^5}}}{{24s{{( {s - r} )}^3}{{( {s + r} )}^2}}}, \\
		&\tilde{F}_2(s)=\frac{{\left( {{s^3} + 4r{s^2} + 5{r^2}s - 2{r^3}} \right)}}{{8r{s^2}\left( {r + s} \right){{\left( {r - s} \right)}^2}}}\ln \left( {\frac{s}{{r + s}}} \right), \\
		&\tilde{F}_3(s)=\frac{1}{{2{{\left( {s - r} \right)}^3}}}{\rm{Li}}{_2}\left( { - \frac{r}{s}} \right), \\
		&\tilde{F}_4(s)= - \frac{{rs}}{{{{\left( {r - s} \right)}^2}{{\left( {r + s} \right)}^3}}}{\rm{Li}}{_3}\left( { - \frac{r}{s}} \right),
	\end{align}
\end{subequations}
where $\zeta ( 3 )=1.20205...$ is Ap\'ery's constant, and ${\rm{Li}}_n(-\frac{r}{s})$ is the polylogarithm function.
We now take the inverse Laplace transform for each line of Eq.(\ref{eq15.22}) to obtain the moment, 
\begin{eqnarray}\label{eq15.22.1}
\langle {\tilde{t}_m \tilde{t}_{\ell}} \rangle_r=\tilde{f}_1(R)+\tilde{f}_2(R)+\tilde{f}_3(R)+\tilde{f}_4(R)
\end{eqnarray}
where 
\begin{subequations}\label{eq15.23}
	\begin{align}
		\tilde{f}_1(R)= \frac{1}{t^2}F_1(t)&= - \frac{{{e^{ - R}}[ {( {6R - 3} ){e^{2R}} + 8{R^2} + 3} ]\zeta ( 3 )}}{{64{R^2}}}  \nonumber \\ &+ \frac{{{e^{ - R}}[ {{e^{2R}}( {{\pi ^2}( {1 + 2{R^2}} ) - 12} ) + 12{e^R}( {3 + 2R( {2R - 1} )} ) - 2( {{\pi ^2} - 6} )R - 24 - {\pi ^2}} ]}}{{96{R^2}}}, \label{eq15.23a} \\
		\tilde{f}_2(R)= \frac{1}{t^2}F_2(t)&=\frac{{{e^R}}}{{8{R^2}}}[ {2 + ( { - 2 + 4R} ){\rm{Ei}}( { - 2R} ) + ( {2 - 4R} ){\rm{Ei}}( { - R} ) - 4R\ln 2 + 2\ln 2} ] \nonumber \\ &+ \frac{1}{{8{R^2}}}[ {( {3 - 2R} ){\rm{Ei}}( { - R} ) + ( { - 3 + 2R} )\ln R + 2( {{\gamma_E} - 1} )R - 2 - 3{\gamma _E}} ] + \frac{{{e^{ - R}}}}{{8{R^2}}}[ {{\rm{Ei}}( R ) - \ln R - {\gamma_E}} ], \label{eq15.23b}	\\
		\tilde{f}_3(R)= \frac{1}{t^2}F_3(t)&= - \frac{1}{4}\int_0^1 {d\tau \frac{{{\gamma _E} + \Gamma ( {0,R\tau } ) + \ln ( {R\tau } )}}{\tau }} {e^{R( {1 - \tau } )}}{( {1 - \tau } )^2}, \label{eq15.23c} \\
		\tilde{f}_4(R)= \frac{1}{t^2}F_4(t)&= \frac{1}{{16R}}\int_0^1 {d\tau } ( {{e^{ - R\tau }} - {e^{R\tau }} + 2R{e^{R\tau }}\tau  - 2{R^2}{\tau ^2}{e^{ - R\tau }}} ){}_3{F_3}[ {\left\{ {1,1,1} \right\},\left\{ {2,2,2} \right\}, - R( {1 - \tau } )} ],		\label{eq15.23d}
	\end{align}
\end{subequations}
\end{widetext}
where $\gamma_E = 0.577216...$ is Euler's constant, ${\rm{Ei}}(x)$ is the exponential integral function, $\Gamma(0,z)$  is the incomplete gamma function, and ${}_3{F_3}(\cdot)$ is the generalized hypergeometric function.

In the limit $r \to 0$, the combined moment can be reduced to
\begin{eqnarray}\label{eq15.24}
\langle {t_m t_{\ell}} |t\rangle_0=\frac{{12 + 4{\pi ^2} - 21\zeta ( 3 )}}{{96}}{t^2}=0.273284... t^2
\end{eqnarray}
such that the rescaled correlation coefficient between $\tilde{t}_m={t}_m/t$ and $\tilde{t}_{\ell}=t_{\ell}/t$ for the SBM is
\begin{eqnarray}\label{eq15.25}
\tilde{c}_0(\tilde{t}_m,\tilde{t}_{\ell})&=&\langle {\tilde{t}_m \tilde{t}_{\ell}}\rangle_0-\langle {\tilde{t}_m } \rangle_0 \langle { \tilde{t}_{\ell}} \rangle_0 \nonumber \\ &=&\frac{{4{\pi ^2} - 3[ {4 + 7\zeta ( 3 )} ]}}{{96}}=0.0232836... 
\end{eqnarray}
This implies weak anti-correlation between  $t_m$ and $t_{\ell}$ in the absence of resetting. This is a new result for SBM has not been exploited in Ref.\cite{hartmann2025exact}.

In the presence of resetting, the correlation coefficient $\tilde{c}_r(\tilde{t}_m,\tilde{t}_{\ell})$ can be computed in terms of Eqs.(\ref{eq15.22.1}), (\ref{eq15.23}), (\ref{eq7.5}) and (\ref{eq8.2a}), but the integrals in Eq.(\ref{eq15.23}) need to be evaluated numerically. When the resetting rate $r$ is small, we can express the correlation coefficient $\tilde{c}_r(\tilde{t}_m,\tilde{t}_{\ell})$ as the first two leading powers in $R$,
\begin{eqnarray}\label{eq15.26}
\tilde{c}_r(\tilde{t}_m,\tilde{t}_{\ell})&=&\tilde{c}_0(\tilde{t}_m,\tilde{t}_{\ell}) +R \Delta_1\tilde{c}_r(\tilde{t}_m,\tilde{t}_{\ell})+R^2 \Delta_2\tilde{c}_r(\tilde{t}_m,\tilde{t}_{\ell}) \nonumber \\&+&o(R^3)
\end{eqnarray}
where
\begin{subequations}\label{eq15.27}
	\begin{align}
		\Delta_1\tilde{c}_r(\tilde{t}_m,\tilde{t}_{\ell})&=\frac{{4{\pi ^2} + 27\zeta ( 3 ) - 72}}{{288}}=-0.000229326..., \\	
		\Delta_2\tilde{c}_r(\tilde{t}_m,\tilde{t}_{\ell})&=	\frac{{16{\pi ^2} - 90\zeta ( 3 ) - 67}}{{1152}}=-0.0149926... . 
	\end{align}		
\end{subequations}

\begin{figure}
	\centerline{\includegraphics*[width=1.0\columnwidth]{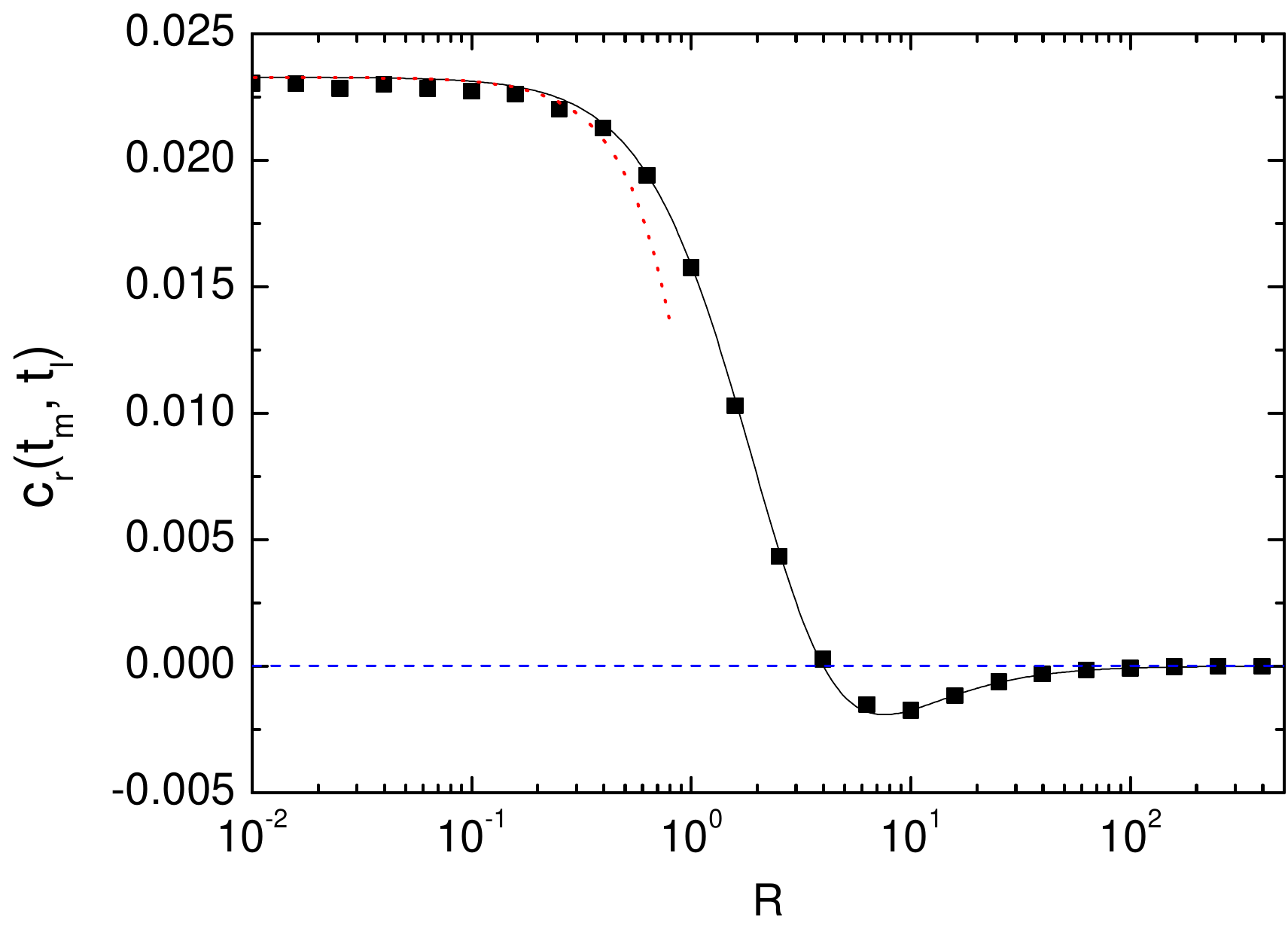}}
	\caption{Correlation coefficient of $t_m$ and $t_{\ell}$ as a function of the resetting rate $r$. The lines and symbols correspond to the theoretical and numerical results, respectively. The red dotted line indicates to the asymptotic behavior of ${c}_r(t_m;t_o) $ for $R \ll 1$, see Eqs.(\ref{eq15.26}) and (\ref{eq15.27}). The horizontal blue dashed line indicates the zero value.    \label{fig_corr_tmtl}}
\end{figure}

Figure \ref{fig_corr_tmtl} presents the correlation coefficient between $t_m$ and $t_{\ell}$ as a function of the parameter $r$. A particularly interesting feature is observed: the correlation coefficient undergoes a sign reversal from positive to negative with increasing $r$, with the transition occurring at a critical value of $R \approx 3.981$. Asymptotically, the correlation diminishes to zero as $r$ approaches infinity. For validation purposes, we have included simulation results (represented by symbols in Fig. \ref{fig_corr_tmtl}), which demonstrate excellent agreement with our analytical predictions.

\section{Conclusions}
In conclusion, we have analytically computed and numerically verified the pairwise joint distributions of three global times—$t_o$ (occupation time), $t_{\ell}$ (the last-passage time), and $t_m$ (time of maximum)—for one-dimensional resetting Brownian motion over a duration $t$. Most of these computations are based on a renewal framework for Poissonian resetting, which establishes a connection with their counterparts derived  via the Feynman-Kac formula and an $\epsilon$-path decomposition method. We demonstrate that the pairwise correlations among the three global times exhibit rich, non-trivial relationships with the resetting rate $r$. Our key findings include: (i) while all moments of $t_{\ell}$ (i.e., $t_{\ell}^m$ with any positive integer $m$) are uncorrelated with $t_o$, it anti-correlated with the higher-order moment of $t_o$. We have shown that the correlation between $t_{\ell}^m$ and $t_o^2$ weakens as $r$ increase and approaches zero in the limit $r \to \infty$; (ii) $t_o$ is positively correlated with $t_m$, and this correlation decreases with $r$, tending to zero as $r \to \infty$ in a power-law way of exponent $-2$ (with a logarithm correction); (iii) The correlation between $t_m$ and $t_{\ell}$ undergoes a sign reversal from positive to negative as $r$ increases. As a byproduct, we obtained several new results for SBM, such as the non-resetting correlation coefficient between $t_{\ell}^m$ and $t_o^2$ for general values of $m$ (see Eq.(\ref{eq11.7})), and the non-resetting correlation coefficient between $t_m$ and $t_{\ell}$ (see Eq.(\ref{eq15.25})), which have not exploited previously  \cite{hartmann2025exact}. Furthermore, the moments of $t_o$ for resetting Brownian bridge \cite{de2022optimal}  derived here (see Eq.(\ref{eq10.13})) have not reported elsewhere either.

Joint distributions provide far more system information than marginals, making these studies highly valuable. Notably, this knowledge enables practical applications, such as inferring hidden properties from easily measurable quantities through correlation analysis. In the future, it would be interesting to address similar questions for other stochastic processes, such as Ornstein-Uhlenbeck process \cite{majumdar2002large,mori2021distribution,mori2022time} and constrained Brownian motions \cite{majumdar2008time}. This remains open questions worthy of further investigations.

\begin{acknowledgments}
	This work is supported by the National Natural Science Foundation of China (11875069), the Key Scientific Research Fund of Anhui Provincial Education Department (2023AH050116) and Anhui Project (Grant No. 2022AH020009).
\end{acknowledgments}

\appendix
\section{Statistics of the last passage-time for RBM}\label{app_tl_rbm}
This section presents a derivation of the probability distribution $P_r(t_{\ell}|t)$ for the last passage-time $t_{\ell}$ of RBM. Our analysis considers two distinct  scenarios over the entire time interval $\left[0, t \right]$. On the one hand, if no reset events occur, with probability  $e^{-rt}$, the distribution of $t_{\ell}$ reduces to the non-resetting case, $P_0(t_{\ell}|t)$. On the other hand, if at least one reset happens, let us denote the last reset time by $\tau$, with probability $r d\tau$. The subsequent reset-free period $\left[ \tau, t\right] $ has probability $e^{-r(t-\tau)}$, and thus the last-passage time decomposes as $t_{\ell}=\tau+t_z$, where $t_{z}$ is a random variable drawn from the non-resetting distribution $P_0(t_{z}|t-\tau)$. Combining these cases , we obtain
\begin{eqnarray}\label{eq8.1}
{P_r}\left( {{t_{\ell}}|t} \right) = {e^{ - rt}}{P_0}\left( {{t_{\ell}}|t} \right) + r\int_0^{t} {d\tau } {e^{ - r\left( {t - \tau } \right)}}{P_0}\left( {{t_{\ell}} - \tau |t - \tau } \right), \nonumber \\
\end{eqnarray}
where $P_0(t_{\ell}|t)$ denotes the last-passage time distribution for non-resetting Brownian motion. Using the arcsine law for $t_{\ell}$, i.e.,  $P_0(t_{\ell}|t)=\frac{1}{{\pi \sqrt {{t_{\ell}}\left( {t - {t_{\ell}}} \right)} }}$, Eq.(\ref{eq8.1}) can be rewritten in terms of the rescaled variable ${\tilde t}_{\ell}={ t}_{\ell}/t$, 
\begin{eqnarray}\label{eq8.1.1}
{P_r}( {{t_{\ell}}|t} ) = \frac{1}{t}{{\tilde P}_r}\left( {{{\tilde t}_{\ell}} = \frac{{{t_{\ell}}}}{t}} \right), 
\end{eqnarray}
with the scaled distribution:
\begin{eqnarray}\label{eq8.1.2}
{{\tilde P}_r}( {{{\tilde t}_{\ell}}} ) = \frac{{{e^{ - R}}}}{{\pi \sqrt {{{\tilde t}_{\ell}}( {1 - {{\tilde t}_{\ell}}} )} }} + \frac{{\sqrt R {e^{ - R( {1 - {{\tilde t}_{\ell}}} )}}{\rm{erf}}( {R{{\tilde t}_{\ell}}} )}}{{\sqrt {\pi ( {1 - {{\tilde t}_{\ell}}} )} }},
\end{eqnarray}
where $R=rt$ represents the average number of resets over the time $t$, and ${\rm{erf}}(x)$ denotes the error function.

The moments of  $t_{\ell}$ can be directly computed from Eq.(\ref{eq8.1}),
\begin{eqnarray}\label{eq8.1.3}
{\langle {t_{\ell}^n} |t\rangle _r} &=& {e^{ - rt}}\int_0^t {d{t_{\ell}}} t_{\ell}^n{P_0}( {{t_{\ell}}|t} ) \nonumber \\&+& r\int_\tau ^t {d{t_{\ell}}} \int_0^t {d\tau } t_{\ell}^n{e^{ - r( {t - \tau } )}}{P_0}( {{t_{\ell}} - \tau |t - \tau } ) \nonumber  \\ & =& {e^{ - rt}}{\langle {t_{\ell}^n} \rangle _0} + r\int_0^t {d\tau } {e^{ - r( {t - \tau } )}}{\langle {{{\left( {{t_{\ell}} + \tau } \right)}^n}|t - \tau } \rangle _0}.\nonumber \\
\end{eqnarray}
Here,  ${\langle { {t_{\ell}^n} |t} \rangle _0} = {t^n}\left( {2n} \right)!/\left[ {\left( {2n} \right)!!} \right]^2$ are the moments of $t_{\ell}$ for SBM, readily derived from the arcsine law.

The first three scaled moments of $t_{\ell}$ for RBM read
\begin{subequations}\label{eq8.2} 
	\begin{align}
		\langle \tilde{t}_{\ell} \rangle_r& =1 + \frac{{{e^{ - R}} - 1}}{{2R}}, \label{eq8.2a}\\
		\langle \tilde{t}_{\ell}^2 \rangle_r&=1 + \frac{{{e^{ - R}}( {R - 3} ) - 4R + 3}}{{4{R^2}}},  \label{eq8.2b} \\
		\langle \tilde{t}_{\ell}^3 \rangle_r&=1 + \frac{3{e^{ - R}}[ {R( {R - 2} ) + 10} ] }{{16{R^3}}} -\frac{  6( {4{R^2} - 6R + 5} )}{{16{R^3}}}. \label{eq8.2c} 
	\end{align}
\end{subequations}
These moments demonstrate monotonic convergence to unity as $R\rightarrow\infty$. We note that these results have been independently reported in recent work \cite{tazbierski2025arcsine}.

\section{Occupation time statistics for resetting Brownian bridges}\label{app_to_rbb}
We analyze the occupation time statistics of a resetting Brownian bridge by introducing the joint probability density $G_r(x,t,t_o|x_0;x_r)$, which describes a resetting Brownian particle starting at $x_0$ reaching position $x$ at time $t$ with occupation time $t_o$ above the origin. Through renewal theory, we establish the governing equation:
\begin{eqnarray}\label{eq10.1}
&&G_r(x,t,t_o|x_0;x_r)=e^{-rt}G_0(x,t,t_o|x_0) +r\int_{0}^{t_o}d{t'_o} \int_{0}^{t}d\tau \nonumber \\&\times& \int_{-\infty}^{\infty} dx' e^{-r\tau}  G_r(x',t-\tau,t'_o|x_0;x_r)G_0(x,\tau,t_o-t'_o|x_r).\nonumber \\
\end{eqnarray}
The interpretation of Eq.(\ref{eq10.1}) is analogous to that of Eq.(\ref{eq5.3}), with the key difference being the removal of the absorbing boundary here. We now perform a double Laplace transform with respect to $t_o$ and $t$, defined as $\tilde{G}_r(x,s,p|x_0;x_r)=\int_{0}^{\infty} dt e^{-st} \int_{0}^{t} dt_o e^{-p t_o} G_r(x,t,t_o|x_0;x_r)$. This transfroms Eq.(\ref{eq10.1}) into:
\begin{eqnarray}\label{eq10.2}
{\tilde G_r}( {x,s,p|{x_0};{x_r}} ) &=& {\tilde G_0}( {x,r + s,p|{x_0}} ) + r{\tilde G_0}( {x,r + s,p|{x_r}} ) \nonumber \\ &\times&\int_{ - \infty }^\infty  {dx'} {\tilde G_r}( {x',s,p|{x_0};{x_r}} ).
\end{eqnarray}
Integrating Eq.(\ref{eq10.2}) over $x$ yields
\begin{eqnarray}\label{eq10.20}
\int_{ - \infty }^\infty  {dx} {{\tilde G}_r}( {x,s,p|{x_0};{x_r}} ) = \frac{{\int_{ - \infty }^\infty  {dx} {{\tilde G}_0}( {x,r + s,p|{x_0}} )}}{{1 - r\int_{ - \infty }^\infty  {dx} {{\tilde G}_0}( {x,r + s,p|{x_r}} )}}.\nonumber \\
\end{eqnarray}
Notably, for $x_0=x_r=0$, it was known from \cite{hartmann2025exact} that,
\begin{eqnarray}\label{eq10.21}
{{\tilde G}_0}( {x,s,p|0} )= \left\{ \begin{array}{llc}
\frac{{{e^{\sqrt {s/D} x}}}}{{\sqrt D ( {\sqrt s  + \sqrt {s + p} } )}}, &     x<0,  \\
\frac{{{e^{ - \sqrt {\left( {s + p} \right)/D} x}}}}{{\sqrt D ( {\sqrt s  + \sqrt {s + p} } )}}, &    x>0,  \\ 
\end{array}  \right.
\end{eqnarray}
and thus
\begin{eqnarray}\label{eq10.22}
\int_{ - \infty }^\infty  {dx} {{\tilde G}_0}( {x,s,p|0} ) = \frac{1}{{\sqrt {s\left( {s + p} \right)} }}.
\end{eqnarray}
Using Eq.(\ref{eq10.22}) and then inserting Eq.(\ref{eq10.20}) into Eq.(\ref{eq10.2}), we obtain
\begin{eqnarray}\label{eq10.23}
{{\tilde G}_r}( {x,s,p|0;0} ) = {{\tilde G}_0}( {x,r + s,p|0} )\frac{{\sqrt {( {r + s} )( {r + s + p} )} }}{{\sqrt {( {r + s} )( {r + s + p} )}  - r}}. \nonumber \\
\end{eqnarray}
For $x=0$, i.e., resetting Brownian bridge \cite{de2022optimal}, Eq.(\ref{eq10.23}) simplifies to
\begin{eqnarray}\label{eq10.24}
{{\tilde G}_r}( {0,s,p|0;0} ) &=& \frac{1}{{\sqrt D ( {\sqrt {r + s}  + \sqrt {r + s + p} } )}} \nonumber \\ & \times &\frac{{\sqrt {( {r + s} )( {r + s + p} )} }}{{\sqrt {( {r + s} )( {r + s + p} )}  - r}}.
\end{eqnarray}
The moments of $t_o$ for the resetting Brownian bridge can be conveniently computed as
\begin{eqnarray}\label{eq10.12}
\left\langle {t_o^n|t} \right\rangle _r^{bridge} = {\left( { - 1} \right)^n}\mathop {\lim }\limits_{p \to 0} \frac{{\partial _p^n{{\tilde g}_r}( {0,t,p|0;0} )}}{{{{\tilde g}_r}( {0,t,p|0;0} )}},
\end{eqnarray}
where ${{\tilde g}_r}( {0,t,p|0;0} )$ is the inverse Laplace transform of ${{\tilde G}_r}( {0,s,p|0;0} )$ with respect to $s$.

\begin{figure}
	\centerline{\includegraphics*[width=1.0\columnwidth]{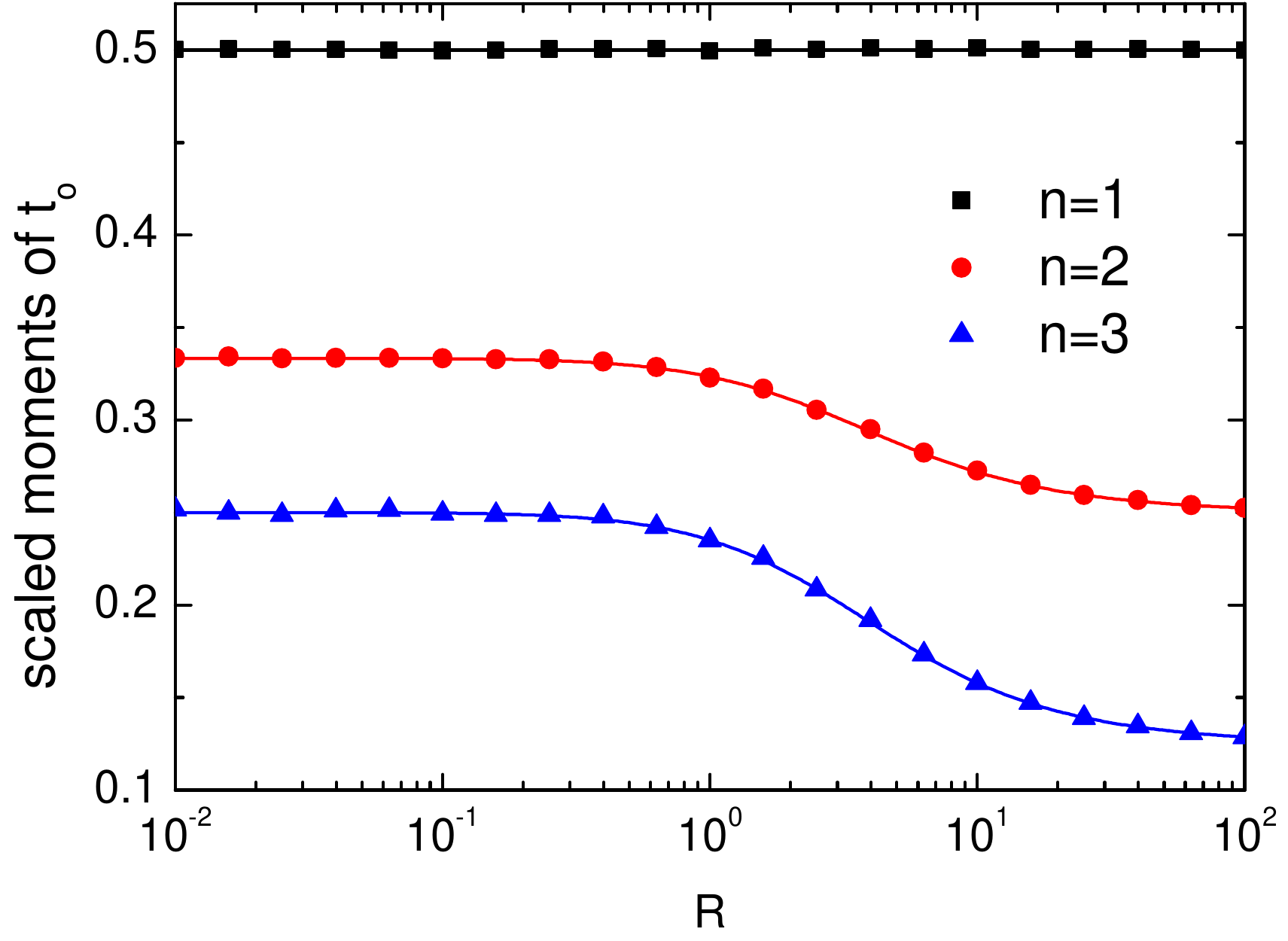}}
	\caption{The scaled moments of the occupation time $t_o$ spent on the positive semi-axis, $\langle {\tilde{t}_o^n} \rangle _r^{bridge}$ for the resetting Brownian bridge, with $n=1$, 2, and 3. Lines and symbols represent the theoretical (Eq.(\ref{eq10.13})) and simulation results, respectively. \label{fig_rbbto}}
\end{figure}

The first three scaled moments are given explicitly by,
\begin{widetext}
\begin{subequations}\label{eq10.13} 
	\begin{align}
		\langle {\tilde{t}_o} \rangle _r^{bridge}&=\frac{1}{2},\label{eq10.13a}\\
		\langle {\tilde{t}_o^2} \rangle _r^{bridge}&=\frac{{2R\left( {2R + 5} \right) + {e^R}\sqrt {\pi R} \left[ {4R\left( {R + 1} \right) - 5} \right]{\rm{erf}}( {\sqrt R } )}}{{16{R^2}\left[ {1 + {e^R}\sqrt {\pi R} {\rm{erf}}( {\sqrt R } )} \right]}}, \label{eq10.13b} \\
		\langle {\tilde{t}_o^3} \rangle _r^{bridge}&=\frac{{2R\left( {2R + 15} \right) + {e^R}\sqrt {\pi R} \left[ {4R\left( {3 + R} \right) - 15} \right]{\rm{erf}}( {\sqrt R } )}}{{32{R^2}\left[ {1 + {e^R}\sqrt {\pi R} {\rm{erf}}( {\sqrt R } )} \right]}},\label{eq10.13c} 
	\end{align}
\end{subequations}
\end{widetext}
where again $R=rt$ is the mean number of resets. The scaled first moment of $t_o$ is a constant, regardless of resetting. In the limit $r \to 0$, the scaled second and third moments reduce to their non-resetting counterparts, $\langle {\tilde{t}_o^2} \rangle _0^{bridge}=1/3$ and $\langle {\tilde{t}_o^3} \rangle _0^{bridge}=1/4$. In the opposite limiting case $r \to \infty$, they approach $\langle {\tilde{t}_o^2} \rangle _r^{bridge} \to 1/4$ and $\langle {\tilde{t}_o^3} \rangle _r^{bridge} \to 1/8$. Between these two limits, both the second and third moments decrease monotonically with the resetting rate (see Fig.\ref{fig_rbbto}).

To our knowledge, our results in Eq.(\ref{eq10.13}) have not reported yet. To validate the analytical findings, we perform simulations for the resetting Brownian bridge, where each bridge is generated numerically by an effective Langevin equation (see \cite{de2022optimal} for details). The simulation results (symbols) are plotted in Fig.\ref{fig_rbbto} alongside the theoretical predictions (lines), showing excellent agreement and thus confirming the validity of our theory.

\section{Survival probability of RBM in a semi-infinite space and with an occupation time weight}\label{sec_app_sp}
This appendix derives the survival probability $Q_r^M(x_0,t,t_o;x_r)$ for RBM 
in a semi-infinite domain with an absorbing boundary at $x=M$, while simultaneously tracking the occupation time $t_o$ above the origin. Formally, $Q_r^M(x_0,t,t_o;x_r)$ is the probability that a path of RBM has not touched an absorbing wall at $x=M$ up to time $t$, with carrying an occupation time $t_o$, providing the particle starts from $x_0$. To do this, we first derive non-resetting counterpart $Q_0^M(x_0,t,t_o)$, and then extends the solution to include stochastic resetting through renewal theory \cite{chechkin2018random,evans2020stochastic}.

\subsection{Non-Resetting survival probability in a semi-infinite space with an occupation time weight}
As mentioned before, let us define $Q_0^M(x_0,t,t_o)$ as the probability that a free Brownian path, starting at $x_0<M$ at time 0, stays below $M$ up to time
$t$ and carries an occupation time $t_o$ in $\left[0, t \right] $. Taking the Laplace transform for $Q_0^M(x_0,t,t_o)$ with respect to $t_o$, 
\begin{eqnarray}\label{eq4.1}
\tilde{q}_0^M(x_0,t,p)=\int_{0}^{t} dt_o e^{-p t_o}Q_0^M(x_0,t,t_o),
\end{eqnarray}  
The backward equation for $\tilde{q}^M_0(x_0,t,p)$ reads \cite{hartmann2025exact}
\begin{eqnarray}\label{eq4.2}
\frac{\partial \tilde{q}_0^M}{\partial t}=D  \frac{\partial^2 \tilde{q}_0^M}{\partial x_0^2}-p \theta(x_0) \tilde{q}_0^M, 
\end{eqnarray} 
subject to an absorbing boundary, $\tilde{q}_0^M(M,t,p)=0$, and initial condition, $\tilde{q}_0^M(x_0,0,p)=1$ for $x_0<M$. Performing a further Laplace transform with respect to $t$, $\tilde{Q}_0^M(x_0,s,p)=\int_{0}^{\infty} dt e^{-s t}\tilde{q}_0^M(x_0,t,p)$, Eq.(\ref{eq4.2}) becomes
\begin{eqnarray}\label{eq4.3}
D \frac{\partial^2 \tilde{Q}_0^M}{\partial x_0^2}-[p \theta(x_0)+s ] \tilde{Q}_0^M=-1.  
\end{eqnarray} 
Eq.(\ref{eq4.3}) can be solved subject to the absorbing boundary condition at $x_0=M$, which yields
\begin{widetext}
\begin{eqnarray}\label{eq4.4}
\tilde{Q}_0^M(x_0,s,p) = \left\{ \begin{array}{llc}
\frac{1}{s} - \frac{{{e^{\sqrt {s/D} {x_0}}}}}{{s\sqrt {p + s} }}\frac{{s + p\cosh \left( {M\sqrt {\frac{{p + s}}{D}} } \right)}}{{\sqrt s \sinh \left( {M\sqrt {\frac{{p + s}}{D}} } \right) + \sqrt {p + s} \cosh \left( {M\sqrt {\frac{{p + s}}{D}} } \right)}}, &   x_0<0,  \\
\frac{1}{{p + s}} - \frac{{\sqrt s \sqrt {p + s} \cosh \left( {{x_0}\sqrt {\frac{{p + s}}{D}} } \right) - p\sinh \left[ {( {M - {x_0}} )\sqrt {\frac{{p + s}}{D}} } \right] + s\sinh \left( {{x_0}\sqrt {\frac{{p + s}}{D}} } \right)}}{{\sqrt s ( {p + s} )\left[ {\sqrt {p + s} \cosh \left( {M\sqrt {\frac{{p + s}}{D}} } \right) + \sqrt s \sinh \left( {M\sqrt {\frac{{p + s}}{D}} } \right)} \right]}}, &   0<x_0\leq M.  \\ 
\end{array}  \right. 
\end{eqnarray}
\end{widetext}

\subsection{Resetting survival probability in a semi-infinite space with an occupation time weight}
For RBM with resetting to $x_r$ at rate $r$, the survival probability $Q_r^M(x_0,t,t_o;x_r)$ obeys the renewal equation \cite{chechkin2018random,evans2020stochastic}, 
\begin{eqnarray}\label{eq5.1}
&&{Q_r^M}( {{x_0},t,{t_o};{x_r}} ) = {e^{ - rt}}{Q_0^M}( {{x_0},t,{t_o}} ) + r\int_0^{{t_o}} {d{{t'}_o}} \int_0^t d\tau \nonumber \\ &\times&{e^{ - r\tau }} {Q_r^M}( {{x_0},t - \tau ,{{t'}_o};{x_r}} ){Q_0^M}( {{x_r},\tau ,{t_o} - {{t'}_o}} ).
\end{eqnarray}
Here, the first term on the right-hand side of Eq.(\ref{eq5.1}) implies that the particle survives until time $t$ without experiencing any reset event. The second term considers the possibility when there is at least one reset event. One considers that the last reset event has occurred at time $t-\tau$, and after that there has been no reset for the duration $\tau$. This probability is given by $r d\tau e^{-r\tau}$, but then this has to be multiplied by $Q_r^M(x_0,t-\tau,{t'}_o;x_r)$, i.e., the probability that the particle survives until time $t-\tau$ with carrying an occupation time ${t'}_o$, and $Q_0^M(x_r,\tau,t_o-{t'}_o)$, i.e., the probability of the particle for the last non-resetting interval $\tau$ and carrying an occupation time $t_o-{t'}_o$. 

To utilize the convolutional structure of Eq.(\ref{eq5.1}), we take a double Laplace transform with respect to $t_o$ and $t$, $\tilde{Q}_r^M(x_0,s,p;x_r)=\int_{0}^{\infty} dt e^{-st} \int_{0}^{t} dt_o e^{-p t_o} Q_r(x_0,t,t_o;x_r)$, and thus  Eq.(\ref{eq5.1}) leads to
\begin{eqnarray}\label{eq5.2}
{{\tilde Q}_r^M}( {{x_0},s,p;{x_r}} ) = \frac{{{{\tilde Q}_0^M}( {{x_0},r + s,p} )}}{{1 - r{{\tilde Q}_0^M}( {{x_r},r + s,p} )}}.
\end{eqnarray}

\section{Propagator of RBM in a semi-infinite space with an occupation time weight}
\label{sec_app_pg}
This appendix presents the derivation of the propagator $G_r^M(x,t,t_o|x_0;x_r)$ for
RBM in a in a semi-infinite domain with an absorbing boundary at $x=M$, while carrying an occupation time $t_o$ above the origin. Formally,  $G_r^M(x,t,t_o|x_0;x_r)$ is the probability density of finding a resetting Brownian particle at $x$ at time $t$ in the presence of an absorbing wall at $x=M$, and simultaneously carry an occupation time $t_o$ over the duration. Similar to the last appendix, our methodology proceeds in two stages: first establishing the non-resetting propagator $G_0^M(x,t,t_o|x_0)$, then incorporating resetting effects through renewal theory \cite{chechkin2018random,evans2020stochastic}.

\subsection{Non-Resetting propagator in a semi-infinite space with an occupation time weight}

We first perform the Laplace transform for the occupation-time-weighted propagator $G_0^M(x,t,t_o|x_0)$ of SBM with respect to $t_o$, 
\begin{eqnarray}\label{eq2.1}
\tilde{g}_0^M(x,t,p|x_0)=\int_{0 }^{t} d t_o e^{-p t_o} G_0^M(x,t,t_o|x_0),
\end{eqnarray} 
which satisfies the Feynman-Kac equation \cite{hartmann2025exact}
\begin{eqnarray}\label{eq2.2}
\frac{\partial \tilde{g}_0^M}{\partial t}=D \frac{\partial^2 \tilde{g}_0^M}{\partial x^2}-p \theta(x)  \tilde{g}_0^M, 
\end{eqnarray} 
subject to the absorbing boundary at $x=M$, $\tilde{g}_0^M(M,t,p|x_0)=0$, and initial condition, $\tilde{g}_0^M(x,0,p|x_0)=\delta(x-x_0)$. Furthermore, we perform a Laplace transform with respect to $t$, 
\begin{eqnarray}\label{eq2.3}
\tilde{G}_0^M(x,s,p|x_0)=\int_{0 }^{\infty} d t e^{-s t} \tilde{g}_0^M(x,t,p|x_0),
\end{eqnarray} 
and have
\begin{eqnarray}\label{eq2.4}
D \frac{d^2 \tilde{G}_0^M}{d x^2}- [  p \theta(x)+s ]  \tilde{G}_0^M =-\delta(x-x_0).
\end{eqnarray} 
Setting $x_0=0$, which will be the initial position considered below, Eq.(\ref{eq2.4}) can be solved for $x<0$ and $0<x<M$, separately, in combination with the boundary conditions, $\tilde{G}_0^M(M,s,p|0)=\tilde{G}_0^M(-\infty,s,p|0)=0$, and  
continuity conditions at $x=0$, $\tilde{G}_0^M(0^+,s,p|0)=\tilde{G}_0^M(0^-,s,p|0)$ and ${ {{\partial _x}\tilde G_0^M\left( {x,s,p} |0\right)} |_{x = {0^ + }}} - { {{\partial _x}\tilde G_0^M\left( {x,s,p}|0 \right)} |_{x = {0^ - }}} =  - 1/D$. The solution is
\begin{widetext}
\begin{eqnarray}\label{eq2.5}
\tilde{G}_0^M(x,s,p|0) = \left\{ \begin{array}{llc}
\frac{{\sinh \left( {M\sqrt {\frac{{p + s}}{D}} } \right)\left[ {\sinh \left( {x\sqrt {\frac{s}{D}} } \right) + \cosh \left( {x\sqrt {\frac{s}{D}} } \right)} \right]}}{{\sqrt D \left[ {\sqrt s \sinh \left( {M\sqrt {\frac{{p + s}}{D}} } \right) + \sqrt {p + s} \cosh \left( {M\sqrt {\frac{{p + s}}{D}} } \right)} \right]}}, &   x<0,  \\
\frac{{\sinh \left[ {\left( {M - x} \right)\sqrt {\frac{{p + s}}{D}} } \right]}}{{\sqrt D \left[ {\sqrt s \sinh \left( {M\sqrt {\frac{{p + s}}{D}} } \right) + \sqrt {p + s} \cosh \left( {M\sqrt {\frac{{p + s}}{D}} } \right)} \right]}}, &   0<x<M.  \\ 
\end{array}  \right. 
\end{eqnarray}
\end{widetext}

Taking the limit $M \to \infty$ for Eq.(\ref{eq2.5}), we can obtain the result when the absorbing boundary is absent, i.e., the occupation-time-weighted propagator in an infinite domain
\begin{eqnarray}\label{eq2.6}
\tilde{G}_0(x,s,p|0) = \left\{ \begin{array}{llc}
\frac{{{e^{x\sqrt {s/D} }}}}{{\sqrt D \left( {\sqrt s  + \sqrt {p + s} } \right)}}, &   x<0,  \\
\frac{{{e^{ - x\sqrt {\left( {p + s} \right)/D} }}}}{{\sqrt D \left( {\sqrt s  + \sqrt {p + s} } \right)}}, &   x>0.  \\ 
\end{array}  \right. 
\end{eqnarray}

\subsection{Resetting propagator in a semi-infinite space with an occupation time weight}
For RBM with resetting to $x_r$ at rate $r$, the propagator satisfies the renewal equation \cite{chechkin2018random,evans2020stochastic},
\begin{eqnarray}\label{eq5.3}
&&{G_r^M}( {x,t,{t_o}|{x_0};{x_r}} ) = {e^{ - rt}}{G_0^M}( {x,t,{t_o}|{x_0}} ) + r\int_0^{{t_o}} {d{{t'}_o}} \nonumber \\ &\times& \int_0^t d\tau  {e^{ - r\tau }} {Q_r^M}( {{x_0},t - \tau ,{{t'}_o};{x_r}} ){G_0^M}( {x,\tau ,{t_o} - {{t'}_o}|{x_r}} ),\nonumber \\ 
\end{eqnarray}
where ${Q_r^M}(x_0,t,t_0;x_r)$ is survival probability of RBM with carrying an information about occupation time (see appendix \ref{sec_app_sp} for details), and the interpretations for the right-hand side of Eq.(\ref{eq5.3}) are similar to those in Eq.(\ref{eq5.1}). Also, we take a double Laplace transform with respect to $t_o$ and $t$, $\tilde{G}_r^M(x,s,p|x_0;x_r)=\int_{0}^{\infty} dt e^{-st} \int_{0}^{t} dt_o e^{-p t_o} G_r^M(x,t,t_o|x_0;x_r)$, and thus Eq.(\ref{eq5.3}) becomes
\begin{eqnarray}\label{eq5.4}
{{\tilde G}_r^M}( {x,s,p|{x_0};{x_r}} )& =& {{\tilde G}_0^M}( {x,r + s,p|{x_0}} ) + r{{\tilde Q}_r^M}( {x_0,s,p;{x_r}} )\nonumber \\  &\times &{{\tilde G}_0^M}( {x,r + s,p|{x_r}} ).
\end{eqnarray}
Setting $x_0=x_r=0$, Eq.(\ref{eq5.4}) can be reduced to 
\begin{eqnarray}\label{eq5.4.1}
{{\tilde G}_r^M}( {x,s,p|{0};{0}} ) = {{\tilde G}_0^M}( {x,r + s,p|{0}} )[ 1+ r{{\tilde Q}_r^M}( {0,s,p;{0}} )],\nonumber \\ 
\end{eqnarray}
where ${{\tilde G}_0^M}( {x, s,p|{0}} )$ is given by Eq.(\ref{eq2.5}), and ${{\tilde Q}_r^M}( {0,s,p;{0}} )$ by Eq.(\ref{eq5.2}).

\section{Survival probability of RBM in a semi-infinite space or in a box}\label{sec_app_rmbsp}
This section presents the survival probability analysis for RBM in two scenarios: (i) a semi-infinite domain with an absorbing boundary and (ii) a finite interval with absorbing boundaries. We first derive the survival probabilities for free Brownian motion (without resetting) and then extend these results to RBM using renewal theory.

\subsection{Survival probability of SBM in a semi-infinite space with an absorbing wall}
or a free Brownian particle in the presence of an absorbing boundary at 
$x=M$, the propagator can be obtained using the method of images \cite{redner2001guide}, 
\begin{eqnarray}\label{eqa1.1}
G_0^M(x,t|x_0)=\frac{1}{\sqrt{4{\pi}Dt}} [  e^{-\frac{(x-x_0)^2}{4Dt}} -  e^{-\frac{(x-2M+x_0)^2}{4Dt}} ] .
\end{eqnarray}
The survival probability, representing the likelihood that the particle has not yet touched the absorbing wall up to time $t$, starting from $x_0<M$, is given by integrating the propagator over all positions below $M$
\begin{eqnarray}\label{eqa1.2}
Q_0^M(x_0,t)&=&\int_{-\infty}^{M} dx G_0^M(x,t|x_0) \nonumber \\  &=&{\rm{erf}}\left( \frac{|M-x_0|}{\sqrt{4Dt}}\right) .
\end{eqnarray}
For later convenience, we also provide the Laplace transforms of the propagator and survival probability:
\begin{eqnarray}\label{eqa1.4}
\tilde{G}_0^M(x,s|x_0)&=&\int_{0}^{\infty} dt e^{-st} G_0^M(x,t|x_0) \nonumber \\ &=& \frac{e^{-\alpha_0 |x-x_0|}-e^{-\alpha_0 |x-2M+x_0|}}{2\sqrt{Ds}},
\end{eqnarray}
and
\begin{eqnarray}\label{eqa1.3}
\tilde{Q}_0^M(x_0,s)&=&\int_{0}^{\infty} dt e^{-st} Q_0^M(x_0,t)\nonumber \\ &=& \frac{1-e^{-\alpha_0 |M-x_0|}}{s},
\end{eqnarray}
where $\alpha_0=\sqrt{s/D}$. 

\subsection{Survival probability of SBM in a box with absorbing boundaries}
For a Brownian particle confined within an interval $\left[0, M \right] $ with absorbing boundaries at both ends, the propagator can be derived using separation of variables \cite{redner2001guide},
\begin{eqnarray}\label{eqa2.1}
G_0^{{\rm{box}}}(x,t|x_0)=\frac{2}{M}\sum_{n=1}^{\infty} \sin\left( \frac{n{\pi}x}{M} \right)  \sin\left( \frac{n{\pi}x_0}{M} \right) e^{-\frac{n^2 {\pi}^2 D t}{M^2}}. \nonumber \\ 
\end{eqnarray}
The corresponding survival probability is obtained by integrating the propagator over the interval,  
\begin{eqnarray}\label{eqa2.2}
&&Q_0^{{\rm{box}}}(x_0,t)=\int_{0}^{M} dx G_0^{{\rm{box}}}(x,t|x_0)\nonumber \\  &=&\frac{4}{\pi}\sum_{n=0}^{\infty} 
\frac{1}{2n+1} \sin\left[  \frac{(2n+1){\pi}x_0}{M}\right]    e^{-\frac{(2n+1)^2 {\pi}^2 D t}{M^2}}.
\end{eqnarray}
The Laplace transforms of these quantities are:
\begin{eqnarray}\label{eqa2.4}
&&\tilde{G}_0^{{\rm{box}}}(x,s|x_0)=\int_{0}^{\infty} dt e^{-st} G_0^{{\rm{box}}}(x,t|x_0)\nonumber \\  &=&\frac{{\cosh [ \alpha_0 {( {M - \left| {x - {x_0}} \right|} ) } ] - \cosh [ \alpha_0 {( {M - x - {x_0}} ) } ]}}{{2\sqrt {sD} \sinh ( \alpha_0 {M } )}},\nonumber \\
\end{eqnarray}
and
\begin{eqnarray}\label{eqa2.3}
\tilde{Q}_0^{{\rm{box}}}(x_0,s)&=&\int_{0}^{\infty} dt e^{-st} Q_0^{{\rm{box}}}(x_0,t)\nonumber \\ &=&\frac{{1 + {e^{{\alpha _0}M}} - {e^{{\alpha _0}{x_0}}} - {e^{{\alpha _0}( {M - {x_0}} )}}}}{{s( {1 + {e^{{\alpha _0}M}}} )}},
\end{eqnarray}
where again $\alpha_0=\sqrt{s/D}$. 

\subsection{Survival probability of RBM with an absorbing wall}
Let $Q_r^M(x_0,t;x_r)$ denote the survival probability for a RBM particle starting from $x_0$, with resetting rate $r$ and resetting position $x_r$, ensuring the particle remains below 
$M$ up to time $t$. Using renewal theory, this probability can be expressed in terms of the survival probability without resetting: 
\begin{eqnarray}\label{eqb1.1}
Q_r^M(x_0,t;x_r)&=&e^{-rt} Q_0^M(x_0,t)+r\int_{0}^{t}dt e^{-r\tau} \nonumber \\ &\times& Q_r^M(x_0,t-\tau;x_r)Q_0^M(x_0,\tau).
\end{eqnarray}
Here, the first term on the right-hand side of Eq.(\ref{eqb1.1}) implies that the particle survives until time $t$ without experiencing any reset event. The second term considers the possibility when there is at least one reset event. One considers that the last reset event has occurred at time $t-\tau$, and after that there has been no reset for the duration $\tau$. This probability is given by $r d\tau e^{-r\tau}$, but then this has to be multiplied by $Q_r^M(x_0,t-\tau;x_r)$, i.e., the probability that the particle survives until time $t-\tau$, and $Q_0^M(x_r,\tau)$, i.e., the probability of the particle surviving for the last non-resetting interval $\tau$. Taking tha Laplace transform for Eq.(\ref{eqb1.1}) with respect to $t$, which yields
\begin{eqnarray}\label{eqb1.2}
\tilde{Q}_r^M(x_0,s;x_r)=\frac{\tilde{Q}_0^M(x_0,r+s)}{1-r\tilde{Q}_0^M(x_r,r+s)}.
\end{eqnarray}
Substituting Eq.(\ref{eqa1.3}) into Eq.(\ref{eqb1.2}), and setting $x_r=0$, we obtain
\begin{eqnarray}\label{eqb1.3}
\tilde{Q}_r^M(x_0,s;0)=\frac{e^{{\alpha} M}-e^{{\alpha} x_0}}{r+s e^{{\alpha}M}},
\end{eqnarray}
where $\alpha=\sqrt{(r+s)/D}$.

\subsection{Survival probability of RBM in a box with absorbing boundaries}
For RBM in a finite interval $\left[0, M \right]$ with absorbing boundaries at both ends, the survival probability $Q_r^{\rm{box}}(x_0,t;x_r)$ can similarly be derived using renewal theory \cite{chechkin2018random,evans2020stochastic}, 
\begin{eqnarray}\label{eqb4.1}
\tilde{Q}_r^{\rm{box}}(x_0,s;x_r)=\frac{\tilde{Q}_0^{\rm{box}}(x_0,r+s)}{1-r\tilde{Q}_0^{\rm{box}}(x_r,r+s)}.
\end{eqnarray}
Using the result in Eq.(\ref{eqa2.3}) for the non-resetting case, we arrive at 
\begin{widetext}
\begin{eqnarray}\label{eqb4.2}
{{\tilde Q}_r^{\rm{box}}}( {{x_0},s};x_r ) =\frac{{\sinh ( {\alpha M} ) - \sinh ( {\alpha {x_0}} ) - \sinh \left[ {\alpha ( {M - {x_0}} )} \right]}}{{s\sinh ( {\alpha M} ) + r\sinh ( {\alpha {x_r}} ) + r\sinh \left[ {\alpha ( {M - {x_r}} )} \right]}},
\end{eqnarray}
\end{widetext}
where $\alpha=\sqrt{(r+s)/D}$ again.

\section{Propagator of RBM in a semi-infinite space or in a box}\label{sec_app_rmbpr}

\subsection{Propagator of RBM  with an absorbing wall}
We define $G_r^M(x,t|x_0;x_r)$ as the probability density function of RBM (with a resetting position $x_r$ and a resetting rate $r$) being the position $x$ at time $t$, starting from $x_0$ at time 0, in the presence of an absorbing boundary at $x=M>0$. To compute the propagator $G_r^M(x,t|x_0;x_r)$, we again use a renewal framework \cite{chechkin2018random,evans2020stochastic},  
\begin{eqnarray}\label{eqb2.1}
G_r^M(x,t|x_0;x_r) &=& {e^{ - rt}}{G_0^M}( {x,t|{x_0}} ) + r \int_0^t {d\tau {e^{ - r\tau }}} \nonumber \\ &\times& Q_r(x_0,t-\tau;x_r)  G_0^M(x,\tau|x_r).
\end{eqnarray}
Here, the first term accounts for trajectories with no resetting events, while the second term incorporates the effect of resetting through a convolution over all possible last resetting times $\tau$. Taking the Laplace transform for Eq.(\ref{eqb2.1}) with respect to $t$, we obtain
\begin{eqnarray}\label{eqb2.2}
\tilde{G}_r^M(x,s|x_0;x_r) &=& \tilde{G}_0^M(x,r+s|x_0) \nonumber \\ &+& r\tilde{Q}_r^M(x_0,s;x_r)\tilde{G}_0^M(x,r+s|x_r).
\end{eqnarray}
Using Eqs.(\ref{eqa1.4}) and (\ref{eqb1.2}), we can derive the explicit expression for $\tilde{G}_r^M(x,s|x_0;x_r)$; the resulting expression is lengthy and not shown here for brevity.

\subsection{Propagator of RBM  in a box with absorbing boundaries }
For a Brownian particle with resetting confined to an interval 
$\left[0, M \right] $ with absorbing boundaries, the propagator follows a similar renewal structure,
\begin{eqnarray}\label{eqb5.1}
\tilde{G}_r^{\rm{box}}(x,s|x_0;x_r) &= & \tilde{G}_0^{\rm{box}}(x,r+s|x_0)  + r\tilde{Q}_r^{\rm{box}}(x_0,s;x_r) \nonumber \\ &\times &\tilde{G}_0^{\rm{box}}(x,r+s|x_r). 
\end{eqnarray}
Again, using Eqs.(\ref{eqa2.4}) and (\ref{eqb4.2}), we can derive the explicit expression for $\tilde{G}_r^{\rm{box}}(x,s|x_0;x_r)$, though we omit the detailed form due to its complexity.


\begin{thebibliography}{143}
	\expandafter\ifx\csname natexlab\endcsname\relax\def\natexlab#1{#1}\fi
	\expandafter\ifx\csname bibnamefont\endcsname\relax
	\def\bibnamefont#1{#1}\fi
	\expandafter\ifx\csname bibfnamefont\endcsname\relax
	\def\bibfnamefont#1{#1}\fi
	\expandafter\ifx\csname citenamefont\endcsname\relax
	\def\citenamefont#1{#1}\fi
	\expandafter\ifx\csname url\endcsname\relax
	\def\url#1{\texttt{#1}}\fi
	\expandafter\ifx\csname urlprefix\endcsname\relax\def\urlprefix{URL }\fi
	\providecommand{\bibinfo}[2]{#2}
	\providecommand{\eprint}[2][]{\url{#2}}
	
	\bibitem[{\citenamefont{Feller}(1971)}]{feller1971introduction}
	\bibinfo{author}{\bibfnamefont{W.}~\bibnamefont{Feller}},
	\bibinfo{journal}{Vols. I \& II, Wiley I} \textbf{\bibinfo{volume}{968}}
	(\bibinfo{year}{1971}).
	
	\bibitem[{\citenamefont{M{\"o}rters and Peres}(2010)}]{morters2010brownian}
	\bibinfo{author}{\bibfnamefont{P.}~\bibnamefont{M{\"o}rters}} \bibnamefont{and}
	\bibinfo{author}{\bibfnamefont{Y.}~\bibnamefont{Peres}},
	\emph{\bibinfo{title}{Brownian motion}}, vol.~\bibinfo{volume}{30}
	(\bibinfo{publisher}{Cambridge University Press}, \bibinfo{year}{2010}).
	
	\bibitem[{\citenamefont{Klafter and Sokolov}(2011)}]{klafter2011first}
	\bibinfo{author}{\bibfnamefont{J.}~\bibnamefont{Klafter}} \bibnamefont{and}
	\bibinfo{author}{\bibfnamefont{I.~M.} \bibnamefont{Sokolov}},
	\emph{\bibinfo{title}{First steps in random walks: from tools to
			applications}} (\bibinfo{publisher}{Oxford University Press},
	\bibinfo{year}{2011}).
	
	\bibitem[{\citenamefont{Mazo}(2008)}]{mazo2008brownian}
	\bibinfo{author}{\bibfnamefont{R.~M.} \bibnamefont{Mazo}},
	\emph{\bibinfo{title}{Brownian motion: fluctuations, dynamics, and
			applications}}, vol. \bibinfo{volume}{112} (\bibinfo{publisher}{OUP Oxford},
	\bibinfo{year}{2008}).
	
	\bibitem[{\citenamefont{Doyle and Snell}(1984)}]{doyle1984random}
	\bibinfo{author}{\bibfnamefont{P.~G.} \bibnamefont{Doyle}} \bibnamefont{and}
	\bibinfo{author}{\bibfnamefont{J.~L.} \bibnamefont{Snell}},
	\emph{\bibinfo{title}{Random walks and electric networks}},
	vol.~\bibinfo{volume}{22} (\bibinfo{publisher}{American Mathematical Soc.},
	\bibinfo{year}{1984}).
	
	\bibitem[{\citenamefont{Baz and Chacko}(2004)}]{baz2004financial}
	\bibinfo{author}{\bibfnamefont{J.}~\bibnamefont{Baz}} \bibnamefont{and}
	\bibinfo{author}{\bibfnamefont{G.}~\bibnamefont{Chacko}},
	\emph{\bibinfo{title}{Financial derivatives: pricing, applications, and
			mathematics}} (\bibinfo{publisher}{Cambridge University Press},
	\bibinfo{year}{2004}).
	
	\bibitem[{\citenamefont{Schweitzer}(2003)}]{schweitzer2003brownian}
	\bibinfo{author}{\bibfnamefont{F.}~\bibnamefont{Schweitzer}},
	\emph{\bibinfo{title}{Brownian agents and active particles: collective
			dynamics in the natural and social sciences}} (\bibinfo{publisher}{Springer},
	\bibinfo{year}{2003}).
	
	\bibitem[{\citenamefont{Majumdar}(2007)}]{majumdar2007brownian}
	\bibinfo{author}{\bibfnamefont{S.~N.} \bibnamefont{Majumdar}}, in
	\emph{\bibinfo{booktitle}{The Legacy Of Albert Einstein: A Collection of
			Essays in Celebration of the Year of Physics}} (\bibinfo{publisher}{World
		Scientific}, \bibinfo{year}{2007}), pp. \bibinfo{pages}{93--129}.
	
	\bibitem[{\citenamefont{L\'evy}(1940)}]{Levy1940ArcsineLaw}
	\bibinfo{author}{\bibfnamefont{P.}~\bibnamefont{L\'evy}},
	\bibinfo{journal}{Compos. Math.} \textbf{\bibinfo{volume}{7}},
	\bibinfo{pages}{283} (\bibinfo{year}{1940}).
	
	\bibitem[{\citenamefont{Dornic and Godreche}(1998)}]{dornic1998large}
	\bibinfo{author}{\bibfnamefont{I.}~\bibnamefont{Dornic}} \bibnamefont{and}
	\bibinfo{author}{\bibfnamefont{C.}~\bibnamefont{Godreche}},
	\bibinfo{journal}{J. Phys. A: Math. Theor.} \textbf{\bibinfo{volume}{31}},
	\bibinfo{pages}{5413} (\bibinfo{year}{1998}).
	
	\bibitem[{\citenamefont{Drouffe and
			Godr{\`e}che}(1998)}]{drouffe1998stationary}
	\bibinfo{author}{\bibfnamefont{J.}~\bibnamefont{Drouffe}} \bibnamefont{and}
	\bibinfo{author}{\bibfnamefont{C.}~\bibnamefont{Godr{\`e}che}},
	\bibinfo{journal}{J. Phys. A: Math. Theor.} \textbf{\bibinfo{volume}{31}},
	\bibinfo{pages}{9801} (\bibinfo{year}{1998}).
	
	\bibitem[{\citenamefont{Baldassarri et~al.}(1999)\citenamefont{Baldassarri,
			Bouchaud, Dornic, and Godr{\`e}che}}]{baldassarri1999statistics}
	\bibinfo{author}{\bibfnamefont{A.}~\bibnamefont{Baldassarri}},
	\bibinfo{author}{\bibfnamefont{J.-P.} \bibnamefont{Bouchaud}},
	\bibinfo{author}{\bibfnamefont{I.}~\bibnamefont{Dornic}}, \bibnamefont{and}
	\bibinfo{author}{\bibfnamefont{C.}~\bibnamefont{Godr{\`e}che}},
	\bibinfo{journal}{Phys. Rev. E} \textbf{\bibinfo{volume}{59}},
	\bibinfo{pages}{R20} (\bibinfo{year}{1999}).
	
	\bibitem[{\citenamefont{Bouchaud and Potters}(2003)}]{bouchaud2003theory}
	\bibinfo{author}{\bibfnamefont{J.-P.} \bibnamefont{Bouchaud}} \bibnamefont{and}
	\bibinfo{author}{\bibfnamefont{M.}~\bibnamefont{Potters}},
	\emph{\bibinfo{title}{Theory of financial risk and derivative pricing: from
			statistical physics to risk management}} (\bibinfo{publisher}{Cambridge
		university press}, \bibinfo{year}{2003}).
	
	\bibitem[{\citenamefont{Brokmann et~al.}(2003)\citenamefont{Brokmann, Hermier,
			Messin, Desbiolles, Bouchaud, and Dahan}}]{brokmann2003statistical}
	\bibinfo{author}{\bibfnamefont{X.}~\bibnamefont{Brokmann}},
	\bibinfo{author}{\bibfnamefont{J.-P.} \bibnamefont{Hermier}},
	\bibinfo{author}{\bibfnamefont{G.}~\bibnamefont{Messin}},
	\bibinfo{author}{\bibfnamefont{P.}~\bibnamefont{Desbiolles}},
	\bibinfo{author}{\bibfnamefont{J.-P.} \bibnamefont{Bouchaud}},
	\bibnamefont{and} \bibinfo{author}{\bibfnamefont{M.}~\bibnamefont{Dahan}},
	\bibinfo{journal}{Phys. Rev. Lett.} \textbf{\bibinfo{volume}{90}},
	\bibinfo{pages}{120601} (\bibinfo{year}{2003}).
	
	\bibitem[{\citenamefont{Stefani et~al.}(2009)\citenamefont{Stefani, Hoogenboom,
			and Barkai}}]{stefani2009beyond}
	\bibinfo{author}{\bibfnamefont{F.~D.} \bibnamefont{Stefani}},
	\bibinfo{author}{\bibfnamefont{J.~P.} \bibnamefont{Hoogenboom}},
	\bibnamefont{and} \bibinfo{author}{\bibfnamefont{E.}~\bibnamefont{Barkai}},
	\bibinfo{journal}{Physics Today} \textbf{\bibinfo{volume}{62}},
	\bibinfo{pages}{34} (\bibinfo{year}{2009}).
	
	\bibitem[{\citenamefont{Barato et~al.}(2018)\citenamefont{Barato, Rold{\'a}n,
			Mart{\'\i}nez, and Pigolotti}}]{barato2018arcsine}
	\bibinfo{author}{\bibfnamefont{A.~C.} \bibnamefont{Barato}},
	\bibinfo{author}{\bibfnamefont{{\'E}.}~\bibnamefont{Rold{\'a}n}},
	\bibinfo{author}{\bibfnamefont{I.~A.} \bibnamefont{Mart{\'\i}nez}},
	\bibnamefont{and}
	\bibinfo{author}{\bibfnamefont{S.}~\bibnamefont{Pigolotti}},
	\bibinfo{journal}{Phys. Rev. Lett.} \textbf{\bibinfo{volume}{121}},
	\bibinfo{pages}{090601} (\bibinfo{year}{2018}).
	
	\bibitem[{\citenamefont{Dey et~al.}(2022)\citenamefont{Dey, Kundu, Das, and
			Banerjee}}]{dey2022experimental}
	\bibinfo{author}{\bibfnamefont{R.}~\bibnamefont{Dey}},
	\bibinfo{author}{\bibfnamefont{A.}~\bibnamefont{Kundu}},
	\bibinfo{author}{\bibfnamefont{B.}~\bibnamefont{Das}}, \bibnamefont{and}
	\bibinfo{author}{\bibfnamefont{A.}~\bibnamefont{Banerjee}},
	\bibinfo{journal}{Phys. Rev. E} \textbf{\bibinfo{volume}{106}},
	\bibinfo{pages}{054113} (\bibinfo{year}{2022}).
	
	\bibitem[{\citenamefont{Ramesh et~al.}(2024)\citenamefont{Ramesh, Peters, and
			Rodriguez}}]{ramesh2024arcsine}
	\bibinfo{author}{\bibfnamefont{V.}~\bibnamefont{Ramesh}},
	\bibinfo{author}{\bibfnamefont{K.}~\bibnamefont{Peters}}, \bibnamefont{and}
	\bibinfo{author}{\bibfnamefont{S.}~\bibnamefont{Rodriguez}},
	\bibinfo{journal}{Phys. Rev. Lett.} \textbf{\bibinfo{volume}{132}},
	\bibinfo{pages}{133801} (\bibinfo{year}{2024}).
	
	\bibitem[{\citenamefont{Nyawo and Touchette}(2017)}]{nyawo2017minimal}
	\bibinfo{author}{\bibfnamefont{P.~T.} \bibnamefont{Nyawo}} \bibnamefont{and}
	\bibinfo{author}{\bibfnamefont{H.}~\bibnamefont{Touchette}},
	\bibinfo{journal}{Europhys. Lett.} \textbf{\bibinfo{volume}{116}},
	\bibinfo{pages}{50009} (\bibinfo{year}{2017}).
	
	\bibitem[{\citenamefont{Nyawo and Touchette}(2018)}]{nyawo2018dynamical}
	\bibinfo{author}{\bibfnamefont{P.~T.} \bibnamefont{Nyawo}} \bibnamefont{and}
	\bibinfo{author}{\bibfnamefont{H.}~\bibnamefont{Touchette}},
	\bibinfo{journal}{Phys. Rev. E} \textbf{\bibinfo{volume}{98}},
	\bibinfo{pages}{052103} (\bibinfo{year}{2018}).
	
	\bibitem[{\citenamefont{Majumdar and Bray}(2002)}]{majumdar2002large}
	\bibinfo{author}{\bibfnamefont{S.~N.} \bibnamefont{Majumdar}} \bibnamefont{and}
	\bibinfo{author}{\bibfnamefont{A.~J.} \bibnamefont{Bray}},
	\bibinfo{journal}{Phys. Rev. E} \textbf{\bibinfo{volume}{65}},
	\bibinfo{pages}{051112} (\bibinfo{year}{2002}).
	
	\bibitem[{\citenamefont{Ehrhardt et~al.}(2004)\citenamefont{Ehrhardt, Majumdar,
			and Bray}}]{ehrhardt2004persistence}
	\bibinfo{author}{\bibfnamefont{G.}~\bibnamefont{Ehrhardt}},
	\bibinfo{author}{\bibfnamefont{S.~N.} \bibnamefont{Majumdar}},
	\bibnamefont{and} \bibinfo{author}{\bibfnamefont{A.~J.} \bibnamefont{Bray}},
	\bibinfo{journal}{Phys. Rev. E} \textbf{\bibinfo{volume}{69}},
	\bibinfo{pages}{016106} (\bibinfo{year}{2004}).
	
	\bibitem[{\citenamefont{Sadhu et~al.}(2018)\citenamefont{Sadhu, Delorme, and
			Wiese}}]{sadhu2018generalized}
	\bibinfo{author}{\bibfnamefont{T.}~\bibnamefont{Sadhu}},
	\bibinfo{author}{\bibfnamefont{M.}~\bibnamefont{Delorme}}, \bibnamefont{and}
	\bibinfo{author}{\bibfnamefont{K.~J.} \bibnamefont{Wiese}},
	\bibinfo{journal}{Phys. Rev. Lett.} \textbf{\bibinfo{volume}{120}},
	\bibinfo{pages}{040603} (\bibinfo{year}{2018}).
	
	\bibitem[{\citenamefont{Sadhu and Wiese}(2021)}]{sadhu2021functionals}
	\bibinfo{author}{\bibfnamefont{T.}~\bibnamefont{Sadhu}} \bibnamefont{and}
	\bibinfo{author}{\bibfnamefont{K.~J.} \bibnamefont{Wiese}},
	\bibinfo{journal}{Phys. Rev. E} \textbf{\bibinfo{volume}{104}},
	\bibinfo{pages}{054112} (\bibinfo{year}{2021}).
	
	\bibitem[{\citenamefont{Singh and Kundu}(2019)}]{SinghArcsinelaws_RTP}
	\bibinfo{author}{\bibfnamefont{P.}~\bibnamefont{Singh}} \bibnamefont{and}
	\bibinfo{author}{\bibfnamefont{A.}~\bibnamefont{Kundu}}, \bibinfo{journal}{J.
		Stat. Mech.} \textbf{\bibinfo{volume}{2019}}, \bibinfo{pages}{083205}
	(\bibinfo{year}{2019}).
	
	\bibitem[{\citenamefont{Bressloff}(2020)}]{bressloff2020occupation}
	\bibinfo{author}{\bibfnamefont{P.~C.} \bibnamefont{Bressloff}},
	\bibinfo{journal}{Phys. Rev. E} \textbf{\bibinfo{volume}{102}},
	\bibinfo{pages}{042135} (\bibinfo{year}{2020}).
	
	\bibitem[{\citenamefont{Mukherjee et~al.}(2024)\citenamefont{Mukherjee,
			Le~Doussal, and Smith}}]{mukherjee2024large}
	\bibinfo{author}{\bibfnamefont{S.}~\bibnamefont{Mukherjee}},
	\bibinfo{author}{\bibfnamefont{P.}~\bibnamefont{Le~Doussal}},
	\bibnamefont{and} \bibinfo{author}{\bibfnamefont{N.~R.} \bibnamefont{Smith}},
	\bibinfo{journal}{Phys. Rev. E} \textbf{\bibinfo{volume}{110}},
	\bibinfo{pages}{024107} (\bibinfo{year}{2024}).
	
	\bibitem[{\citenamefont{Boutcheng et~al.}(2016)\citenamefont{Boutcheng,
			Bouetou, Burkhardt, Rosso, Zoia, and Crepin}}]{boutcheng2016occupation}
	\bibinfo{author}{\bibfnamefont{H.~J.~O.} \bibnamefont{Boutcheng}},
	\bibinfo{author}{\bibfnamefont{T.~B.} \bibnamefont{Bouetou}},
	\bibinfo{author}{\bibfnamefont{T.~W.} \bibnamefont{Burkhardt}},
	\bibinfo{author}{\bibfnamefont{A.}~\bibnamefont{Rosso}},
	\bibinfo{author}{\bibfnamefont{A.}~\bibnamefont{Zoia}}, \bibnamefont{and}
	\bibinfo{author}{\bibfnamefont{K.~T.} \bibnamefont{Crepin}},
	\bibinfo{journal}{J. Stat. Mech.} \textbf{\bibinfo{volume}{2016}},
	\bibinfo{pages}{053213} (\bibinfo{year}{2016}).
	
	\bibitem[{\citenamefont{Godreche and Luck}(2001)}]{godreche2001statistics}
	\bibinfo{author}{\bibfnamefont{C.}~\bibnamefont{Godreche}} \bibnamefont{and}
	\bibinfo{author}{\bibfnamefont{J.}~\bibnamefont{Luck}}, \bibinfo{journal}{J.
		Stat. Phys.} \textbf{\bibinfo{volume}{104}}, \bibinfo{pages}{489}
	(\bibinfo{year}{2001}).
	
	\bibitem[{\citenamefont{Burov and Barkai}(2011)}]{PhysRevLett.107.170601}
	\bibinfo{author}{\bibfnamefont{S.}~\bibnamefont{Burov}} \bibnamefont{and}
	\bibinfo{author}{\bibfnamefont{E.}~\bibnamefont{Barkai}},
	\bibinfo{journal}{Phys. Rev. Lett.} \textbf{\bibinfo{volume}{107}},
	\bibinfo{pages}{170601} (\bibinfo{year}{2011}).
	
	\bibitem[{\citenamefont{Burov and Barkai}(2007)}]{burov2007occupation}
	\bibinfo{author}{\bibfnamefont{S.}~\bibnamefont{Burov}} \bibnamefont{and}
	\bibinfo{author}{\bibfnamefont{E.}~\bibnamefont{Barkai}},
	\bibinfo{journal}{Phys. Rev. Lett.} \textbf{\bibinfo{volume}{98}},
	\bibinfo{pages}{250601} (\bibinfo{year}{2007}).
	
	\bibitem[{\citenamefont{Bel and Barkai}(2005)}]{bel2005weak}
	\bibinfo{author}{\bibfnamefont{G.}~\bibnamefont{Bel}} \bibnamefont{and}
	\bibinfo{author}{\bibfnamefont{E.}~\bibnamefont{Barkai}},
	\bibinfo{journal}{Phys. Rev. Lett.} \textbf{\bibinfo{volume}{94}},
	\bibinfo{pages}{240602} (\bibinfo{year}{2005}).
	
	\bibitem[{\citenamefont{Barkai}(2006)}]{barkai2006residence}
	\bibinfo{author}{\bibfnamefont{E.}~\bibnamefont{Barkai}}, \bibinfo{journal}{J.
		Stat. Phys.} \textbf{\bibinfo{volume}{123}}, \bibinfo{pages}{883}
	(\bibinfo{year}{2006}).
	
	\bibitem[{\citenamefont{Del~Vecchio and Majumdar}(2025)}]{del2025generalized}
	\bibinfo{author}{\bibfnamefont{G.~D.~V.} \bibnamefont{Del~Vecchio}}
	\bibnamefont{and} \bibinfo{author}{\bibfnamefont{S.~N.}
		\bibnamefont{Majumdar}}, \bibinfo{journal}{J. Stat. Mech.}
	\textbf{\bibinfo{volume}{2025}}, \bibinfo{pages}{023207}
	(\bibinfo{year}{2025}).
	
	\bibitem[{\citenamefont{Singh}(2022)}]{singh2022extreme}
	\bibinfo{author}{\bibfnamefont{P.}~\bibnamefont{Singh}},
	\bibinfo{journal}{Phys. Rev. E} \textbf{\bibinfo{volume}{105}},
	\bibinfo{pages}{024113} (\bibinfo{year}{2022}).
	
	\bibitem[{\citenamefont{Majumdar and Comtet}(2002)}]{PhysRevLett.89.060601}
	\bibinfo{author}{\bibfnamefont{S.~N.} \bibnamefont{Majumdar}} \bibnamefont{and}
	\bibinfo{author}{\bibfnamefont{A.}~\bibnamefont{Comtet}},
	\bibinfo{journal}{Phys. Rev. Lett.} \textbf{\bibinfo{volume}{89}},
	\bibinfo{pages}{060601} (\bibinfo{year}{2002}).
	
	\bibitem[{\citenamefont{Sabhapandit et~al.}(2006)\citenamefont{Sabhapandit,
			Majumdar, and Comtet}}]{PhysRevE.73.051102}
	\bibinfo{author}{\bibfnamefont{S.}~\bibnamefont{Sabhapandit}},
	\bibinfo{author}{\bibfnamefont{S.~N.} \bibnamefont{Majumdar}},
	\bibnamefont{and} \bibinfo{author}{\bibfnamefont{A.}~\bibnamefont{Comtet}},
	\bibinfo{journal}{Phys. Rev. E} \textbf{\bibinfo{volume}{73}},
	\bibinfo{pages}{051102} (\bibinfo{year}{2006}).
	
	\bibitem[{\citenamefont{Grebenkov}(2007)}]{grebenkov2007residence}
	\bibinfo{author}{\bibfnamefont{D.}~\bibnamefont{Grebenkov}},
	\bibinfo{journal}{Phys. Rev. E} \textbf{\bibinfo{volume}{76}},
	\bibinfo{pages}{041139} (\bibinfo{year}{2007}).
	
	\bibitem[{\citenamefont{Kaldasch and Engel}(2022)}]{kaldasch2022stiffness}
	\bibinfo{author}{\bibfnamefont{S.}~\bibnamefont{Kaldasch}} \bibnamefont{and}
	\bibinfo{author}{\bibfnamefont{A.}~\bibnamefont{Engel}},
	\bibinfo{journal}{Phys. Rev. E} \textbf{\bibinfo{volume}{105}},
	\bibinfo{pages}{034132} (\bibinfo{year}{2022}).
	
	\bibitem[{\citenamefont{Kay and Giuggioli}(2023)}]{kay2023extreme}
	\bibinfo{author}{\bibfnamefont{T.}~\bibnamefont{Kay}} \bibnamefont{and}
	\bibinfo{author}{\bibfnamefont{L.}~\bibnamefont{Giuggioli}},
	\bibinfo{journal}{J. Phys. A: Math. Theor.} \textbf{\bibinfo{volume}{56}},
	\bibinfo{pages}{345002} (\bibinfo{year}{2023}).
	
	\bibitem[{\citenamefont{Huang and
			Chen}(2024{\natexlab{a}})}]{huang2024extremal}
	\bibinfo{author}{\bibfnamefont{F.}~\bibnamefont{Huang}} \bibnamefont{and}
	\bibinfo{author}{\bibfnamefont{H.}~\bibnamefont{Chen}},
	\bibinfo{journal}{Physica A} \textbf{\bibinfo{volume}{633}},
	\bibinfo{pages}{129389} (\bibinfo{year}{2024}{\natexlab{a}}).
	
	\bibitem[{\citenamefont{Den~Hollander et~al.}(2019)\citenamefont{Den~Hollander,
			Majumdar, Meylahn, and Touchette}}]{den2019properties}
	\bibinfo{author}{\bibfnamefont{F.}~\bibnamefont{Den~Hollander}},
	\bibinfo{author}{\bibfnamefont{S.~N.} \bibnamefont{Majumdar}},
	\bibinfo{author}{\bibfnamefont{J.~M.} \bibnamefont{Meylahn}},
	\bibnamefont{and}
	\bibinfo{author}{\bibfnamefont{H.}~\bibnamefont{Touchette}},
	\bibinfo{journal}{J. Phys. A: Math. Theor.} \textbf{\bibinfo{volume}{52}},
	\bibinfo{pages}{175001} (\bibinfo{year}{2019}).
	
	\bibitem[{\citenamefont{Smith et~al.}(2023)\citenamefont{Smith, Majumdar, and
			Schehr}}]{smith2023striking}
	\bibinfo{author}{\bibfnamefont{N.}~\bibnamefont{Smith}},
	\bibinfo{author}{\bibfnamefont{S.~N.} \bibnamefont{Majumdar}},
	\bibnamefont{and} \bibinfo{author}{\bibfnamefont{G.}~\bibnamefont{Schehr}},
	\bibinfo{journal}{Europhys. Lett.} \textbf{\bibinfo{volume}{142}},
	\bibinfo{pages}{51002} (\bibinfo{year}{2023}).
	
	\bibitem[{\citenamefont{Yan and Chen}(2023)}]{yan2023breakdown}
	\bibinfo{author}{\bibfnamefont{H.}~\bibnamefont{Yan}} \bibnamefont{and}
	\bibinfo{author}{\bibfnamefont{H.}~\bibnamefont{Chen}},
	\bibinfo{journal}{Physica Scripta} \textbf{\bibinfo{volume}{98}},
	\bibinfo{pages}{125226} (\bibinfo{year}{2023}).
	
	\bibitem[{\citenamefont{Ta{\'z}bierski and
			Magdziarz}(2025)}]{tazbierski2025arcsine}
	\bibinfo{author}{\bibfnamefont{K.}~\bibnamefont{Ta{\'z}bierski}}
	\bibnamefont{and}
	\bibinfo{author}{\bibfnamefont{M.}~\bibnamefont{Magdziarz}},
	\bibinfo{journal}{Chaos} \textbf{\bibinfo{volume}{35}}
	(\bibinfo{year}{2025}).
	
	\bibitem[{\citenamefont{Burenev et~al.}(2024)\citenamefont{Burenev, Majumdar,
			and Rosso}}]{burenev2024occupation}
	\bibinfo{author}{\bibfnamefont{I.~N.} \bibnamefont{Burenev}},
	\bibinfo{author}{\bibfnamefont{S.~N.} \bibnamefont{Majumdar}},
	\bibnamefont{and} \bibinfo{author}{\bibfnamefont{A.}~\bibnamefont{Rosso}},
	\bibinfo{journal}{Phys. Rev. E} \textbf{\bibinfo{volume}{109}},
	\bibinfo{pages}{044150} (\bibinfo{year}{2024}).
	
	\bibitem[{\citenamefont{Mukherjee and Smith}(2023)}]{mukherjee2023dynamical}
	\bibinfo{author}{\bibfnamefont{S.}~\bibnamefont{Mukherjee}} \bibnamefont{and}
	\bibinfo{author}{\bibfnamefont{N.~R.} \bibnamefont{Smith}},
	\bibinfo{journal}{Phys. Rev. E} \textbf{\bibinfo{volume}{107}},
	\bibinfo{pages}{064133} (\bibinfo{year}{2023}).
	
	\bibitem[{\citenamefont{Schehr and Le~Doussal}(2010)}]{schehr2010extreme}
	\bibinfo{author}{\bibfnamefont{G.}~\bibnamefont{Schehr}} \bibnamefont{and}
	\bibinfo{author}{\bibfnamefont{P.}~\bibnamefont{Le~Doussal}},
	\bibinfo{journal}{J. Stat. Mech.} \textbf{\bibinfo{volume}{2010}},
	\bibinfo{pages}{P01009} (\bibinfo{year}{2010}).
	
	\bibitem[{\citenamefont{Majumdar et~al.}(2020)\citenamefont{Majumdar, Pal, and
			Schehr}}]{majumdar2020extreme}
	\bibinfo{author}{\bibfnamefont{S.~N.} \bibnamefont{Majumdar}},
	\bibinfo{author}{\bibfnamefont{A.}~\bibnamefont{Pal}}, \bibnamefont{and}
	\bibinfo{author}{\bibfnamefont{G.}~\bibnamefont{Schehr}},
	\bibinfo{journal}{Phys. Rep.} \textbf{\bibinfo{volume}{840}},
	\bibinfo{pages}{1} (\bibinfo{year}{2020}).
	
	\bibitem[{\citenamefont{Majumdar and Schehr}(2024)}]{majumdar2024statistics}
	\bibinfo{author}{\bibfnamefont{S.~N.} \bibnamefont{Majumdar}} \bibnamefont{and}
	\bibinfo{author}{\bibfnamefont{G.}~\bibnamefont{Schehr}},
	\emph{\bibinfo{title}{Statistics of Extremes and Records in Random
			Sequences}} (\bibinfo{publisher}{Oxford University Press},
	\bibinfo{year}{2024}).
	
	\bibitem[{\citenamefont{Redner}(2001)}]{redner2001guide}
	\bibinfo{author}{\bibfnamefont{S.}~\bibnamefont{Redner}},
	\emph{\bibinfo{title}{A guide to first-passage processes}}
	(\bibinfo{publisher}{Cambridge University Press}, \bibinfo{year}{2001}).
	
	\bibitem[{\citenamefont{Majumdar et~al.}(2008)\citenamefont{Majumdar,
			Randon-Furling, Kearney, and Yor}}]{majumdar2008time}
	\bibinfo{author}{\bibfnamefont{S.~N.} \bibnamefont{Majumdar}},
	\bibinfo{author}{\bibfnamefont{J.}~\bibnamefont{Randon-Furling}},
	\bibinfo{author}{\bibfnamefont{M.~J.} \bibnamefont{Kearney}},
	\bibnamefont{and} \bibinfo{author}{\bibfnamefont{M.}~\bibnamefont{Yor}},
	\bibinfo{journal}{J. Phys. A: Math. Theor.} \textbf{\bibinfo{volume}{41}},
	\bibinfo{pages}{365005} (\bibinfo{year}{2008}).
	
	\bibitem[{\citenamefont{Randon-Furling and
			Majumdar}(2007)}]{randon2007distribution}
	\bibinfo{author}{\bibfnamefont{J.}~\bibnamefont{Randon-Furling}}
	\bibnamefont{and} \bibinfo{author}{\bibfnamefont{S.~N.}
		\bibnamefont{Majumdar}}, \bibinfo{journal}{J. Stat. Mech.}
	\textbf{\bibinfo{volume}{2007}}, \bibinfo{pages}{P10008}
	(\bibinfo{year}{2007}).
	
	\bibitem[{\citenamefont{Majumdar
			et~al.}(2010{\natexlab{a}})\citenamefont{Majumdar, Rosso, and
			Zoia}}]{majumdar2010time}
	\bibinfo{author}{\bibfnamefont{S.~N.} \bibnamefont{Majumdar}},
	\bibinfo{author}{\bibfnamefont{A.}~\bibnamefont{Rosso}}, \bibnamefont{and}
	\bibinfo{author}{\bibfnamefont{A.}~\bibnamefont{Zoia}}, \bibinfo{journal}{J.
		Phys. A: Math. Theor.} \textbf{\bibinfo{volume}{43}}, \bibinfo{pages}{115001}
	(\bibinfo{year}{2010}{\natexlab{a}}).
	
	\bibitem[{\citenamefont{Singh and Pal}(2021)}]{singh2021extremal}
	\bibinfo{author}{\bibfnamefont{P.}~\bibnamefont{Singh}} \bibnamefont{and}
	\bibinfo{author}{\bibfnamefont{A.}~\bibnamefont{Pal}},
	\bibinfo{journal}{Phys. Rev. E} \textbf{\bibinfo{volume}{103}},
	\bibinfo{pages}{052119} (\bibinfo{year}{2021}).
	
	\bibitem[{\citenamefont{Mori et~al.}(2021)\citenamefont{Mori, Majumdar, and
			Schehr}}]{mori2021distribution}
	\bibinfo{author}{\bibfnamefont{F.}~\bibnamefont{Mori}},
	\bibinfo{author}{\bibfnamefont{S.~N.} \bibnamefont{Majumdar}},
	\bibnamefont{and} \bibinfo{author}{\bibfnamefont{G.}~\bibnamefont{Schehr}},
	\bibinfo{journal}{Europhys. Lett.} \textbf{\bibinfo{volume}{135}},
	\bibinfo{pages}{30003} (\bibinfo{year}{2021}).
	
	\bibitem[{\citenamefont{Mori et~al.}(2022)\citenamefont{Mori, Majumdar, and
			Schehr}}]{mori2022time}
	\bibinfo{author}{\bibfnamefont{F.}~\bibnamefont{Mori}},
	\bibinfo{author}{\bibfnamefont{S.~N.} \bibnamefont{Majumdar}},
	\bibnamefont{and} \bibinfo{author}{\bibfnamefont{G.}~\bibnamefont{Schehr}},
	\bibinfo{journal}{Phys. Rev. E} \textbf{\bibinfo{volume}{106}},
	\bibinfo{pages}{054110} (\bibinfo{year}{2022}).
	
	\bibitem[{\citenamefont{Huang and Chen}(2024{\natexlab{b}})}]{huang2024extreme}
	\bibinfo{author}{\bibfnamefont{F.}~\bibnamefont{Huang}} \bibnamefont{and}
	\bibinfo{author}{\bibfnamefont{H.}~\bibnamefont{Chen}}, \bibinfo{journal}{J.
		Stat. Mech.} \textbf{\bibinfo{volume}{2024}}, \bibinfo{pages}{093212}
	(\bibinfo{year}{2024}{\natexlab{b}}).
	
	\bibitem[{\citenamefont{Guo et~al.}(2023)\citenamefont{Guo, Yan, and
			Chen}}]{guo2023extremal}
	\bibinfo{author}{\bibfnamefont{W.}~\bibnamefont{Guo}},
	\bibinfo{author}{\bibfnamefont{H.}~\bibnamefont{Yan}}, \bibnamefont{and}
	\bibinfo{author}{\bibfnamefont{H.}~\bibnamefont{Chen}},
	\bibinfo{journal}{Phys. Rev. E} \textbf{\bibinfo{volume}{108}},
	\bibinfo{pages}{044115} (\bibinfo{year}{2023}).
	
	\bibitem[{\citenamefont{Guo et~al.}(2024)\citenamefont{Guo, Yan, and
			Chen}}]{guo2024extremal}
	\bibinfo{author}{\bibfnamefont{W.}~\bibnamefont{Guo}},
	\bibinfo{author}{\bibfnamefont{H.}~\bibnamefont{Yan}}, \bibnamefont{and}
	\bibinfo{author}{\bibfnamefont{H.}~\bibnamefont{Chen}}, \bibinfo{journal}{J.
		Stat. Mech.} \textbf{\bibinfo{volume}{2024}}, \bibinfo{pages}{023209}
	(\bibinfo{year}{2024}).
	
	\bibitem[{\citenamefont{Majumdar
			et~al.}(2010{\natexlab{b}})\citenamefont{Majumdar, Rosso, and
			Zoia}}]{majumdar2010hitting}
	\bibinfo{author}{\bibfnamefont{S.~N.} \bibnamefont{Majumdar}},
	\bibinfo{author}{\bibfnamefont{A.}~\bibnamefont{Rosso}}, \bibnamefont{and}
	\bibinfo{author}{\bibfnamefont{A.}~\bibnamefont{Zoia}},
	\bibinfo{journal}{Phys. Rev. Lett.} \textbf{\bibinfo{volume}{104}},
	\bibinfo{pages}{020602} (\bibinfo{year}{2010}{\natexlab{b}}).
	
	\bibitem[{\citenamefont{Randon-Furling
			et~al.}(2009)\citenamefont{Randon-Furling, Majumdar, and
			Comtet}}]{PhysRevLett.103.140602}
	\bibinfo{author}{\bibfnamefont{J.}~\bibnamefont{Randon-Furling}},
	\bibinfo{author}{\bibfnamefont{S.~N.} \bibnamefont{Majumdar}},
	\bibnamefont{and} \bibinfo{author}{\bibfnamefont{A.}~\bibnamefont{Comtet}},
	\bibinfo{journal}{Phys. Rev. Lett.} \textbf{\bibinfo{volume}{103}},
	\bibinfo{pages}{140602} (\bibinfo{year}{2009}).
	
	\bibitem[{\citenamefont{Dumonteil et~al.}(2013)\citenamefont{Dumonteil,
			Majumdar, Rosso, and Zoia}}]{dumonteil2013spatial}
	\bibinfo{author}{\bibfnamefont{E.}~\bibnamefont{Dumonteil}},
	\bibinfo{author}{\bibfnamefont{S.~N.} \bibnamefont{Majumdar}},
	\bibinfo{author}{\bibfnamefont{A.}~\bibnamefont{Rosso}}, \bibnamefont{and}
	\bibinfo{author}{\bibfnamefont{A.}~\bibnamefont{Zoia}},
	\bibinfo{journal}{Proc. Natl. Acad. Sci. U.S.A.}
	\textbf{\bibinfo{volume}{110}}, \bibinfo{pages}{4239} (\bibinfo{year}{2013}).
	
	\bibitem[{\citenamefont{Chupeau et~al.}(2015)\citenamefont{Chupeau,
			B{\'e}nichou, and Majumdar}}]{chupeau2015convex}
	\bibinfo{author}{\bibfnamefont{M.}~\bibnamefont{Chupeau}},
	\bibinfo{author}{\bibfnamefont{O.}~\bibnamefont{B{\'e}nichou}},
	\bibnamefont{and} \bibinfo{author}{\bibfnamefont{S.~N.}
		\bibnamefont{Majumdar}}, \bibinfo{journal}{Phys. Rev. E}
	\textbf{\bibinfo{volume}{91}}, \bibinfo{pages}{050104}
	(\bibinfo{year}{2015}).
	
	\bibitem[{\citenamefont{Majumdar
			et~al.}(2021{\natexlab{a}})\citenamefont{Majumdar, Mori, Schawe, and
			Schehr}}]{PhysRevE.103.022135}
	\bibinfo{author}{\bibfnamefont{S.~N.} \bibnamefont{Majumdar}},
	\bibinfo{author}{\bibfnamefont{F.}~\bibnamefont{Mori}},
	\bibinfo{author}{\bibfnamefont{H.}~\bibnamefont{Schawe}}, \bibnamefont{and}
	\bibinfo{author}{\bibfnamefont{G.}~\bibnamefont{Schehr}},
	\bibinfo{journal}{Phys. Rev. E} \textbf{\bibinfo{volume}{103}},
	\bibinfo{pages}{022135} (\bibinfo{year}{2021}{\natexlab{a}}).
	
	\bibitem[{\citenamefont{Majumdar
			et~al.}(2021{\natexlab{b}})\citenamefont{Majumdar, Mori, Schawe, and
			Schehr}}]{majumdar2021mean}
	\bibinfo{author}{\bibfnamefont{S.~N.} \bibnamefont{Majumdar}},
	\bibinfo{author}{\bibfnamefont{F.}~\bibnamefont{Mori}},
	\bibinfo{author}{\bibfnamefont{H.}~\bibnamefont{Schawe}}, \bibnamefont{and}
	\bibinfo{author}{\bibfnamefont{G.}~\bibnamefont{Schehr}},
	\bibinfo{journal}{Phys. Rev. E} \textbf{\bibinfo{volume}{103}},
	\bibinfo{pages}{022135} (\bibinfo{year}{2021}{\natexlab{b}}).
	
	\bibitem[{\citenamefont{Bao and Jia}(2006)}]{bao2006last}
	\bibinfo{author}{\bibfnamefont{J.-D.} \bibnamefont{Bao}} \bibnamefont{and}
	\bibinfo{author}{\bibfnamefont{Y.}~\bibnamefont{Jia}}, \bibinfo{journal}{J.
		Stat. Phys.} \textbf{\bibinfo{volume}{123}}, \bibinfo{pages}{861}
	(\bibinfo{year}{2006}).
	
	\bibitem[{\citenamefont{Comtet et~al.}(2020)\citenamefont{Comtet, Cornu, and
			Schehr}}]{comtet2020last}
	\bibinfo{author}{\bibfnamefont{A.}~\bibnamefont{Comtet}},
	\bibinfo{author}{\bibfnamefont{F.}~\bibnamefont{Cornu}}, \bibnamefont{and}
	\bibinfo{author}{\bibfnamefont{G.}~\bibnamefont{Schehr}},
	\bibinfo{journal}{J. Stat. Phys.} \textbf{\bibinfo{volume}{181}},
	\bibinfo{pages}{1565} (\bibinfo{year}{2020}).
	
	\bibitem[{\citenamefont{Fang et~al.}(2021)\citenamefont{Fang, Gan, Holmes,
			Huang, Pek{\"o}z, R{\"o}llin, and Tang}}]{fang2021arcsine}
	\bibinfo{author}{\bibfnamefont{X.}~\bibnamefont{Fang}},
	\bibinfo{author}{\bibfnamefont{H.~L.} \bibnamefont{Gan}},
	\bibinfo{author}{\bibfnamefont{S.}~\bibnamefont{Holmes}},
	\bibinfo{author}{\bibfnamefont{H.}~\bibnamefont{Huang}},
	\bibinfo{author}{\bibfnamefont{E.}~\bibnamefont{Pek{\"o}z}},
	\bibinfo{author}{\bibfnamefont{A.}~\bibnamefont{R{\"o}llin}},
	\bibnamefont{and} \bibinfo{author}{\bibfnamefont{W.}~\bibnamefont{Tang}},
	\bibinfo{journal}{J. Appl. Probab.} \textbf{\bibinfo{volume}{58}},
	\bibinfo{pages}{851} (\bibinfo{year}{2021}).
	
	\bibitem[{\citenamefont{Leung}(2004)}]{leung2004novel}
	\bibinfo{author}{\bibfnamefont{B.~H.} \bibnamefont{Leung}},
	\bibinfo{journal}{IEEE Trans. Circuits Syst. I: Regul. Paper}
	\textbf{\bibinfo{volume}{51}}, \bibinfo{pages}{471} (\bibinfo{year}{2004}).
	
	\bibitem[{\citenamefont{Robson et~al.}(2014)\citenamefont{Robson, Leung, and
			Gong}}]{robson2014truly}
	\bibinfo{author}{\bibfnamefont{S.}~\bibnamefont{Robson}},
	\bibinfo{author}{\bibfnamefont{B.}~\bibnamefont{Leung}}, \bibnamefont{and}
	\bibinfo{author}{\bibfnamefont{G.}~\bibnamefont{Gong}},
	\bibinfo{journal}{IEEE Trans. Circuits Syst. II: Express Briefs}
	\textbf{\bibinfo{volume}{61}}, \bibinfo{pages}{937} (\bibinfo{year}{2014}).
	
	\bibitem[{\citenamefont{Bao and Jia}(2004)}]{bao2004determination}
	\bibinfo{author}{\bibfnamefont{J.-D.} \bibnamefont{Bao}} \bibnamefont{and}
	\bibinfo{author}{\bibfnamefont{Y.}~\bibnamefont{Jia}},
	\bibinfo{journal}{Phys. Rev. C} \textbf{\bibinfo{volume}{69}},
	\bibinfo{pages}{027602} (\bibinfo{year}{2004}).
	
	\bibitem[{\citenamefont{Hwang and Given}(2006)}]{hwang2006last}
	\bibinfo{author}{\bibfnamefont{C.-O.} \bibnamefont{Hwang}} \bibnamefont{and}
	\bibinfo{author}{\bibfnamefont{J.~A.} \bibnamefont{Given}},
	\bibinfo{journal}{Phys. Rev. E} \textbf{\bibinfo{volume}{74}},
	\bibinfo{pages}{027701} (\bibinfo{year}{2006}).
	
	\bibitem[{\citenamefont{Yu et~al.}(2021)\citenamefont{Yu, Lee, and
			Hwang}}]{yu2021last}
	\bibinfo{author}{\bibfnamefont{U.}~\bibnamefont{Yu}},
	\bibinfo{author}{\bibfnamefont{Y.-M.} \bibnamefont{Lee}}, \bibnamefont{and}
	\bibinfo{author}{\bibfnamefont{C.-O.} \bibnamefont{Hwang}},
	\bibinfo{journal}{Journal of Scientific Computing}
	\textbf{\bibinfo{volume}{88}}, \bibinfo{pages}{82} (\bibinfo{year}{2021}).
	
	\bibitem[{\citenamefont{Hartmann and Majumdar}(2025)}]{hartmann2025exact}
	\bibinfo{author}{\bibfnamefont{A.~K.} \bibnamefont{Hartmann}} \bibnamefont{and}
	\bibinfo{author}{\bibfnamefont{S.~N.} \bibnamefont{Majumdar}},
	\bibinfo{journal}{Phys. Rev. E} \textbf{\bibinfo{volume}{111}},
	\bibinfo{pages}{044134} (\bibinfo{year}{2025}).
	
	\bibitem[{\citenamefont{Mori et~al.}(2019)\citenamefont{Mori, Majumdar, and
			Schehr}}]{mori2019time}
	\bibinfo{author}{\bibfnamefont{F.}~\bibnamefont{Mori}},
	\bibinfo{author}{\bibfnamefont{S.~N.} \bibnamefont{Majumdar}},
	\bibnamefont{and} \bibinfo{author}{\bibfnamefont{G.}~\bibnamefont{Schehr}},
	\bibinfo{journal}{Phys. Rev. Lett.} \textbf{\bibinfo{volume}{123}},
	\bibinfo{pages}{200201} (\bibinfo{year}{2019}).
	
	\bibitem[{\citenamefont{Mori et~al.}(2020)\citenamefont{Mori, Majumdar, and
			Schehr}}]{mori2020distribution}
	\bibinfo{author}{\bibfnamefont{F.}~\bibnamefont{Mori}},
	\bibinfo{author}{\bibfnamefont{S.~N.} \bibnamefont{Majumdar}},
	\bibnamefont{and} \bibinfo{author}{\bibfnamefont{G.}~\bibnamefont{Schehr}},
	\bibinfo{journal}{Phys. Rev. E} \textbf{\bibinfo{volume}{101}},
	\bibinfo{pages}{052111} (\bibinfo{year}{2020}).
	
	\bibitem[{\citenamefont{Evans and
			Majumdar}(2011{\natexlab{a}})}]{evans2011diffusion}
	\bibinfo{author}{\bibfnamefont{M.~R.} \bibnamefont{Evans}} \bibnamefont{and}
	\bibinfo{author}{\bibfnamefont{S.~N.} \bibnamefont{Majumdar}},
	\bibinfo{journal}{Phys. Rev. Lett.} \textbf{\bibinfo{volume}{106}},
	\bibinfo{pages}{160601} (\bibinfo{year}{2011}{\natexlab{a}}).
	
	\bibitem[{\citenamefont{Kusmierz et~al.}(2014)\citenamefont{Kusmierz, Majumdar,
			Sabhapandit, and Schehr}}]{kusmierz2014first}
	\bibinfo{author}{\bibfnamefont{L.}~\bibnamefont{Kusmierz}},
	\bibinfo{author}{\bibfnamefont{S.~N.} \bibnamefont{Majumdar}},
	\bibinfo{author}{\bibfnamefont{S.}~\bibnamefont{Sabhapandit}},
	\bibnamefont{and} \bibinfo{author}{\bibfnamefont{G.}~\bibnamefont{Schehr}},
	\bibinfo{journal}{Phys. Rev. Lett.} \textbf{\bibinfo{volume}{113}},
	\bibinfo{pages}{220602} (\bibinfo{year}{2014}).
	
	\bibitem[{\citenamefont{Ku{\'s}mierz and
			Gudowska-Nowak}(2015)}]{kusmierz2015optimal}
	\bibinfo{author}{\bibfnamefont{{\L}.}~\bibnamefont{Ku{\'s}mierz}}
	\bibnamefont{and}
	\bibinfo{author}{\bibfnamefont{E.}~\bibnamefont{Gudowska-Nowak}},
	\bibinfo{journal}{Phys. Rev. E} \textbf{\bibinfo{volume}{92}},
	\bibinfo{pages}{052127} (\bibinfo{year}{2015}).
	
	\bibitem[{\citenamefont{Boyer and Solis-Salas}(2014)}]{PhysRevLett.112.240601}
	\bibinfo{author}{\bibfnamefont{D.}~\bibnamefont{Boyer}} \bibnamefont{and}
	\bibinfo{author}{\bibfnamefont{C.}~\bibnamefont{Solis-Salas}},
	\bibinfo{journal}{Phys. Rev. Lett.} \textbf{\bibinfo{volume}{112}},
	\bibinfo{pages}{240601} (\bibinfo{year}{2014}).
	
	\bibitem[{\citenamefont{Chechkin and Sokolov}(2018)}]{chechkin2018random}
	\bibinfo{author}{\bibfnamefont{A.}~\bibnamefont{Chechkin}} \bibnamefont{and}
	\bibinfo{author}{\bibfnamefont{I.}~\bibnamefont{Sokolov}},
	\bibinfo{journal}{Phys. Rev. Lett.} \textbf{\bibinfo{volume}{121}},
	\bibinfo{pages}{050601} (\bibinfo{year}{2018}).
	
	\bibitem[{\citenamefont{Evans et~al.}(2022)\citenamefont{Evans, Majumdar, and
			Schehr}}]{JPA2022.55.274005}
	\bibinfo{author}{\bibfnamefont{M.~R.} \bibnamefont{Evans}},
	\bibinfo{author}{\bibfnamefont{S.~N.} \bibnamefont{Majumdar}},
	\bibnamefont{and} \bibinfo{author}{\bibfnamefont{G.}~\bibnamefont{Schehr}},
	\bibinfo{journal}{J. Phys. A: Math. Theor.} \textbf{\bibinfo{volume}{55}},
	\bibinfo{pages}{274005} (\bibinfo{year}{2022}).
	
	\bibitem[{\citenamefont{Montanari and
			Zecchina}(2002)}]{montanari2002optimizing}
	\bibinfo{author}{\bibfnamefont{A.}~\bibnamefont{Montanari}} \bibnamefont{and}
	\bibinfo{author}{\bibfnamefont{R.}~\bibnamefont{Zecchina}},
	\bibinfo{journal}{Phys. Rev. Lett.} \textbf{\bibinfo{volume}{88}},
	\bibinfo{pages}{178701} (\bibinfo{year}{2002}).
	
	\bibitem[{\citenamefont{Keidar et~al.}(2025)\citenamefont{Keidar, Blumer,
			Hirshberg, and Reuveni}}]{keidar2025adaptive}
	\bibinfo{author}{\bibfnamefont{T.~D.} \bibnamefont{Keidar}},
	\bibinfo{author}{\bibfnamefont{O.}~\bibnamefont{Blumer}},
	\bibinfo{author}{\bibfnamefont{B.}~\bibnamefont{Hirshberg}},
	\bibnamefont{and} \bibinfo{author}{\bibfnamefont{S.}~\bibnamefont{Reuveni}},
	\bibinfo{journal}{Nat. Commun.} \textbf{\bibinfo{volume}{16}},
	\bibinfo{pages}{7259} (\bibinfo{year}{2025}).
	
	\bibitem[{\citenamefont{Reuveni et~al.}(2014)\citenamefont{Reuveni, Urbakh, and
			Klafter}}]{reuveni2014role}
	\bibinfo{author}{\bibfnamefont{S.}~\bibnamefont{Reuveni}},
	\bibinfo{author}{\bibfnamefont{M.}~\bibnamefont{Urbakh}}, \bibnamefont{and}
	\bibinfo{author}{\bibfnamefont{J.}~\bibnamefont{Klafter}},
	\bibinfo{journal}{Proc. Natl. Acad. Sci. USA} \textbf{\bibinfo{volume}{111}},
	\bibinfo{pages}{4391} (\bibinfo{year}{2014}).
	
	\bibitem[{\citenamefont{Rotbart et~al.}(2015)\citenamefont{Rotbart, Reuveni,
			and Urbakh}}]{rotbart2015michaelis}
	\bibinfo{author}{\bibfnamefont{T.}~\bibnamefont{Rotbart}},
	\bibinfo{author}{\bibfnamefont{S.}~\bibnamefont{Reuveni}}, \bibnamefont{and}
	\bibinfo{author}{\bibfnamefont{M.}~\bibnamefont{Urbakh}},
	\bibinfo{journal}{Phys. Rev. E} \textbf{\bibinfo{volume}{92}},
	\bibinfo{pages}{060101} (\bibinfo{year}{2015}).
	
	\bibitem[{\citenamefont{Evans et~al.}(2020)\citenamefont{Evans, Majumdar, and
			Schehr}}]{evans2020stochastic}
	\bibinfo{author}{\bibfnamefont{M.~R.} \bibnamefont{Evans}},
	\bibinfo{author}{\bibfnamefont{S.~N.} \bibnamefont{Majumdar}},
	\bibnamefont{and} \bibinfo{author}{\bibfnamefont{G.}~\bibnamefont{Schehr}},
	\bibinfo{journal}{J. Phys. A: Math. Theor.} \textbf{\bibinfo{volume}{53}},
	\bibinfo{pages}{193001} (\bibinfo{year}{2020}).
	
	\bibitem[{\citenamefont{Gupta and Jayvar}(2022)}]{Gupta2022Review}
	\bibinfo{author}{\bibfnamefont{S.}~\bibnamefont{Gupta}} \bibnamefont{and}
	\bibinfo{author}{\bibfnamefont{A.~M.} \bibnamefont{Jayvar}},
	\bibinfo{journal}{Front. Phys.} \textbf{\bibinfo{volume}{10}},
	\bibinfo{pages}{789097} (\bibinfo{year}{2022}).
	
	\bibitem[{\citenamefont{Evans and
			Majumdar}(2011{\natexlab{b}})}]{evans2011diffusion2}
	\bibinfo{author}{\bibfnamefont{M.~R.} \bibnamefont{Evans}} \bibnamefont{and}
	\bibinfo{author}{\bibfnamefont{S.~N.} \bibnamefont{Majumdar}},
	\bibinfo{journal}{J. Phys. A: Math. Theor.} \textbf{\bibinfo{volume}{44}},
	\bibinfo{pages}{435001} (\bibinfo{year}{2011}{\natexlab{b}}).
	
	\bibitem[{\citenamefont{Pal et~al.}(2016)\citenamefont{Pal, Kundu, and
			Evans}}]{pal2016diffusion}
	\bibinfo{author}{\bibfnamefont{A.}~\bibnamefont{Pal}},
	\bibinfo{author}{\bibfnamefont{A.}~\bibnamefont{Kundu}}, \bibnamefont{and}
	\bibinfo{author}{\bibfnamefont{M.~R.} \bibnamefont{Evans}},
	\bibinfo{journal}{J. Phys. A: Math. Theor.} \textbf{\bibinfo{volume}{49}},
	\bibinfo{pages}{225001} (\bibinfo{year}{2016}).
	
	\bibitem[{\citenamefont{Eule and Metzger}(2016)}]{NJP2016.18.033006}
	\bibinfo{author}{\bibfnamefont{S.}~\bibnamefont{Eule}} \bibnamefont{and}
	\bibinfo{author}{\bibfnamefont{J.~J.} \bibnamefont{Metzger}},
	\bibinfo{journal}{New J. Phys.} \textbf{\bibinfo{volume}{18}},
	\bibinfo{pages}{033006} (\bibinfo{year}{2016}).
	
	\bibitem[{\citenamefont{Nagar and Gupta}(2016)}]{PhysRevE.93.060102}
	\bibinfo{author}{\bibfnamefont{A.}~\bibnamefont{Nagar}} \bibnamefont{and}
	\bibinfo{author}{\bibfnamefont{S.}~\bibnamefont{Gupta}},
	\bibinfo{journal}{Phys. Rev. E} \textbf{\bibinfo{volume}{93}},
	\bibinfo{pages}{060102} (\bibinfo{year}{2016}).
	
	\bibitem[{\citenamefont{Shkilev}(2017)}]{PhysRevE.96.012126}
	\bibinfo{author}{\bibfnamefont{V.~P.} \bibnamefont{Shkilev}},
	\bibinfo{journal}{Phys. Rev. E} \textbf{\bibinfo{volume}{96}},
	\bibinfo{pages}{012126} (\bibinfo{year}{2017}).
	
	\bibitem[{\citenamefont{Ku\ifmmode~\acute{s}\else \'{s}\fi{}mierz and
			Toyoizumi}(2019)}]{PhysRevE.100.032110}
	\bibinfo{author}{\bibfnamefont{L.}~\bibnamefont{Ku\ifmmode~\acute{s}\else
			\'{s}\fi{}mierz}} \bibnamefont{and}
	\bibinfo{author}{\bibfnamefont{T.}~\bibnamefont{Toyoizumi}},
	\bibinfo{journal}{Phys. Rev. E} \textbf{\bibinfo{volume}{100}},
	\bibinfo{pages}{032110} (\bibinfo{year}{2019}).
	
	\bibitem[{\citenamefont{Chen and Ye}(2022)}]{chen2022random}
	\bibinfo{author}{\bibfnamefont{H.}~\bibnamefont{Chen}} \bibnamefont{and}
	\bibinfo{author}{\bibfnamefont{Y.}~\bibnamefont{Ye}}, \bibinfo{journal}{Phys.
		Rev. E} \textbf{\bibinfo{volume}{106}}, \bibinfo{pages}{044139}
	(\bibinfo{year}{2022}).
	
	\bibitem[{\citenamefont{Evans and Majumdar}(2014)}]{Evans2014_Reset_Highd}
	\bibinfo{author}{\bibfnamefont{M.~R.} \bibnamefont{Evans}} \bibnamefont{and}
	\bibinfo{author}{\bibfnamefont{S.~N.} \bibnamefont{Majumdar}},
	\bibinfo{journal}{J. Phys. A: Math. Theor.} \textbf{\bibinfo{volume}{47}},
	\bibinfo{pages}{285001} (\bibinfo{year}{2014}).
	
	\bibitem[{\citenamefont{Christou and Schadschneider}(2015)}]{Christou2015}
	\bibinfo{author}{\bibfnamefont{C.}~\bibnamefont{Christou}} \bibnamefont{and}
	\bibinfo{author}{\bibfnamefont{A.}~\bibnamefont{Schadschneider}},
	\bibinfo{journal}{J. Phys. A: Math. Theor.} \textbf{\bibinfo{volume}{48}},
	\bibinfo{pages}{285003} (\bibinfo{year}{2015}).
	
	\bibitem[{\citenamefont{Pal and Prasad}(2019)}]{PhysRevE.99.032123}
	\bibinfo{author}{\bibfnamefont{A.}~\bibnamefont{Pal}} \bibnamefont{and}
	\bibinfo{author}{\bibfnamefont{V.~V.} \bibnamefont{Prasad}},
	\bibinfo{journal}{Phys. Rev. E} \textbf{\bibinfo{volume}{99}},
	\bibinfo{pages}{032123} (\bibinfo{year}{2019}).
	
	\bibitem[{\citenamefont{Domazetoski et~al.}(2020)\citenamefont{Domazetoski,
			Mas\'o-Puigdellosas, Sandev, M\'endez, Iomin, and
			Kocarev}}]{PhysRevResearch.2.033027}
	\bibinfo{author}{\bibfnamefont{V.}~\bibnamefont{Domazetoski}},
	\bibinfo{author}{\bibfnamefont{A.}~\bibnamefont{Mas\'o-Puigdellosas}},
	\bibinfo{author}{\bibfnamefont{T.}~\bibnamefont{Sandev}},
	\bibinfo{author}{\bibfnamefont{V.~m.~c.} \bibnamefont{M\'endez}},
	\bibinfo{author}{\bibfnamefont{A.}~\bibnamefont{Iomin}}, \bibnamefont{and}
	\bibinfo{author}{\bibfnamefont{L.}~\bibnamefont{Kocarev}},
	\bibinfo{journal}{Phys. Rev. Res.} \textbf{\bibinfo{volume}{2}},
	\bibinfo{pages}{033027} (\bibinfo{year}{2020}).
	
	\bibitem[{\citenamefont{Bressloff}(2021)}]{BressloffJSTAT2021}
	\bibinfo{author}{\bibfnamefont{P.~C.} \bibnamefont{Bressloff}},
	\bibinfo{journal}{J. Stat. Mech.} \textbf{\bibinfo{volume}{2021}},
	\bibinfo{pages}{063206} (\bibinfo{year}{2021}).
	
	\bibitem[{\citenamefont{Chen and Huang}(2022)}]{PhysRevE.105.034109}
	\bibinfo{author}{\bibfnamefont{H.}~\bibnamefont{Chen}} \bibnamefont{and}
	\bibinfo{author}{\bibfnamefont{F.}~\bibnamefont{Huang}},
	\bibinfo{journal}{Phys. Rev. E} \textbf{\bibinfo{volume}{105}},
	\bibinfo{pages}{034109} (\bibinfo{year}{2022}).
	
	\bibitem[{\citenamefont{Riascos et~al.}(2020)\citenamefont{Riascos, Boyer,
			Herringer, and Mateos}}]{PhysRevE.101.062147}
	\bibinfo{author}{\bibfnamefont{A.~P.} \bibnamefont{Riascos}},
	\bibinfo{author}{\bibfnamefont{D.}~\bibnamefont{Boyer}},
	\bibinfo{author}{\bibfnamefont{P.}~\bibnamefont{Herringer}},
	\bibnamefont{and} \bibinfo{author}{\bibfnamefont{J.~L.}
		\bibnamefont{Mateos}}, \bibinfo{journal}{Phys. Rev. E}
	\textbf{\bibinfo{volume}{101}}, \bibinfo{pages}{062147}
	(\bibinfo{year}{2020}).
	
	\bibitem[{\citenamefont{Huang and Chen}(2021)}]{huang2021random}
	\bibinfo{author}{\bibfnamefont{F.}~\bibnamefont{Huang}} \bibnamefont{and}
	\bibinfo{author}{\bibfnamefont{H.}~\bibnamefont{Chen}},
	\bibinfo{journal}{Phys. Rev. E} \textbf{\bibinfo{volume}{103}},
	\bibinfo{pages}{062132} (\bibinfo{year}{2021}).
	
	\bibitem[{\citenamefont{Ye and Chen}(2022)}]{JSM2022.053201}
	\bibinfo{author}{\bibfnamefont{Y.}~\bibnamefont{Ye}} \bibnamefont{and}
	\bibinfo{author}{\bibfnamefont{H.}~\bibnamefont{Chen}}, \bibinfo{journal}{J.
		Stat. Mech.} \textbf{\bibinfo{volume}{2022}}, \bibinfo{pages}{053201}
	(\bibinfo{year}{2022}).
	
	\bibitem[{\citenamefont{Evans and Majumdar}(2018{\natexlab{a}})}]{EvansJPA2018}
	\bibinfo{author}{\bibfnamefont{M.~R.} \bibnamefont{Evans}} \bibnamefont{and}
	\bibinfo{author}{\bibfnamefont{S.~N.} \bibnamefont{Majumdar}},
	\bibinfo{journal}{J. Phys. A: Math. Theor.} \textbf{\bibinfo{volume}{52}},
	\bibinfo{pages}{01LT01} (\bibinfo{year}{2018}{\natexlab{a}}).
	
	\bibitem[{\citenamefont{Pal et~al.}(2019)\citenamefont{Pal, Ku\'smierz, and
			Reuveni}}]{PalNJP2019}
	\bibinfo{author}{\bibfnamefont{A.}~\bibnamefont{Pal}},
	\bibinfo{author}{\bibfnamefont{L.}~\bibnamefont{Ku\'smierz}},
	\bibnamefont{and} \bibinfo{author}{\bibfnamefont{S.}~\bibnamefont{Reuveni}},
	\bibinfo{journal}{New J. Phys.} \textbf{\bibinfo{volume}{21}},
	\bibinfo{pages}{113024} (\bibinfo{year}{2019}).
	
	\bibitem[{\citenamefont{Bodrova and Sokolov}(2020)}]{PhysRevE.101.052130}
	\bibinfo{author}{\bibfnamefont{A.~S.} \bibnamefont{Bodrova}} \bibnamefont{and}
	\bibinfo{author}{\bibfnamefont{I.~M.} \bibnamefont{Sokolov}},
	\bibinfo{journal}{Phys. Rev. E} \textbf{\bibinfo{volume}{101}},
	\bibinfo{pages}{052130} (\bibinfo{year}{2020}).
	
	\bibitem[{\citenamefont{Gupta et~al.}(2020{\natexlab{a}})\citenamefont{Gupta,
			Plata, Kundu, and Pal}}]{GuptaJPA2020}
	\bibinfo{author}{\bibfnamefont{D.}~\bibnamefont{Gupta}},
	\bibinfo{author}{\bibfnamefont{C.~A.} \bibnamefont{Plata}},
	\bibinfo{author}{\bibfnamefont{A.}~\bibnamefont{Kundu}}, \bibnamefont{and}
	\bibinfo{author}{\bibfnamefont{A.}~\bibnamefont{Pal}}, \bibinfo{journal}{J.
		Phys. A: Math. Theor.} \textbf{\bibinfo{volume}{54}}, \bibinfo{pages}{025003}
	(\bibinfo{year}{2020}{\natexlab{a}}).
	
	\bibitem[{\citenamefont{Mercado-V{\'a}squez
			et~al.}(2020)\citenamefont{Mercado-V{\'a}squez, Boyer, Majumdar, and
			Schehr}}]{mercado2020intermittent}
	\bibinfo{author}{\bibfnamefont{G.}~\bibnamefont{Mercado-V{\'a}squez}},
	\bibinfo{author}{\bibfnamefont{D.}~\bibnamefont{Boyer}},
	\bibinfo{author}{\bibfnamefont{S.~N.} \bibnamefont{Majumdar}},
	\bibnamefont{and} \bibinfo{author}{\bibfnamefont{G.}~\bibnamefont{Schehr}},
	\bibinfo{journal}{J. Stat. Mech.} \textbf{\bibinfo{volume}{2020}},
	\bibinfo{pages}{113203} (\bibinfo{year}{2020}).
	
	\bibitem[{\citenamefont{Radice}(2021)}]{radice2021one}
	\bibinfo{author}{\bibfnamefont{M.}~\bibnamefont{Radice}},
	\bibinfo{journal}{Phys. Rev. E} \textbf{\bibinfo{volume}{104}},
	\bibinfo{pages}{044126} (\bibinfo{year}{2021}).
	
	\bibitem[{\citenamefont{Santra et~al.}(2021)\citenamefont{Santra, Das, and
			Nath}}]{santra2021brownian}
	\bibinfo{author}{\bibfnamefont{I.}~\bibnamefont{Santra}},
	\bibinfo{author}{\bibfnamefont{S.}~\bibnamefont{Das}}, \bibnamefont{and}
	\bibinfo{author}{\bibfnamefont{S.~K.} \bibnamefont{Nath}},
	\bibinfo{journal}{J. Phys. A: Math. Theor.} \textbf{\bibinfo{volume}{54}},
	\bibinfo{pages}{334001} (\bibinfo{year}{2021}).
	
	\bibitem[{\citenamefont{Pal}(2015)}]{pal2015diffusion}
	\bibinfo{author}{\bibfnamefont{A.}~\bibnamefont{Pal}}, \bibinfo{journal}{Phys.
		Rev. E} \textbf{\bibinfo{volume}{91}}, \bibinfo{pages}{012113}
	(\bibinfo{year}{2015}).
	
	\bibitem[{\citenamefont{Ahmad et~al.}(2019)\citenamefont{Ahmad, Nayak, Bansal,
			Nandi, and Das}}]{ahmad2019first}
	\bibinfo{author}{\bibfnamefont{S.}~\bibnamefont{Ahmad}},
	\bibinfo{author}{\bibfnamefont{I.}~\bibnamefont{Nayak}},
	\bibinfo{author}{\bibfnamefont{A.}~\bibnamefont{Bansal}},
	\bibinfo{author}{\bibfnamefont{A.}~\bibnamefont{Nandi}}, \bibnamefont{and}
	\bibinfo{author}{\bibfnamefont{D.}~\bibnamefont{Das}},
	\bibinfo{journal}{Phys. Rev. E} \textbf{\bibinfo{volume}{99}},
	\bibinfo{pages}{022130} (\bibinfo{year}{2019}).
	
	\bibitem[{\citenamefont{Ray and Reuveni}(2020)}]{ray2020diffusion}
	\bibinfo{author}{\bibfnamefont{S.}~\bibnamefont{Ray}} \bibnamefont{and}
	\bibinfo{author}{\bibfnamefont{S.}~\bibnamefont{Reuveni}},
	\bibinfo{journal}{The Journal of chemical physics}
	\textbf{\bibinfo{volume}{152}} (\bibinfo{year}{2020}).
	
	\bibitem[{\citenamefont{Evans and Majumdar}(2018{\natexlab{b}})}]{evans2018run}
	\bibinfo{author}{\bibfnamefont{M.~R.} \bibnamefont{Evans}} \bibnamefont{and}
	\bibinfo{author}{\bibfnamefont{S.~N.} \bibnamefont{Majumdar}},
	\bibinfo{journal}{J. Phys. A: Math. Theor.} \textbf{\bibinfo{volume}{51}},
	\bibinfo{pages}{475003} (\bibinfo{year}{2018}{\natexlab{b}}).
	
	\bibitem[{\citenamefont{Santra et~al.}(2020)\citenamefont{Santra, Basu, and
			Sabhapandit}}]{santra2020run}
	\bibinfo{author}{\bibfnamefont{I.}~\bibnamefont{Santra}},
	\bibinfo{author}{\bibfnamefont{U.}~\bibnamefont{Basu}}, \bibnamefont{and}
	\bibinfo{author}{\bibfnamefont{S.}~\bibnamefont{Sabhapandit}},
	\bibinfo{journal}{J. Stat. Mech.} \textbf{\bibinfo{volume}{2020}},
	\bibinfo{pages}{113206} (\bibinfo{year}{2020}).
	
	\bibitem[{\citenamefont{Scacchi and Sharma}(2018)}]{scacchi2018mean}
	\bibinfo{author}{\bibfnamefont{A.}~\bibnamefont{Scacchi}} \bibnamefont{and}
	\bibinfo{author}{\bibfnamefont{A.}~\bibnamefont{Sharma}},
	\bibinfo{journal}{Mol. Phys.} \textbf{\bibinfo{volume}{116}},
	\bibinfo{pages}{460} (\bibinfo{year}{2018}).
	
	\bibitem[{\citenamefont{Kumar et~al.}(2020)\citenamefont{Kumar, Sadekar, and
			Basu}}]{kumar2020active}
	\bibinfo{author}{\bibfnamefont{V.}~\bibnamefont{Kumar}},
	\bibinfo{author}{\bibfnamefont{O.}~\bibnamefont{Sadekar}}, \bibnamefont{and}
	\bibinfo{author}{\bibfnamefont{U.}~\bibnamefont{Basu}},
	\bibinfo{journal}{Phys. Rev. E} \textbf{\bibinfo{volume}{102}},
	\bibinfo{pages}{052129} (\bibinfo{year}{2020}).
	
	\bibitem[{\citenamefont{De~Bruyne et~al.}(2022)\citenamefont{De~Bruyne,
			Majumdar, and Schehr}}]{de2022optimal}
	\bibinfo{author}{\bibfnamefont{B.}~\bibnamefont{De~Bruyne}},
	\bibinfo{author}{\bibfnamefont{S.~N.} \bibnamefont{Majumdar}},
	\bibnamefont{and} \bibinfo{author}{\bibfnamefont{G.}~\bibnamefont{Schehr}},
	\bibinfo{journal}{Phys. Rev. Lett.} \textbf{\bibinfo{volume}{128}},
	\bibinfo{pages}{200603} (\bibinfo{year}{2022}).
	
	\bibitem[{\citenamefont{Basu et~al.}(2019)\citenamefont{Basu, Kundu, and
			Pal}}]{basu2019symmetric}
	\bibinfo{author}{\bibfnamefont{U.}~\bibnamefont{Basu}},
	\bibinfo{author}{\bibfnamefont{A.}~\bibnamefont{Kundu}}, \bibnamefont{and}
	\bibinfo{author}{\bibfnamefont{A.}~\bibnamefont{Pal}},
	\bibinfo{journal}{Phys. Rev. E} \textbf{\bibinfo{volume}{100}},
	\bibinfo{pages}{032136} (\bibinfo{year}{2019}).
	
	\bibitem[{\citenamefont{Wang et~al.}(2021)\citenamefont{Wang, Cherstvy, Kantz,
			Metzler, and Sokolov}}]{wang2021time}
	\bibinfo{author}{\bibfnamefont{W.}~\bibnamefont{Wang}},
	\bibinfo{author}{\bibfnamefont{A.~G.} \bibnamefont{Cherstvy}},
	\bibinfo{author}{\bibfnamefont{H.}~\bibnamefont{Kantz}},
	\bibinfo{author}{\bibfnamefont{R.}~\bibnamefont{Metzler}}, \bibnamefont{and}
	\bibinfo{author}{\bibfnamefont{I.~M.} \bibnamefont{Sokolov}},
	\bibinfo{journal}{Phys. Rev. E} \textbf{\bibinfo{volume}{104}},
	\bibinfo{pages}{024105} (\bibinfo{year}{2021}).
	
	\bibitem[{\citenamefont{Vinod et~al.}(2022{\natexlab{a}})\citenamefont{Vinod,
			Cherstvy, Metzler, and Sokolov}}]{vinod2022time}
	\bibinfo{author}{\bibfnamefont{D.}~\bibnamefont{Vinod}},
	\bibinfo{author}{\bibfnamefont{A.~G.} \bibnamefont{Cherstvy}},
	\bibinfo{author}{\bibfnamefont{R.}~\bibnamefont{Metzler}}, \bibnamefont{and}
	\bibinfo{author}{\bibfnamefont{I.~M.} \bibnamefont{Sokolov}},
	\bibinfo{journal}{Phys. Rev. E} \textbf{\bibinfo{volume}{106}},
	\bibinfo{pages}{034137} (\bibinfo{year}{2022}{\natexlab{a}}).
	
	\bibitem[{\citenamefont{Vinod et~al.}(2022{\natexlab{b}})\citenamefont{Vinod,
			Cherstvy, Wang, Metzler, and Sokolov}}]{vinod2022nonergodicity}
	\bibinfo{author}{\bibfnamefont{D.}~\bibnamefont{Vinod}},
	\bibinfo{author}{\bibfnamefont{A.~G.} \bibnamefont{Cherstvy}},
	\bibinfo{author}{\bibfnamefont{W.}~\bibnamefont{Wang}},
	\bibinfo{author}{\bibfnamefont{R.}~\bibnamefont{Metzler}}, \bibnamefont{and}
	\bibinfo{author}{\bibfnamefont{I.~M.} \bibnamefont{Sokolov}},
	\bibinfo{journal}{Phys. Rev. E} \textbf{\bibinfo{volume}{105}},
	\bibinfo{pages}{L012106} (\bibinfo{year}{2022}{\natexlab{b}}).
	
	\bibitem[{\citenamefont{Wang et~al.}(2022)\citenamefont{Wang, Cherstvy,
			Metzler, and Sokolov}}]{wang2022restoring}
	\bibinfo{author}{\bibfnamefont{W.}~\bibnamefont{Wang}},
	\bibinfo{author}{\bibfnamefont{A.~G.} \bibnamefont{Cherstvy}},
	\bibinfo{author}{\bibfnamefont{R.}~\bibnamefont{Metzler}}, \bibnamefont{and}
	\bibinfo{author}{\bibfnamefont{I.~M.} \bibnamefont{Sokolov}},
	\bibinfo{journal}{Phys. Rev. Res.} \textbf{\bibinfo{volume}{4}},
	\bibinfo{pages}{013161} (\bibinfo{year}{2022}).
	
	\bibitem[{\citenamefont{Barkai et~al.}(2023)\citenamefont{Barkai,
			Flaquer-Galmes, and M{\'e}ndez}}]{barkai2023ergodic}
	\bibinfo{author}{\bibfnamefont{E.}~\bibnamefont{Barkai}},
	\bibinfo{author}{\bibfnamefont{R.}~\bibnamefont{Flaquer-Galmes}},
	\bibnamefont{and}
	\bibinfo{author}{\bibfnamefont{V.}~\bibnamefont{M{\'e}ndez}},
	\bibinfo{journal}{Phys. Rev. E} \textbf{\bibinfo{volume}{108}},
	\bibinfo{pages}{064102} (\bibinfo{year}{2023}).
	
	\bibitem[{\citenamefont{Reuveni}(2016)}]{PhysRevLett.116.170601}
	\bibinfo{author}{\bibfnamefont{S.}~\bibnamefont{Reuveni}},
	\bibinfo{journal}{Phys. Rev. Lett.} \textbf{\bibinfo{volume}{116}},
	\bibinfo{pages}{170601} (\bibinfo{year}{2016}).
	
	\bibitem[{\citenamefont{Pal and Reuveni}(2017)}]{pal2017first}
	\bibinfo{author}{\bibfnamefont{A.}~\bibnamefont{Pal}} \bibnamefont{and}
	\bibinfo{author}{\bibfnamefont{S.}~\bibnamefont{Reuveni}},
	\bibinfo{journal}{Phys. Rev. Lett.} \textbf{\bibinfo{volume}{118}},
	\bibinfo{pages}{030603} (\bibinfo{year}{2017}).
	
	\bibitem[{\citenamefont{Pal et~al.}(2022)\citenamefont{Pal, Kostinski, and
			Reuveni}}]{JPA2022.55.021001}
	\bibinfo{author}{\bibfnamefont{A.}~\bibnamefont{Pal}},
	\bibinfo{author}{\bibfnamefont{S.}~\bibnamefont{Kostinski}},
	\bibnamefont{and} \bibinfo{author}{\bibfnamefont{S.}~\bibnamefont{Reuveni}},
	\bibinfo{journal}{J. Phys. A: Math. Theor.} \textbf{\bibinfo{volume}{55}},
	\bibinfo{pages}{021001} (\bibinfo{year}{2022}).
	
	\bibitem[{\citenamefont{Fuchs et~al.}(2016)\citenamefont{Fuchs, Goldt, and
			Seifert}}]{fuchs2016stochastic}
	\bibinfo{author}{\bibfnamefont{J.}~\bibnamefont{Fuchs}},
	\bibinfo{author}{\bibfnamefont{S.}~\bibnamefont{Goldt}}, \bibnamefont{and}
	\bibinfo{author}{\bibfnamefont{U.}~\bibnamefont{Seifert}},
	\bibinfo{journal}{EPL (Europhys. Lett.)} \textbf{\bibinfo{volume}{113}},
	\bibinfo{pages}{60009} (\bibinfo{year}{2016}).
	
	\bibitem[{\citenamefont{Pal and Rahav}(2017)}]{pal2017integral}
	\bibinfo{author}{\bibfnamefont{A.}~\bibnamefont{Pal}} \bibnamefont{and}
	\bibinfo{author}{\bibfnamefont{S.}~\bibnamefont{Rahav}},
	\bibinfo{journal}{Phys. Rev. E} \textbf{\bibinfo{volume}{96}},
	\bibinfo{pages}{062135} (\bibinfo{year}{2017}).
	
	\bibitem[{\citenamefont{Gupta et~al.}(2020{\natexlab{b}})\citenamefont{Gupta,
			Plata, and Pal}}]{gupta2020work}
	\bibinfo{author}{\bibfnamefont{D.}~\bibnamefont{Gupta}},
	\bibinfo{author}{\bibfnamefont{C.~A.} \bibnamefont{Plata}}, \bibnamefont{and}
	\bibinfo{author}{\bibfnamefont{A.}~\bibnamefont{Pal}},
	\bibinfo{journal}{Phys. Rev. Lett.} \textbf{\bibinfo{volume}{124}},
	\bibinfo{pages}{110608} (\bibinfo{year}{2020}{\natexlab{b}}).
	
	\bibitem[{\citenamefont{Mori et~al.}(2023)\citenamefont{Mori, Olsen, and
			Krishnamurthy}}]{mori2023entropy}
	\bibinfo{author}{\bibfnamefont{F.}~\bibnamefont{Mori}},
	\bibinfo{author}{\bibfnamefont{K.~S.} \bibnamefont{Olsen}}, \bibnamefont{and}
	\bibinfo{author}{\bibfnamefont{S.}~\bibnamefont{Krishnamurthy}},
	\bibinfo{journal}{Phys. Rev. Res.} \textbf{\bibinfo{volume}{5}},
	\bibinfo{pages}{023103} (\bibinfo{year}{2023}).
	
	\bibitem[{\citenamefont{Godr\'eche and Luck}(2022)}]{JStatMech2022.063202}
	\bibinfo{author}{\bibfnamefont{C.}~\bibnamefont{Godr\'eche}} \bibnamefont{and}
	\bibinfo{author}{\bibfnamefont{J.-M.} \bibnamefont{Luck}},
	\bibinfo{journal}{J. Stat. Mech.} \textbf{\bibinfo{volume}{2022}},
	\bibinfo{pages}{063202} (\bibinfo{year}{2022}).
	
	\bibitem[{\citenamefont{Majumdar et~al.}(2022)\citenamefont{Majumdar, Mounaix,
			Sabhapandit, and Schehr}}]{JPA2022.55.034002}
	\bibinfo{author}{\bibfnamefont{S.~N.} \bibnamefont{Majumdar}},
	\bibinfo{author}{\bibfnamefont{P.}~\bibnamefont{Mounaix}},
	\bibinfo{author}{\bibfnamefont{S.}~\bibnamefont{Sabhapandit}},
	\bibnamefont{and} \bibinfo{author}{\bibfnamefont{G.}~\bibnamefont{Schehr}},
	\bibinfo{journal}{J. Phys. A: Math. Theor.} \textbf{\bibinfo{volume}{55}},
	\bibinfo{pages}{034002} (\bibinfo{year}{2022}).
	
	\bibitem[{\citenamefont{Kumar and Pal}(2023)}]{kumar2023universal}
	\bibinfo{author}{\bibfnamefont{A.}~\bibnamefont{Kumar}} \bibnamefont{and}
	\bibinfo{author}{\bibfnamefont{A.}~\bibnamefont{Pal}},
	\bibinfo{journal}{Phys. Rev. Lett.} \textbf{\bibinfo{volume}{130}},
	\bibinfo{pages}{157101} (\bibinfo{year}{2023}).
	
	\bibitem[{\citenamefont{De~Bruyne and Mori}(2023)}]{de2023resetting}
	\bibinfo{author}{\bibfnamefont{B.}~\bibnamefont{De~Bruyne}} \bibnamefont{and}
	\bibinfo{author}{\bibfnamefont{F.}~\bibnamefont{Mori}},
	\bibinfo{journal}{Phys. Rev. Res.} \textbf{\bibinfo{volume}{5}},
	\bibinfo{pages}{013122} (\bibinfo{year}{2023}).
	
	\bibitem[{\citenamefont{Gupta et~al.}(2014)\citenamefont{Gupta, Majumdar, and
			Schehr}}]{gupta2014fluctuating}
	\bibinfo{author}{\bibfnamefont{S.}~\bibnamefont{Gupta}},
	\bibinfo{author}{\bibfnamefont{S.~N.} \bibnamefont{Majumdar}},
	\bibnamefont{and} \bibinfo{author}{\bibfnamefont{G.}~\bibnamefont{Schehr}},
	\bibinfo{journal}{Phys. Rev. Lett.} \textbf{\bibinfo{volume}{112}},
	\bibinfo{pages}{220601} (\bibinfo{year}{2014}).
	
	\bibitem[{\citenamefont{Tal-Friedman et~al.}(2020)\citenamefont{Tal-Friedman,
			Pal, Sekhon, Reuveni, and Roichman}}]{tal2020experimental}
	\bibinfo{author}{\bibfnamefont{O.}~\bibnamefont{Tal-Friedman}},
	\bibinfo{author}{\bibfnamefont{A.}~\bibnamefont{Pal}},
	\bibinfo{author}{\bibfnamefont{A.}~\bibnamefont{Sekhon}},
	\bibinfo{author}{\bibfnamefont{S.}~\bibnamefont{Reuveni}}, \bibnamefont{and}
	\bibinfo{author}{\bibfnamefont{Y.}~\bibnamefont{Roichman}},
	\bibinfo{journal}{J. Phys. Chem. Lett.} \textbf{\bibinfo{volume}{11}},
	\bibinfo{pages}{7350} (\bibinfo{year}{2020}).
	
	\bibitem[{\citenamefont{Besga et~al.}(2020)\citenamefont{Besga, Bovon,
			Petrosyan, Majumdar, and Ciliberto}}]{besga2020optimal}
	\bibinfo{author}{\bibfnamefont{B.}~\bibnamefont{Besga}},
	\bibinfo{author}{\bibfnamefont{A.}~\bibnamefont{Bovon}},
	\bibinfo{author}{\bibfnamefont{A.}~\bibnamefont{Petrosyan}},
	\bibinfo{author}{\bibfnamefont{S.~N.} \bibnamefont{Majumdar}},
	\bibnamefont{and}
	\bibinfo{author}{\bibfnamefont{S.}~\bibnamefont{Ciliberto}},
	\bibinfo{journal}{Phys. Rev. Res.} \textbf{\bibinfo{volume}{2}},
	\bibinfo{pages}{032029} (\bibinfo{year}{2020}).
	
	\bibitem[{\citenamefont{Faisant et~al.}(2021)\citenamefont{Faisant, Besga,
			Petrosyan, Ciliberto, and Majumdar}}]{faisant2021optimal}
	\bibinfo{author}{\bibfnamefont{F.}~\bibnamefont{Faisant}},
	\bibinfo{author}{\bibfnamefont{B.}~\bibnamefont{Besga}},
	\bibinfo{author}{\bibfnamefont{A.}~\bibnamefont{Petrosyan}},
	\bibinfo{author}{\bibfnamefont{S.}~\bibnamefont{Ciliberto}},
	\bibnamefont{and} \bibinfo{author}{\bibfnamefont{S.~N.}
		\bibnamefont{Majumdar}}, \bibinfo{journal}{J. Stat. Mech.}
	\textbf{\bibinfo{volume}{2021}}, \bibinfo{pages}{113203}
	(\bibinfo{year}{2021}).
	
	\bibitem[{\citenamefont{Majumdar and Comtet}(2004)}]{majumdar2004exact}
	\bibinfo{author}{\bibfnamefont{S.~N.} \bibnamefont{Majumdar}} \bibnamefont{and}
	\bibinfo{author}{\bibfnamefont{A.}~\bibnamefont{Comtet}},
	\bibinfo{journal}{Phys. Rev. Lett.} \textbf{\bibinfo{volume}{92}},
	\bibinfo{pages}{225501} (\bibinfo{year}{2004}).
	
	\bibitem[{\citenamefont{Majumdar and Comtet}(2005)}]{majumdar2005airy}
	\bibinfo{author}{\bibfnamefont{S.~N.} \bibnamefont{Majumdar}} \bibnamefont{and}
	\bibinfo{author}{\bibfnamefont{A.}~\bibnamefont{Comtet}},
	\bibinfo{journal}{J. Stat. Phys.} \textbf{\bibinfo{volume}{119}},
	\bibinfo{pages}{777} (\bibinfo{year}{2005}).
	
\end{thebibliography}

\end{document}